\documentclass{article}

\usepackage{amssymb,amsmath,mathtools,xcolor,graphicx,xspace,colortbl,ragged2e,rotating} %
\usepackage{amsmath}  
\usepackage{amssymb,amsmath,mathtools,xcolor,graphicx,xspace,colortbl,ragged2e,rotating}  
\usepackage{boxedminipage}  
\usepackage{graphics}  
\usepackage{graphicx}  
\usepackage{ragged2e}  
\usepackage{wrapfig}  
\usepackage{xcolor}  
\graphicspath{{publicatie_graphics/}{publicatie_tcache/}{publicatie_gcache/}}
\DeclareGraphicsExtensions{.pdf,.eps,.ps,.png,.jpg,.jpeg}
\graphicspath{{publicatie_graphics/}{publicatie_tcache/}{publicatie_gcache/}}
\DeclareGraphicsExtensions{.pdf,.eps,.ps,.png,.jpg,.jpeg}
\graphicspath{{publicatie_graphics/}{publicatie_tcache/}{publicatie_gcache/}}
\DeclareGraphicsExtensions{.pdf,.eps,.ps,.png,.jpg,.jpeg}
\graphicspath{{publicatie_graphics/}{publicatie_tcache/}{publicatie_gcache/}}
\DeclareGraphicsExtensions{.pdf,.eps,.ps,.png,.jpg,.jpeg}
\graphicspath{{publicatie_graphics/}{publicatie_tcache/}{publicatie_gcache/}}
\DeclareGraphicsExtensions{.pdf,.eps,.ps,.png,.jpg,.jpeg}
\graphicspath{{publicatie_graphics/}{publicatie_tcache/}{publicatie_gcache/}}
\DeclareGraphicsExtensions{.pdf,.eps,.ps,.png,.jpg,.jpeg}
\graphicspath{{publicatie_graphics/}{publicatie_tcache/}{publicatie_gcache/}}
\DeclareGraphicsExtensions{.pdf,.eps,.ps,.png,.jpg,.jpeg}
\graphicspath{{publicatie_graphics/}{publicatie_tcache/}{publicatie_gcache/}}
\DeclareGraphicsExtensions{.pdf,.eps,.ps,.png,.jpg,.jpeg}
\graphicspath{{publicatie_graphics/}{publicatie_tcache/}{publicatie_gcache/}}
\DeclareGraphicsExtensions{.pdf,.eps,.ps,.png,.jpg,.jpeg}
\graphicspath{{publicatie_graphics/}{publicatie_tcache/}{publicatie_gcache/}}
\DeclareGraphicsExtensions{.pdf,.eps,.ps,.png,.jpg,.jpeg}
\graphicspath{{publicatie_graphics/}{publicatie_tcache/}{publicatie_gcache/}}
\DeclareGraphicsExtensions{.pdf,.eps,.ps,.png,.jpg,.jpeg}
\graphicspath{{publicatie_graphics/}{publicatie_tcache/}{publicatie_gcache/}}
\DeclareGraphicsExtensions{.pdf,.eps,.ps,.png,.jpg,.jpeg}
\graphicspath{{publicatie_graphics/}{publicatie_tcache/}{publicatie_gcache/}}
\DeclareGraphicsExtensions{.pdf,.eps,.ps,.png,.jpg,.jpeg}

\begin{document}

\section*{General weak segregation theory with an application to monodisperse semi-flexible diblock copolymers}
\ \ \ \ \ P.M.
Jager \footnote{EY, Centre for Tax \& Legal Knowledge, Boompjes 258, P.O. Box 2295, 3011 XZ Rotterdam, The Netherlands, menno.jager@nl.ey.com}, W.J. Briels \footnote{MESA+ Institute for Nanotechnology, University of Twente, P.O. Box 217, 7500 AE Enschede, The Netherlands, w.j.briels@utwente.nl}\textsuperscript{,}\footnote{Forschungszentrum Jülich, IBI-4, D-52425 Jülich, Germany, w.briels@fz-juelich.de} and J.J.M. Slot \footnote{Department of Mathematics and Computer Science, Applied Analysis, Eindhoven University of Technology, PO Box 513, 5600 MB Eindhoven, The Netherlands; j.j.m.slot@tue.nl \par \textsuperscript{ *}Corresponding author}\textsuperscript{,*}

\section*{Abstract }

A general theory has been developed for a polydisperse semi-flexible multi-block copolymer melt. Using the Bawendi-Freed approach to model semi-flexible chains an expression for the Landau free energy is derived in the
weak segregation regime which includes the density and orientation order-parameters. The orientation order-parameter is described in the smectic phase and in more complicated structures such as the hexagonal phase. The Landau free energy contains contributions of two kinds of interactions. The first
kind is the Flory-Huggins interaction which describes the incompatibility of chemically different blocks and may induce microphase separation. The second
kind is the Maier-Saupe interaction which may induce nematic ordening. In the framework of the weak segragation limit the Landau theory allows to predict
phase structures in the melt as a function of the composition, persistence length and the strength of the Flory-Huggins and Maier-Saupe interaction. The general theory is applied to a simple system of monodisperse semi-flexible diblock copolymers. In several phase diagrams a number of possible phase structures is predicted such as the bcc, hexagonal, smectic-A, smectic-C and nematic phase. The influence
of the Maier-Saupe interaction on the microphase structure is thoroughly discussed. 

\section{Introduction }
In the past decades much attention has been paid to the phase behaviour of diblock copolymers and other kinds of block copolymers both theoretically (Ref. \cite{Slot} till \cite{Ohta})
and experimentally
(Ref. \cite{Bradley1} till \cite{Bradley4}). This
phase behaviour determines to a large extent the mechanical, optical and electrical properties of these materials. For example block copolymers are applied
in thermoplastic elastomers which form an interesting class of synthetic materials. Their industrial interest primarily stems from the fact that they combine
the mechanical properties of elastomers with the processability of thermoplastics, in other words they share their elasticity with rubbers but in contrast
with the latter they can be re-processed simply by re-melting. At least one kind of block should be able to crystallize sufficiently above room temperature,
while the non-crystalizable blocks provide the elasticity to the system. In these systems the crystallizable blocks form crosslinks between chains so that
a network is formed. Without crosslinks the chains can move freely with respect to each other so that the system behaves as a liquid with a high viscosity.
However due to the partial crystallization the freedom of each chain to move is restricted. 

In this paper we develop a general theory of a polydisperse semi-flexible multi-block copolymer melt which is an extension of the theory in Ref. \cite{Slot}. The theory is made very general so that it can not only be applied to simple systems but also more complex systems which are more realistic. In this paper it is applied to the simplest system of monodisperse diblocks, but the application can be extended to more complex systems such as monodisperse triblocks and polydisperse diblocks. The general theory describes a melt which is an arbitrary mixture of multi-block copolymers. The main parameters are the mixture composition determined by the number of chains $n_{s}$ for each chain species $s$, the chain composition of species $s$ determined by Ising-like variables $\sigma _{s}^{\alpha }(l)$ such that $\sigma_{s}^{\alpha }(l)\equiv 1$\textbf{\ }when segment\textbf{\ }$l$ is in a block of kind $\alpha $ and $\sigma _{s}^{\alpha }(l)\equiv 0$ otherwise, the bending stiffness $\lambda_{\alpha}$ of a block of kind $\alpha$, the strength of the Flory-Huggins $\tilde{\chi}_{\alpha\beta}$ and Maier-Saupe $\omega_{\alpha\beta}$ interaction between blocks of kind $\alpha$ and $\beta$, the wavenumber $q_{*}$ of a phase structure, the symmetry properties  of a phase structure defined by the set of wave vectors $H=\{\pm\underline{q}_1, \pm\underline{q}_2, ...\}$ with a fixed wavenumber $q_{1}=q_{2}=...=q_{*}$ and the strength of the density $\Psi_{\alpha}$ and orientation $\Upsilon_{\alpha}$ order-parameter of block $\alpha$. The general theory is applied to a simple system of monodisperse semi-flexible diblocks to calculate the complete phase diagram. In this way the effect of stiffness on the phase behaviour of block copolymers is investigated.   

The effect of stiffness on monodisperse semi-flexible diblocks has also been investigated in Ref. \cite{Reenders}, \cite{Friedel}, \cite{Singh}, \cite{Holyst} and  \cite{Matsen}. In Ref. \cite{Reenders} and \cite{Holyst} a rod-coil system is investigated and in Ref. \cite{Friedel}, \cite{Singh} and \cite{Matsen} the diblock is semi-flexible. In all these papers it appeared
that both the spinodal $\chi _{s} L$ and $\omega _{s} L$ are lowered when one or both blocks in the diblock copolymer are made stiffer. The semi-flexiblity is described in different ways in the above mentioned papers. In the general theory in the next section of this paper the Bawendi-Freed version of the freely rotating chain model is applied to describe semi-flexibility, see also \cite{Lagowski},
\cite{Bixon} and \cite{Bawendi}.
In Ref. \cite{Friedel} the Kratky-Porod model is applied and Ref. \cite{Matsen}
applies the Sait{\^o} model introduced in Ref. \cite{Saito}. In Singh's paper
\cite{Singh} the freely jointed chain model is used to describe semi-flexible
chains. In this discrete model rods are connected to each other which can be rotated freely with respect to the connection points between them. The rod
length is a measure of the stiffness. The rod-part or semi-flexible blocks in the diblock chain in Ref. \cite{Reenders}, \cite{Friedel}, \cite{Singh}, \cite{Holyst} and  \cite{Matsen} may have a certain orientation in a microphase or nematic
phase. In Ref. \cite{Friedel} and \cite{Matsen}
the orientation is ignored, but in Ref. \cite{Reenders}
and \cite{Singh} the orientation tensor has the form,
\begin{equation}Q_{i j}^{\alpha } (\underline{q}) =(n_{i}^{\alpha } n_{j}^{\alpha } -\frac{1}{3} \delta _{i j}) Q^{\alpha } (\underline{q}) \qquad \text{with}\qquad
\alpha = A, B, \label{form}
\end{equation}in which it has also been assumed that the orientation vectors $\underline{n}^{A}$ and $\underline{n}^{B}$ of the A- and B-block are equal, so $\underline{n}^{A} =\underline{n}^{B} =\underline{n}$. In the paper of Holyst \cite{Holyst} this form has
not been used. The six components of the orientation tensor of the rod-part $Q_{i j} (\underline{q})$ with $i j =x x$, $y y$, $z z$, $x y$, $x z$ and $y z$ are treated as independent variables in the expression of the Landau free energy. In both Ref. \cite{Reenders}
and \cite{Holyst} the orientation of the coil-part in the rod-coil diblock
copolymer is neglected.

In Ref. \cite{Friedel} till \cite{Hernandez} the weak segregation theory is restricted to the calculation
of the spinodal line, but in Ref. \cite{Leibler}, \cite{Reenders} and this paper the complete phase diagram
is calculated. Reenders et. al. \cite{Reenders} calculated the phase diagrams
of monodisperse rod-coil diblocks which displayed the various phases such as the bcc, hexagonal, lamellar, nematic, smectic-A and smectic-C state. As mentioned before
the orientation tensor of the rod part has the same form as Eq. (\ref{form}). Reenders et. al. \cite{Reenders}
made the ansatz that the wave vector $\underline{q}$ of the orientation tensor $Q_{i j}^{\alpha } (\underline{q})$ can only be zero, so that there is only global nematic ordering. A consequence of this ansatz is that a possible orientation is negligibly
small if $\chi  >\chi _{s}$ and $\omega  =0$. Therefore the tensor $Q_{i j} (\underline{q}) =Q_{i j} (\underline{0})$ is neglected in the part of the phase diagram in which $\omega  <\omega _{s}$. In the description of the smectic state a posible local orientation $Q_{i j} (\underline{q} \neq \underline{0})$ is not taken into account. However, in Ref. \cite{Singh},
\cite{Holyst} and this paper
a possible contribution of orientation tensors with nonzero $\underline{q}$ are not excluded if there is microphase separation. In a certain microphase structure the wave vectors in the orientation tensor $Q_{i j}^{\alpha } (\underline{q})$ must be the same as those in the scalar order-parameter $\Psi  (\underline{q})$, because in that case the symmetry of the structure is conserved. In this way the orientation may play a more important role when $\chi  >\chi _{s}$ and $\omega  =0$. However, in addition to the set of local orientions $Q_{i j}^{\alpha } (\underline{q})$ global orientions $Q_{i j}^{\alpha } (\underline{0})$ in the same direction are still possible in a certain microphase. 

In Ref. \cite{Slot}
till \cite{Aliev2} the phase behaviour is described by the weak segregation
theory which is only valid just below to the critical temperature.  Further away from the phase transition
at much smaller temperatures the phase structure becomes more markedly. Here the weak segregation approximation is not valid. Other approaches
must be applied in the intermediate and strong segregation regime such as the self-consistent field theory in Ref. \cite{Matsen}
till \cite{Osipov} or computer simulations in for example Ref. \cite{Osipov2}
till \cite{Ohta}. In weak segregation some results of the phase behaviour
found in Ref. \cite{Slot} till \cite{Aliev2}
are observed in experiments, see also Ref. \cite{Bradley1} till \cite{Bradley4}.
In this paper we restrict to the weak segregation approximation. In the past this approximation has been applied in several kinds of block copolymers systems in
which bending stiffness of blocks and/or polydispersity in block length are included. In Ref. \cite{JohnFriedel} till \cite{Aliev2}
the effect of polydispersity is investigated for different kinds of block copolymers, but the work of Ref. \cite{Leibler} till \cite{Hernandez} is restricted to monodisperse systems in the framework of the weak segregation limit. The monodisperse systems in Ref. \cite{Leibler} till \cite{Wang}
are diblock copolymers. Monodisperse triblock copolymers are considered in Ref. \cite{Werner} and Ref. \cite{Aksimentiev}. In Ref. \cite{Hernandez} a melt of monodisperse side-chain liquid-crystalline
polymers is investigated. 

In this paper we show and discuss the results
of the calculated phase diagrams of monodisperse semi-flexible diblocks.
Before discussing the results we explain how we developed a general theory of a polydisperse semi-flexible multi-block copolymer melt which is, as mentioned earlier, an extension of the theory in Ref. \cite{Slot}.
In that paper the polydisperse multi-block copolymers are totally flexible. In this paper the chains are made semi-flexible by adding a persistence length or bending stiffness $\lambda _{a}$ to each block $\alpha $. If $\lambda _{a} \rightarrow 0$ block $a$ becomes totally flexible and if $\lambda _{a} \rightarrow \infty $\ it can be regarded as a rigid rod. Because of the bending stiffness in the general theory the blocks may become oriented due to global nematic ordening caused by the Maier-Saupe interaction. The Flory-Huggins interaction might cause a space dependent
alignment in a microphase state in addition to a space dependent density order-parameter. The theory in Ref. \cite{Slot}
is further extended by taking this additional orientation into account by means of the tensor $Q_{i j}^{\alpha }$ with $i ,j =x ,y ,z$ and $\alpha  =a ,b ,\ldots $. This extension is realised by writing the Landau free energy as a power series expansion of both density order-parameters $\Psi ^{\alpha }$ and orientation tensors $Q_{i j}^{\alpha }$. 

To describe the semi-flexible chains we apply the Bawendi-Freed
version of the freely rotating chain model see \cite{Lagowski}, \cite{Bixon}
and \cite{Bawendi}. In this continuous chain model the chain length $L$ may fluctuate and is only on average constant. Therefore we use a special parameter $l$ to label the monomers. It is the average distance between a certain monomer and one of the chain ends measured along the contour. If
$l$ would be the actual contour length, then the tangent vector $\underline{u} (l) =\frac{d \underline{r} (l)}{d l}$ must be a unit vector. Because $l$ is the average distance the squared length $u^{2} (l)$ is only on average equal to one, $ \langle u^{2} (l) \rangle  =1$. This less restrictive condition makes the calculation of the single-chain correlation functions possible which can be found in Supplemental material I. Here the Bawendi-Freed approach is applied and explained in more detail. This also makes the explicit derivation of the Landau free energy possible, because it depends on the single-chain correlation functions. This derivation can be found in section "The model" and Appendix A. In Appendix B the general theory of a melt of polydisperse semi-flexible block copolymers is applied to a simpler system of monodisperse semi-flexible diblock copolymers. Here the second order terms in the Landau free energy are written in matrix form and can be rewritten in terms of eigenvalues and eigenvectors. The three dimensional symmetric matrix has three positive eigenvalues in the disordered state, but just below the critical
temperature at the order-disorder phase transition one of the three eigenvalues becomes negative. The corresponding eigenvector is called the primary eigenvector.
The other eigenvectors are secondary eigenvectors. By minimizing the Landau free energy with respect to the secondary eigenvectors the Landau free energy
can be written as a power series expansion in only the primary eigenvector. This expression is used to minimize the free energy in the bcc, hexagonal, lamellar,
nematic, smectic-A and smectic-C state. The state with the lowest minimum is the most probable state. In this way the phase diagram is calculated. As mentioned earlier in the expression of the free energy a space dependent orientation tensor is included, because it is necessary
to take into account the possibility of local alignment in a certain microphase structure. In Appendix C it is explained how a space dependent orientation is described in a smectic state and in more complicated microphase structures such as the hexagonal state. The orientation tensor of the hexagonal structure is described as a superposition of three orientation tensors of smectic-A states
in such a way that the symmetry of the hexagonal strucure is conserved. In the same way the bcc structure is a superposition of six orientation tensors
of smectic-A states.

\section{The model}
In this section we develop a general theory of a polydisperse semi-flexible multi-block copolymer melt. Certain steps in the derivation of final form of the the Landau free energy given by Eq. (\ref{vrijeenergie}) are not described thorougly. For more details we refer to Appendix A and Supplemental material I till IV. The general theory in this paper is based to a large extent on earlier work which can be found in Ref. \cite{SlotJager}.

To describe a general melt of semi-flexible multi-block copolymers we will
employ the typical \textit{coarse graining} one usually encounters in
polymer physics \cite{LifGroKho}. Consider a melt of $n_{c}$ copolymer
chains in a volume $V$. As these chains consist of arbitrary sequences
(blocks) of up-to $M$ chemically different types of monomers, the number of
possible chains is astronomically large. To denote the various different
species of chains present in the system, we will use the label $s$. Each
chain belonging to species $s$, of which there are $n_{s}$ present, will
consist of $N_{s}+1$ monomers also referred to as segments, connected by $%
N_{s}$ deformable bonds of average length $a$ having fixed bond angles
between them. The idea being that each of the $M$ chemically different
blocks of segments will have their own fixed angle between subsequent bonds.
We will label these $M$ chemically different segments by Greek lowercase
symbols $\alpha $\textbf{, }$\beta $ etc. running from $1$ to $M$. In the
sequal we will use a continuous notation, i.e. chains will be represented by
continuous curves obtained via a proper continuum limit ($%
N_{s}\longrightarrow \infty $, $a\longrightarrow 0$ such that $L_{s}\equiv
N_{s}a$ remains constant etc.). To specify a given chain species $s$,
Ising-like variables $\sigma _{s}^{\alpha }(l)$ will be introduced with $%
\alpha =1,...,M$ and $l\,\in \,[0,L_{s}]$ in such a way that $\sigma
_{s}^{\alpha }(l)\equiv 1$\textbf{\ }when segment\textbf{\ }$l$ is of type $%
\alpha $ and $\sigma _{s}^{\alpha }(l)\equiv 0$ otherwise. The continuous parameter $l$ can not only be regarded as a parameter to label segments, but also as a measure of distance between a certain segment and one of the chain ends measured along the contour. According to Supplemental material I $l$ is not the actual contour length, but the average distance, because in the Bawendi-Freed model the segments are connected by springs. The conformations
of the $n_{s}$ chains belonging to species $s$ will be specified by the
positions of the corresponding segments that make up these chains and the
set of tangent vectors along the chains, i.e. the set of curves $\{%
\underline{R}_{m}^{s}(l)\,\mid \,0\leqslant l\leqslant L_{s},\textbf{ }m=1,...,n_{s}\}_{s}$ defined with respect to some origin $O$ in $V$ , and the set
of tangent vectors to these curves $\{\underline{u}_{m}^{s}(l)\equiv 
\underline{\dot{R}}_{m}^{s}(l)\equiv \frac{\partial \underline{R}_{m}^{s}(l)%
}{\partial l}\,\mid \,0\leqslant l\leqslant L_{s},\textbf{ }m=1,...,n_{s}\}_{s}$. As the bonds are
deformable these tangent vectors will not be of unit-length at every point
along the contours of the chains. Nevertheless, it will turn out that they
will be unit vectors in an averaged sense, as will become clear later-on in Supplemental material I.
The total set $\{\underline{R}_{m}^{s}(l)\,,\underline{u}_{m}^{s}(l)\mid
\,0\leqslant l\leqslant L_{s}\}_{sm}$ defines a configurational \textit{%
micro-state} of the whole system. Such a micro-state will be denoted by $%
\gamma $. A variable $G$ which is a function of these micro-states, i.e. a
so-called \textit{state-variable}, will be written as $\hat{G}\equiv
G(\gamma )$. Examples of important state-variables which we will need
later-on are the \textit{microscopic }$\alpha $-\textit{segment density} $%
\hat{\rho}^{\alpha }(\underline{x})$, defined for each $\underline{x}\in V$
by, 
\begin{equation}
\hat{\rho}^{\alpha }(\underline{x})\equiv
\sum_{sm}\int_{0}^{L_{s}}dl\,\sigma _{s}^{\alpha }(l)\,\delta (\underline{x}-%
\underline{R}_{m}^{s}(l))  \label{1}
\end{equation}%
and the \textit{overall microscopic segment density} $\hat{\rho}(\underline{x%
})$ through, 
\begin{equation}
\hat{\rho}(\underline{x})\equiv \sum_{\alpha }\hat{\rho}^{\alpha }(%
\underline{x})=\sum_{sm}\int_{0}^{L_{s}}dl\,\delta (\underline{x}-\underline{%
R}_{m}^{s}(l)).  \label{2}
\end{equation}%
By integrating these densities over $V$ we obtain respectively the total
number of $\alpha $-segments $N^{\alpha }$ and the overall number of
segments $N$ in the system. This last number can be either written as $%
\sum_{\alpha }N^{\alpha }$ or as $\sum_{s}n_{s}N_{s}$. Thus the fraction of $%
\alpha $-segments is given by $f^{\alpha }\equiv \frac{N^{\alpha }}{N}$.
Without loss of generality we will choose our length-scale in such a way
that each segment has a unit volume and therefore that $N\equiv V$. In that
case it follows that, 
\begin{equation}
\frac{1}{V}\int_{V}d^{3}x\,\hat{\rho}(\underline{x})\equiv 1  \label{3}
\end{equation}%
and that $f^{\alpha }$ can be written as 
\begin{equation}
f^{\alpha }\equiv \frac{1}{V}\int_{V}d^{3}x\,\hat{\rho}^{\alpha }(\underline{%
x}).  \label{4}
\end{equation}%
The description of semi-flexibility and the tendency to locally align will
require the introduction of the \textit{microscopic }$\alpha $\textit{%
-segment orientation-tensor density} $\underline{\underline{\hat{S}}}%
^{\alpha }(\underline{x})$, defined by, 
\begin{equation}
\underline{\underline{\hat{S}}}^{\alpha }(\underline{x})\equiv
\sum_{sm}\int_{0}^{L_{s}}dl\,\sigma _{s}^{\alpha }(l)\,\underline{u}%
_{m}^{s}(l)\,\underline{u}_{m}^{s}(l)\,\delta (\underline{x}-\underline{R}%
_{m}^{s}(l)).  \label{5}
\end{equation}%
In order to account approximately for the effect of excluded volume due to
the repulsive nature of the intra-chain and inter-chain potentials at short
distances, we will assume that the system is \textit{incompressible }, i.e.
that the overall microscopic segment density is not only equal to $1$
''globally'', as in (\ref{3}), but also ''locally'', that is $\hat{\rho}(%
\underline{x})\equiv 1$, $\forall \underline{x}\in V$. As our aim is to
derive a Landau free energy for this copolymer melt of $M$ \textit{%
quasi-components }\cite{PanyKu1}, we need to define a set of $2M$ \textit{%
order-parameters} or actually $2M$ \textit{order- parameter fields} to
describe the possible \textit{inhomogeneous and anisotropic phases} of the
system and to be able to calculate their free energy. These order-parameter
fields can be defined by coarse graining the following set of \textit{%
microscopic order-parameter fields}, 
\begin{equation}
\hat{\psi}^{\alpha }(\underline{x})\equiv \hat{\rho}^{\alpha }(\underline{x}%
)-f^{\alpha }\qquad (\alpha =1,...,M)  \label{8}
\end{equation}%
and 
\begin{equation}
\underline{\underline{\hat{Q}}}\,^{\alpha }(\underline{x})\equiv \underline{%
\underline{\hat{S}}}^{\alpha }(\underline{x})-\frac{1}{3}\hat{\rho}^{\alpha
}(\underline{x})\underline{\underline{\,I}}\qquad (\alpha =1,...,M).
\label{9}
\end{equation}%
It is clear that due to the incompressibility assumption only $M-1$ of the
scalar fields will be independent, as it follows that, 
\begin{equation}
\sum_{\alpha }\hat{\psi}^{\alpha }(\underline{x})\equiv 0.  \label{10}
\end{equation}%
The interactions between the various segments in this copolymer melt can be
described in terms of these microscopic order-parameter fields. This can be
shown in the following way. Under the assumption that segment- segment
interactions are pairwise additive, the total interaction energy $\hat{W}$
of the system will be of the form, 
\begin{eqnarray}
\hat{W} &\equiv &\frac{1}{2}\sum_{\alpha \beta }\sum_{sm}\sum_{s^{\prime
}m^{\prime }}\int_{0}^{L_{s}}dl\int_{0}^{L_{s^{\prime }}}dl^{\prime
}\,\sigma _{s}^{\alpha }(l)\,\sigma _{s^{\prime }}^{\beta }(l^{\prime
})\,\times  \notag \\
&&\times w_{\alpha \beta }(\underline{R}_{m}^{s}(l)-\underline{R}_{m^{\prime
}}^{s^{\prime }}(l^{\prime })\,;\underline{u}_{m}^{s}(l)\cdot \,\underline{u}%
_{m^{\prime }}^{s^{\prime }}(l^{\prime })).  \label{11}
\end{eqnarray}%
where $w_{\alpha \beta }(\underline{x}\,;\underline{\,u}\cdot \underline{u}%
^{\prime })$ is the interaction potential between a segment of type $\alpha $
and a segment of type $\beta $, which is assumed to be short-ranged in
space, i.e. $w_{\alpha \beta }(\underline{x}\,;\underline{\,u}\cdot 
\underline{u}^{\prime })\simeq \widetilde{w}_{\alpha \beta }(\underline{\,u}%
\cdot \underline{u}^{\prime })\,\delta (\underline{x})$ and $\widetilde{w}%
_{\alpha \beta }(\underline{\,u}\cdot \underline{u}^{\prime })$ can be
expanded in the following way, 
\begin{equation}
\widetilde{w}_{\alpha \beta }(\underline{\,u}\cdot \underline{u}^{\prime
})=\varepsilon _{\alpha \beta }-\omega _{\alpha \beta }\,(\underline{\,u}%
\cdot \underline{u}^{\prime })^{2}+...  \label{12}
\end{equation}%
This expansion does not contain the term $\underline{\,u}\cdot \underline{u}%
^{\prime }$ or any odd power of it for that matter because of the fore-aft
symmetry of the segments. Although formally $\hat{W}$ contains
''self-energy''\ terms, i.e. terms with $\alpha =\beta $, $s=s^{\prime }$, $%
m=m^{\prime }$ and $l=l^{\prime }$, we will not bother to explicitly exclude
them in the notation used in (\ref{11}). By substituting (\ref{12}) into (%
\ref{11}) it is easy to see that $\hat{W}$ can be written in terms of the
specific microscopic segment densities $\hat{\rho}^{\alpha }(\underline{x})$
(\ref{1}) and $\underline{\underline{\hat{S}}}^{\alpha }(\underline{x})$ (%
\ref{5}) as, 
\begin{equation}
\hat{W}\simeq \frac{1}{2}\sum_{\alpha \beta }\varepsilon _{\alpha \beta
}\int_{V}d^{3}x\,\,\hat{\rho}^{\alpha }(\underline{x})\,\hat{\rho}^{\beta }(%
\underline{x})-\frac{1}{2}\sum_{\alpha \beta }\omega _{\alpha \beta
}\int_{V}d^{3}x\,\underline{\underline{\hat{S}}}^{\alpha }(\underline{x}):%
\underline{\underline{\hat{S}}}^{\beta }(\underline{x}).\,  \label{13}
\end{equation}%
By eliminating one of the $\hat{\rho}$ 's, say $\hat{\rho}^{M}$, from (\ref%
{13}) using $\sum_{\alpha }\hat{\rho}^{\alpha }(\underline{x})\equiv 1$ and
substituting $\hat{\rho}^{\alpha }(\underline{x})=$ $\hat{\psi}^{\alpha }(%
\underline{x})+f^{\alpha }$ and $\underline{\underline{\hat{S}}}^{\alpha }(%
\underline{x})=\underline{\underline{\hat{Q}}}\,^{\alpha }(\underline{x})+%
\frac{1}{3}\hat{\rho}^{\alpha }(\underline{x})\underline{\underline{\,I}}$,
one ends up with, 
\begin{gather}
\hat{W}\simeq \frac{1}{2}\,\sum_{\alpha \beta }\text{\/}^{^{\prime
}}E_{\alpha \beta }\int_{V}d^{3}x\,\hat{\psi}^{\alpha }(\underline{x})\hat{%
\psi}^{\beta }(\underline{x})-\frac{1}{2}\sum_{\alpha \beta }\omega _{\alpha
\beta }\int_{V}d^{3}x\,\underline{\underline{\hat{Q}}}^{\alpha }(\underline{x%
}):\underline{\underline{\hat{Q}}}^{\beta }(\underline{x})  \notag \\
-\frac{1}{3}\sum_{\alpha \beta }\omega _{\alpha \beta }\int_{V}d^{3}x\,(\hat{%
\psi}^{\alpha }(\underline{x})+f^{\alpha })Tr\underline{\underline{\hat{Q}}}%
^{\beta }(\underline{x})  \label{14}
\end{gather}%
with $E_{\alpha \beta }\equiv \varepsilon _{\alpha \beta }^{\prime
}-\varepsilon _{\alpha M}^{\prime }-\varepsilon _{\beta M}^{\prime
}+\varepsilon _{MM}^{\prime }$ and $\varepsilon _{\alpha \beta }^{\prime
}=\varepsilon _{\alpha \beta }-\frac{1}{3}\omega _{\alpha \beta }$. The
prime in the first term of $\hat{W}$ implies that both sums run from $1$ to 
$M-1$. In terms of the set of Flory $\chi $-parameters \cite{Flory} between
the different segments, i.e.,
\begin{equation}
\chi _{\alpha \beta }\equiv \varepsilon _{\alpha \beta }-\frac{\varepsilon
_{\alpha \alpha }+\varepsilon _{\beta \beta }}{2}\qquad \text{with}\qquad
\chi _{\alpha \alpha }=0\quad ,\forall \alpha ,  \label{15}
\end{equation}%
this $E_{\alpha \beta }$ can be written as, 
\begin{equation}
E_{\alpha \beta }=\chi _{\alpha \beta }-\chi _{\alpha M}-\chi _{\beta
M}\equiv -2\,\tilde{\chi}_{\alpha \beta }  \label{16}
\end{equation}%
and therefore finally $\hat{W}$ becomes, 
\begin{gather}
\hat{W}\simeq -\sum_{\alpha \beta }\text{\/}^{^{\prime }}\,\tilde{\chi}%
_{\alpha \beta }\int_{V}d^{3}x\,\hat{\psi}^{\alpha }(\underline{x})\hat{\psi}%
^{\beta }(\underline{x})-\frac{1}{2}\sum_{\alpha \beta }\omega _{\alpha
\beta }\int_{V}d^{3}x\,\underline{\underline{\hat{Q}}}^{\alpha }(\underline{x%
}):\underline{\underline{\hat{Q}}}^{\beta }(\underline{x})  \notag \\
-\frac{1}{3}\sum_{\alpha \beta }\omega _{\alpha \beta }\int_{V}d^{3}x\,(\hat{%
\psi}^{\alpha }(\underline{x})+f^{\alpha })Tr\underline{\underline{\hat{Q}}}%
^{\beta }(\underline{x}).  \label{17}
\end{gather}%
For the binary case ($M=2$), the only remaining $\tilde{\chi}$-parameter, $\,%
\tilde{\chi}_{11}$, reduces to the more familiar $\chi _{12}$. With this
interaction energy of the copolymer melt we can formulate the system's 
\textit{partition function} $Z$, but for this it is necessary to specify the 
\textit{unnormalized statistical weight} $e^{-\hat{H}_{0}}$ (we use units
such that $k_{B}T=1$) of the system in absence of these interactions. As we
want to allow chains locally to have an arbitrary degree of flexibility, say
ranging from fully flexible to rigid-rod like behaviour, we will describe
the \textit{unperturbed }semi-flexible melt by an ensemble of locally
persistent Gaussian chains in the spirit of the Bawendi-Freed approach \cite%
{Bixon}, \cite{Bawendi}. The \textit{unperturbed Hamiltonian} $\hat{H}_{0}$
in this approach is given by,
\begin{eqnarray}
\hat{H}_{0} &\equiv &\frac{3}{4\,}\sum_{sm}\left\{ \left[ \underline{u}%
_{m}^{s}(0)\,\right] ^{2}+\left[ \underline{u}_{m}^{s}(L_{s})\,\right]
^{2}\right\} +  \notag \\
&&+\frac{3}{4\,}\sum_{sm}\int_{0}^{L_{s}}dl\,\left\{ \frac{1}{\lambda _{s}(l)%
}\,\left[ \underline{u}_{m}^{s}(l)\,\right] ^{2}+\lambda _{s}(l)\,\left[ 
\underline{\dot{u}}_{m}^{s}(l)\,\right] ^{2}\right\} ,  \label{18}
\end{eqnarray}%
where $\lambda _{s}(l)\,$denotes the local \textit{persistence length} of
chains of type $s$ . For a multiblock copolymer this $\lambda _{s}(l)$ will
have the following form,
\begin{equation}
\lambda _{s}(l)\equiv \sum\limits_{i=1}^{N_{s}^{b}}\lambda _{s}^{(i)}\left[
\theta (l-\sum\limits_{j=1}^{i-1}L_{s}^{(j)})-\theta
(l-\sum\limits_{j=1}^{i}L_{s}^{(j)})\right] ,  \label{19}
\end{equation}%
where $L_{s}^{(j)}$and $\lambda _{s}^{(i)}$ respectively denote the block length of the $j$-th block
and the persistence length of the $i$-th block in a chain of type $s$ which
consists of $N_{s}^{b}$ blocks. The function $\theta (...)$ in (\ref{19}) between the square brackets is the Heaviside step function. The first term in (\ref{18}) containing the
orientation vectors of the endpoints of the chains and the particular choice
of the coefficients in the second term of (\ref{18}) ensure homogeneity of
the chains and Porod-Kratky like behavior in an averaged sense, i.e. $%
\langle \,\,\left[ \underline{u}_{m}^{s}(l)\right] ^{2}\,\rangle
_{0}=1,\quad \forall l\in \lbrack 0,L_{s}]$. $\langle \,\,...\,\rangle _{0}$
is an average over all possible chain configurations of one single free
chain. It is defined in Eq. (\ref{33}) in Appendix A.

In Supplemental material I it is explained that in the Bawendi-Freed approach the semi-flexible chain can be regarded as a chain of harmonic oscillators in which the segments are connected by springs. Due to the springs the length of the tangent vector is only on average a unit vector, so as mentioned earlier  $%
\langle \,\,\left[ \underline{u}_{m}^{s}(l)\right] ^{2}\,\rangle
_{0}=1,\quad \forall l\in \lbrack 0,L_{s}]$. This less strict condition makes it possible to derive the single-chain correlation functions. This derivation is described in detail in Supplemental material I. The single-chain correlation functions are applied in Appendix A in which the final form of the Landau free energy given by Eq. (\ref{vrijeenergie}) is derived.  

The total Hamiltonian $\hat{H}$ for a certain melt configuration contains
the unperturbed Hamiltonian $\hat{H}_{0}$ and the total interaction energy $%
\hat{W}$, 
\begin{equation}
\hat{H}=\hat{H}_{0}+\hat{W}.  \label{20}
\end{equation}%
Every configuration gives a contribution $\exp (-\hat{H})$ to the partion
function $Z$. In Appendix A a functional integral $Z$ over all possible
configurations is evaluated. From the partition function $Z$ the Landau free
energy $F_{L}$ is derived by considering the most dominant contribution to $%
Z $. This contribution corresponds to the most probable i.e. equilibrium
state of the melt. The final result (in units of $k_{B}T$) is the Landau free energy written in terms of the Fourier transform of the order-parameter fields, 
\begin{gather}
\frac{F_{L}}{V}=\underset{\underline{\Psi },\overline{\Upsilon }}{\min }%
\{(\Gamma _{ab}^{(2)}-\widetilde{\chi }_{ab})\Psi ^{a}\Psi ^{b}+2\Gamma _{a%
\overline{b}}^{(2)}\Psi ^{a}\Upsilon ^{\overline{b}}+  \notag \\
(\Gamma _{\overline{a}\overline{b}}^{(2)}-\frac{1}{2}\omega _{\overline{a}%
\overline{b}})\Upsilon ^{\overline{a}}\Upsilon ^{\overline{b}}-\frac{1}{3}%
\omega _{ab}\Upsilon ^{a,ij}\delta _{ij}(\Psi ^{b}+f^{b})+  \notag \\
\Gamma _{abc}^{(3)}\Psi ^{a}\Psi ^{b}\Psi ^{c}+3\Gamma _{ab\overline{c}%
}^{(3)}\Psi ^{a}\Psi ^{b}\Upsilon ^{\overline{c}}+3\Gamma _{a\overline{b}%
\overline{c}}^{(3)}\Psi ^{a}\Upsilon ^{\overline{b}}\Upsilon ^{\overline{c}%
}+\Gamma _{\overline{a}\overline{b}\overline{c}}^{(3)}\Upsilon ^{\overline{a}%
}\Upsilon ^{\overline{b}}\Upsilon ^{\overline{c}}+  \notag \\
\Gamma _{abcd}^{(4)}\Psi ^{a}\Psi ^{b}\Psi ^{c}\Psi ^{d}+4\Gamma _{abc%
\overline{d}}^{(4)}\Psi ^{a}\Psi ^{b}\Psi ^{c}\Upsilon ^{\overline{d}%
}+6\Gamma _{ab\overline{c}\overline{d}}^{(4)}\Psi ^{a}\Psi ^{b}\Upsilon ^{%
\overline{c}}\Upsilon ^{\overline{d}}+  \notag \\
4\Gamma _{a\overline{b}\overline{c}\overline{d}}^{(4)}\Psi ^{a}\Upsilon ^{%
\overline{b}}\Upsilon ^{\overline{c}}\Upsilon ^{\overline{d}}+\Gamma _{%
\overline{a}\overline{b}\overline{c}\overline{d}}^{(4)}\Upsilon ^{\overline{a%
}}\Upsilon ^{\overline{b}}\Upsilon ^{\overline{c}}\Upsilon ^{\overline{d}}\}
\label{vrijeenergie}
\end{gather}
with $\tilde{\chi}_{ab}\equiv \tilde{\chi}_{\alpha \beta }\,\delta (%
\underline{q}_{1}+\underline{q}_{2})$, $\omega _{\overline{a}\overline{b}%
}\equiv \omega _{\alpha \beta }\,|2\delta _{ii^{\prime }}\delta _{jj^{\prime
}}-\delta _{ij}\delta _{i^{\prime }j^{\prime }}|\,\delta (\underline{q}_{1}+%
\underline{q}_{2})$, $\Psi ^{a}$ $\equiv \frac{\psi ^{\alpha }(-\underline{q}%
_{1})}{V}$, $\Upsilon ^{\overline{a}}\equiv \frac{Q_{ij}^{\alpha }(-%
\underline{q}_{1})}{V}$ and $f^{b}=f^{\beta }\delta (\underline{q}_{2})$. In
this expression of the free energy we use the \textit{composite labels} $%
a\equiv (\underline{q}_{1}\neq \underline{0},\alpha )$, $b\equiv (\underline{%
q}_{2}\neq \underline{0},\beta )$ etc. and $\overline{a}\equiv (\underline{q}%
_{1},ij,\alpha )$, $\overline{b}\equiv (\underline{q}_{2},i^{\prime
}j^{\prime },\beta )$ etc. in which the
pairs $ij$ and $i^{\prime }j^{\prime }$ are one of the six unique pairs $xx$%
, $yy$, $zz$, $xy$, $xz$ and $yz$. The coefficient functions ($\Gamma $'s)
are called \textit{vertices} and are defined by Eq. (\ref{vertex2}), (\ref%
{vertex3}) and (\ref{vertex4}) in Supplemental material II. In each term in Eq. (\ref{vrijeenergie}) we see that each composite label in a subscript matches with a composite label in a superscript in one of the order-parameters. In such a match the Einstein summation convention is used. The summation over the wave vectors $\underline{q}$ is limited over a set $H=\{\pm\underline{q}_1, \pm\underline{q}_2, ...\}$ in which all wave vectors have the same fixed magnitude $q_*$. This set $H$ defines the symmetry properties and wavelength of a phase structure. 

In Appendix B Eq. (\ref{vrijeenergie}) is applied to a melt of monodisperse semi-flexible diblock copolymers in which there is only one Flory-Huggins interaction parameter $\chi_{AB} = \chi$ between A- and B-blocks. A microphase structure is formed when $\chi$ is strong enough and above the spindal $\chi_s$. According to Appendix B the earlier mentioned magnitude $q_*$ follows from minimizing the spinodal expression given by Eq. (\ref{spinodalChi}) with respect to the magnitude  $q_*$ of the wave vectors in set $H$. When we know $q_*$ we can calculate the complete expression of the Landau free energy for a certain phase structure with a set of wave vectors $H$, for example the lamellar state in which $H=\{\pm\underline{q}\}$. This complete expression is minimized with respect to the order-parameters. In the same way the minimum free energy is calculated for other phase structures $H$. The phase structure with the lowest minimum is the most probable state of the melt. In this way the phase diagram is calculated. When the persistence length of $\lambda_A$ and $\lambda_B$ are getting close to zero, then the diblocks become totally flexible. In that case the expression of the Landau free energy given by Eq. (\ref{vrijeenergie}) can be reduced to the expression derived in Ref. \cite{Leibler}.

\section{Results and discussion}
The theory in the previous section and Appendix B is applied to calculate the phase diagram of a melt of semi-flexible diblock copolymers. In this section
the influence of the stiffness on the phase behaviour is investigated in different systems. In each system the stiffness of the A- and B-block is fixed.
We have considered and compared the following possibilities: (a) rod-coil ($\lambda _{A}/L_{A} =10^{2} \gg \lambda _{B}/L_{B} =10^{ -4}$); (b) semi-coil ($\lambda _{A}/L_{A} =10^{ -1} \gg \lambda _{B}/L_{B} =10^{ -4}$); (c) semi-semi ($\lambda _{A}/L_{A} =10^{ -1} =\lambda _{B}/L_{B}$). 

Before presenting the results one important point has to be clarified. In our model the
Maier-Saupe interaction and bending stiffness are described by independent parameters $\omega _{\alpha  \beta }$\ and $\lambda _{\alpha }/L_{\alpha }$ in which $\alpha  ,\beta  =A$ or $B$. However, in reality these parameters are related. The Maier-Saupe interaction between two blocks becomes stronger if the blocks become
stiffer. In a rod-rod system in which$\;\lambda _{A}/L_{A} \gg 1$ and $\lambda _{B}/L_{B} \gg 1$ the parameters $\omega _{\alpha  \beta }$ cannot be chosen small, because such a melt would be an unphysical system according to Ref. \cite{Semenov}
and \cite{Kokhlov}. In these papers the Onsager model is applied to investigate
a solution or melt of liquid-crystalline polymers. According to the Onsager model the isotropic state is not possible in a melt of rigid rods. Due to the
excluded volume effect the rods will always align in a melt. In our theory the steric interaction is described by assuming that the melt is incompressible.
Due to steric interactions the chains tend to align. The contibution of this alligment force is included in the Maier-Saupe parameters $\omega _{\alpha  \beta }$. These parameters also contain contributions of attractive van der Waals forces. According to our theory the isotropic state is possible
for every stiffness $\lambda _{A}/L_{A}$ and $\lambda _{B}/L_{B}$ if the parameters $\omega _{\alpha  \beta }$ are zero. This means that there is no alignment force which is not possible. Due the excluded volume effect there is always an alignment
force so that $\omega _{\alpha  \beta } >0$. If the A- or B-block becomes stiffer and stiffer a weaker force is necessary to induce nematic ordening. Then at least one of the parameters
$\omega _{\alpha  \beta }$ may become larger than the spinodal $\omega _{\alpha  \beta  ,s} =\omega _{\alpha  \beta  ,s} (q_{ \ast } =0)$, so that the melt is always in the nematic state. In that case our theory cannot be used to make further predictions of the phase behaviour.
Our expression of the Landau free energy can only be applied in the neighbourhood of a phase transition at which the isotropic state is converted into an
ordered state. However, according to Ref. \cite{Semenov} and \cite{Kokhlov}
in a melt of semi-flexible chains the isotropic state becomes possible if the persistence length $\lambda $ of the chain is small enough. $\lambda $ must satify the condition $\frac{\lambda }{d} <50$ in which $d$ is the diameter of the chain. In our theory the length-scale is chosen such that one monomer occupies one unit of volume, so that the
order of magnitude of $d$ is one, $d \approx 1$. In system (c) in which ($\lambda _{A}/L_{A} =10^{ -1} =\lambda _{B}/L_{B}$) this condition is satified, because the diblock length $L$ can be chosen such that $\lambda _{A} ,\lambda _{B} \leq 10^{ -1} L <50$. Then it is allowed to choose the parameters $\omega _{\alpha  \beta }$ such that each $\omega _{\alpha  \beta }$ is smaller than the spinodal $\omega _{\alpha  \beta  ,s} =\omega _{\alpha  \beta  ,s} (q_{ \ast } =0)$.

\includegraphics[scale=0.37]{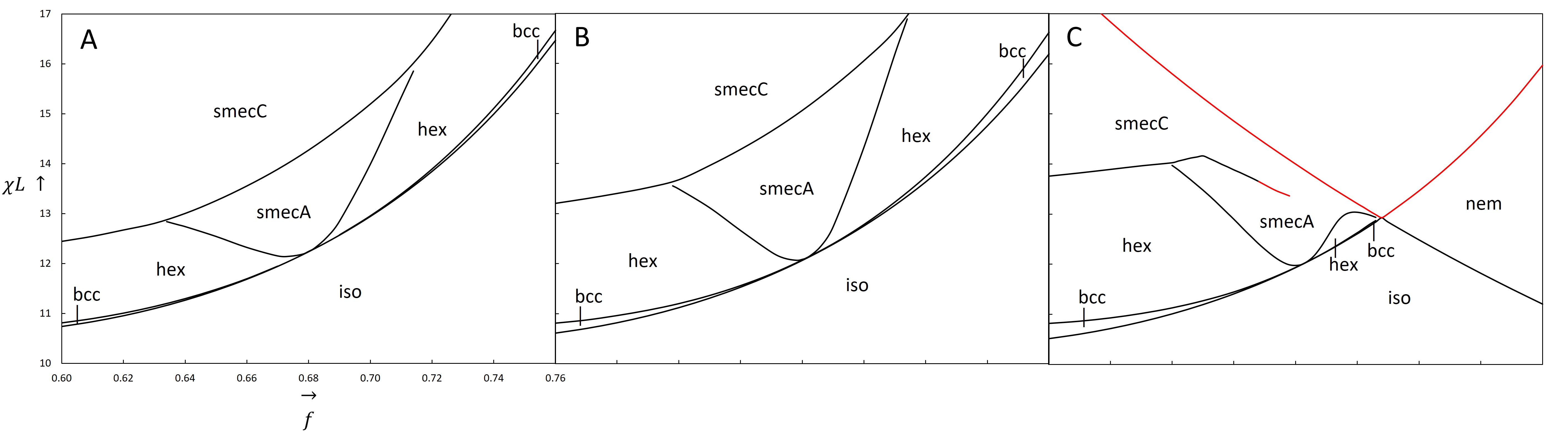} \newline

In Fig. (1) three phase diagrams of a melt of monodisperse diblock copolymers have been calculated
in which the persistence length of the A-block $\lambda _{A}$ is much larger than the persistence length of the B-block, $\lambda _{A}$ $ \gg \lambda _{B}$. $\lambda _{B}/L_{B}$ is taken equal to $\lambda _{B}/L_{B} =10^{ -4}$ so that the B-block can be regarded as totally flexible. The A-block is made very stiff, $\lambda _{A}/L_{A} =10^{2}$, and behaves as a rigid rod. Because the B-block is very flexible the orientation $\Upsilon _{B}$ has been neglected in the calculation, $\Upsilon _{B} \approx 0$. Also the influence of the Maier-Saupe parameters $\omega _{B B}$ and $\omega _{A B}$\ is neglected. Only the interaction $\omega _{A A} =\omega $ between stiff A-blocks is taken into account. In Fig. (1A), (1B) and (1C) the ratio $r =\frac{\omega _{A A}}{\chi } =\frac{\omega }{\chi }$ is taken equal to $0.0$, $0.4$ and $0.7$, respectively. The fraction of A-blocks $f^{A}$ defined by Eq. (\ref{4}) is denoted as $f$ in Fig. (1) and the rest of this section, because we restrict to diblock copolymers. In all three phase diagrams in Fig. (1) $\chi L$ and $f$ are varied within the same interval. In Fig. (1) we can clearly see that the phase behaviour is influenced in different ways by the strength of the
nematic interaction $\omega $ with respect to $\chi $. The domain of the bcc phase is a very narrow line when $r =0.0$. The bcc line becomes broader at a larger $r$. The domains of the hexagonal and smectic-A phase also become larger. The smectic-C domain is shifted to a larger $\chi $. 

To explain the influence of $r$ on the phase behaviour the absolute value of the components $\vert \Psi _{1}\vert $ and $\vert \Upsilon _{1}^{A}\vert $ of the primary eigenvector $\underline{x}_{1} =(\Psi _{1} ,\Upsilon _{1}^{A} ,\Upsilon _{1}^{B})$ are investigated in both the smectic-C, smectic-A, hexagonal and bcc phase. It appears that the density parameter $\vert \Psi _{1}\vert $\ in the different microphase structures is always ordered according to the sequence,
\begin{equation}\vert \Psi _{1 ,s m e c C}\vert  \geq \vert \Psi _{1 ,s m e c A}\vert  \geq \vert \Psi _{1 ,h e x}\vert  \geq \vert \Psi _{1 ,b c c}\vert  , \label{psiA}
\end{equation}in which $\vert \Psi _{1 ,s m e c C}\vert $, $\vert \Psi _{1 ,s m e c A}\vert $, $\vert \Psi _{1 ,h e x}\vert $ and $\vert \Psi _{1 ,b c c}\vert $ denote the density parameters in the smectic-C, smectic-A, hexagonal and bcc phase, respectively. The orientation strength $\vert \Upsilon _{1}^{A}\vert $ is ordered by,
\begin{equation}\vert \Upsilon _{1 ,s m e c C}^{A}\vert  \geq \vert \Upsilon _{1 ,s m e c A}^{A}\vert  \geq \vert \Upsilon _{1 ,h e x}^{A}\vert  \geq \vert \Upsilon _{1 ,b c c}^{A}\vert  . \label{upsA}
\end{equation}When $r$ increases it appears that both $\vert \Psi _{1}\vert $ and $\vert \Upsilon _{1}^{A}\vert $ become smaller in each microphase structure. So the Maier-Saupe interaction counteracts microphase separation. This effect is also observed
in the phase diagrams in Fig. (1). The bcc and hexagonal domain become larger at a larger $r$ and the domain of smectic-A and smectic-C phase are shifted to a larger $\chi  L$. By means of Eq. (\ref{psiA}) and (\ref{upsA}) it can be concluded
that the melt prefers a microphase with a smaller $\vert \Psi _{1}\vert $ and $\vert \Upsilon _{1}^{A}\vert $ when $r$ increases. The decrease of $\vert \Upsilon _{1}^{A}\vert $ is contrary to what we would expect, because a strong Maier-Saupe interaction will induce a nematic state. The parameter $\vert \Upsilon _{1}^{A}\vert $ is the strength of a space dependent orientation. In the nematic state there is global alignment of stiff A-blocks with strength $\vert \Upsilon _{1 0}^{A}\vert $. A nonzero $\vert \Upsilon _{1 0}^{A}\vert $ is also induced in a microphase, but is very weak. The secondary eigenvector $\underline{x}_{1 0} =(\Upsilon _{1 0}^{A} ,\Upsilon _{1 0}^{B}) \approx (\Upsilon _{1 0}^{A} ,0)$ contains a global oriention $\Upsilon _{1 0}^{A}$. The secondary parameter $x_{1 0} = \pm \vert \underline{x}_{1 0}\vert  \approx  \pm \vert \Upsilon _{1 0}^{A}\vert $ is given by Eq. (\ref{xn}) in which $n =10$,
\begin{equation}x_{1 0} =\frac{ -C_{11 ,10}^{(3)} x_{1}^{2}}{2 \lambda _{1 0}} +O (x_{1}^{3}) . \label{ups10}
\end{equation}If the Maier-Saupe interaction increases the positive eigenvalue $\lambda _{1 0}$ becomes smaller which enhances the global orientation $\vert \Upsilon _{1 0}^{A}\vert $. This parameter increases strongly if $\omega _{\alpha  \beta }$ reaches the spinodal $\omega _{\alpha  \beta  ,s} (q_{ \ast } =0)$, because then $\lambda _{1 0}\downarrow 0$. In the bcc phase a global nematic ordening cannot be induced so that $\vert \Upsilon _{10 ,b c c}^{A}\vert  =0$. 

In Fig. (1B) and (1C) there is a red line at which $\omega _{A A} =\omega  =\omega _{s} (q_{ \ast } =0)$ or $\chi  =\chi _{s} (q_{ \ast } \neq 0)$. In Fig. (1C) at $f =0.71$ both $\omega  =\omega _{s} (q_{ \ast } =0)$ and $\chi  =\chi _{s} (q_{ \ast } \neq 0)$ in the red line. At that point the eigenvalues $\lambda _{1}$ and $\lambda _{1 0}$ in Eq. (\ref{Fx}) are zero so that both microphase separation and nematic ordening become possible.
Then the final form given by Eq. (\ref{Fx1}) cannot be applied at a larger $\chi $. Here only the eigenvalue $\lambda _{1}$ is negative which makes it possible to write the free energy in a power series expansion of only one primary parameter $x_{1}$. If both eigenvalues $\lambda _{1}$\ and\ $\lambda _{1 0}$\ are negative there are two primary parameters $x_{1}$ and $x_{1 0}$. Then the final form of the free energy becomes,
\begin{gather}\frac{F_{L}}{V} =\min_{\left \{x_{1} ,x_{1 0}\right \}}\{\lambda _{1} x_{1}^{2} +\lambda _{1 0} x_{1 0}^{2} +C_{1 1 1}^{(3)} x_{1}^{3} +C_{11 ,10}^{(3)} x_{1}^{2} x_{1 0} +C_{10 ,10 ,10}^{(3)} x_{1 0}^{3} + \nonumber  \\
\widetilde{C}_{1 1 1 1}^{(4)} x_{1}^{4} +\widetilde{C}_{111 ,10}^{(4)} x_{1}^{3} x_{1 0} +\widetilde{C}_{11 ,10 ,10}^{(4)} x_{1}^{2} x_{1 0}^{2} +\widetilde{C}_{10 ,10 ,10 ,10}^{(4)} x_{1 0}^{4}\} , \label{Fx2}\end{gather}in which the coeffients of the fourth order terms contain contributions from secondary eigenvectors $\underline{x}_{2}$, $\underline{x}_{3}$ and $\underline{x}_{2 0}$,
\begin{subequations}
\begin{gather}\widetilde{C}_{1 1 1 1}^{(4)} =C_{1 1 1 1}^{(4)} -\sum _{n \neq 1 ,10}\frac{(C_{1 1 n}^{(3)})^{2}}{4 \lambda _{n}} , \label{C1111} \\
\widetilde{C}_{111 ,10}^{(4)} =C_{111 ,10}^{(4)} -\frac{C_{11 ,2}^{(3)} C_{1 ,10 ,2}^{(3)}}{2 \lambda _{2}} -\frac{C_{11 ,3}^{(3)} C_{1 ,10 ,3}^{(3)}}{2 \lambda _{3}}\text{,} \\
\widetilde{C}_{11 ,10 ,10}^{(4)} =C_{11 ,10 ,10}^{(4)} -\frac{(C_{1 ,10 ,2}^{(3)})^{2}}{4 \lambda _{2}} -\frac{(C_{1 ,10 ,3}^{(3)})^{2}}{4 \lambda _{3}} -\frac{C_{11 ,20}^{(3)} C_{10 ,10 ,20}^{(3)}}{2 \lambda _{2 0}}\text{,} \\
\text{and} \nonumber  \\
\widetilde{C}_{10 ,10 ,10 ,10}^{(4)} =C_{10 ,10 ,10 ,10}^{(4)} -\frac{(C_{10 ,10 ,20}^{(3)})^{2}}{4 \lambda _{2 0}} . \label{C10101010}\end{gather}

Numerically it appears that in each kind of microphase structure the coefficient $C_{1 1 1 1}^{(4)}$ in Eq. (\ref{Fx1}) and (\ref{C1111}) becomes negative if $\chi $ becomes too large. $C_{1 1 1 1}^{(4)}$ is always positive if $\chi $ is close enough to $\chi _{s}$. Here the mean field approximation is more reliable. However, the coefficient $C_{10 ,10 ,10 ,10}^{(4)}$ in Eq. (\ref{C10101010}) is always negative so that in Eq. (\ref{Fx2})
$\widetilde{C}_{10 ,10 ,10 ,10}^{(4)} x_{1 0}^{4} <0$. In that case the free energy given by Eq. (\ref{Fx2}) cannot have a finite minimum. Higher order
terms would be necessary to compensate for the negative contribution. These terms are very complicated and will not be calculated. So above the red line
in Fig. (1B) and (1C) the phase behaviour cannot be predicted by means of Eq. (\ref{Fx2}). Because below the red line
a smectic-C state is found, we also expect a layered structure above that line. In this structure the orientation tensor $\underline{\underline{Q}}_{\alpha } (\underline{x})$ is space dependent, but contains a stronger global component $\underline{\underline{Q}}_{\alpha }^{0}$, because $\lambda _{1 0} <0$. In this way the space average $ \langle \underline{\underline{Q}}_{\alpha } (\underline{x}) \rangle $,
\end{subequations}
\begin{equation} \langle \underline{\underline{Q}}_{\alpha } (\underline{x}) \rangle  =\frac{1}{V} \int _{V}d^{3} x \text{ } \underline{\underline{Q}}_{\alpha } (\underline{x}) =\underline{\underline{Q}}_{\alpha }^{0}\text{,}
\end{equation}is nonzero. If $\omega  <\omega _{s} (q_{ \ast } =0)$ there is also global nematic ordening possible in addition to the space dependent orientation. In that part of the phase diagram the
global orientation $x_{1 0}$ is a secondary parameter which is expressed by Eq. (\ref{ups10}). If $\lambda _{1 0} \gg 0$ this orientation is very weak, but when $\lambda _{1 0}\downarrow 0$ it becomes stronger. If it is too strong the expression of the Landau free energy given by Eq. (\ref{Fx1})
is not reliable, because the fourth order term is getting too close to zero or may even become negative. 

In the right part of Fig. (1C) the boundary line between the smectic-C and hexagonal domain has a discontinuity at $f =0.65$. At this point the angle $\theta $ at the boundary changes discontinuously from $13$ to $89.9$ degrees when $f$ is increasing. In that part of the phase diagram the smectic-C domain is close to the red line. Here the induced global nematic ordening given by Eq. (\ref{ups10}) becomes stronger, because $\lambda _{1 0}\downarrow 0$. In the fourth order term in Eq. (\ref{Fx1}) this nematic ordening gives a large contribution. Due
to this effect the minimum at $\theta  =89.9$ degrees has become the lowest minimum. At $f =0.67$ the black boundary line has been stopped and is extended by a red line. At $f >0.67$ the eigenvalue $\lambda _{1 0}$ in Eq. (\ref{Fx}) is too close to zero so that the fourth order term has become negative. Here the
phase behaviour cannot be predicted. However, reliable predictions can also not be made if the fourth order term is still positive but close to zero. Then
Eq. (\ref{Fx1}) is also not reliable. 

In Ref. \cite{Singh}. the dependence of the spinodal $\chi _{s} L$ and the corresponding wave mode $q_{ \ast }$ on the Maier-Saupe interaction has been investigated for semi-flexible monodisperse diblocks. It appeared that $\chi _{s} L$ is lowered and $q_{ \ast }$ is increased by the Maier-Saupe interaction. The same effect is observed by calculating $\chi _{s} L$ and $q_{ \ast }L$ as a function of  $r =\frac{\omega }{\chi }$ using Eq. (\ref{spinodalChi}) in this paper for monodisperse rod-coil diblocks. The relation between $\chi _{s} L$ and $r$ and also between $q_{ \ast }L$ and $r$ appeared to be approximately linear. The effect is caused by the large difference in bending stiffness between the A- and B-block. If the A- and B-block have the same bending stiffness the influence
of the Maier-Saupe interaction on the spinodal $\chi _{s} L$ and the wave mode $q_{ \ast }$ is negligible. When $r$ becomes larger $q_{ \ast }$ is increasing which means that the size of the A- and B-rich domains becomes smaller. A smaller domain size increases the total contact
area between A- and B-blocks in the whole melt. This could explain why in each microphase structure $\vert \Psi _{1}\vert $ and $\vert \Upsilon _{1}^{A}\vert $ are lowered when $r$ increases. A weaker microphase separation increases the mixing entropy, but the enthalpic contribution $\widetilde{\chi }_{a b} \Psi ^{a} \Psi ^{b}$ in Eq. (\ref{FL}) becomes smaller. At $r =0.0$ the wave length $\frac{2 \pi }{q_{ \ast }}$ is maximal and a microphase cannot be formed at a smaller wave length. If the domain size becomes smaller the enthalpic contribution
$\widetilde{\chi }_{a b} \Psi ^{a} \Psi ^{b}$ becomes too weak and cannot be compensated by an increase of the mixing entropy. However, when the Maier-Saupe interaction is switched
on the enthalpic contribution $\frac{1}{2} \omega _{\overline{a} \overline{b}} \Upsilon ^{\overline{a}} \Upsilon ^{\overline{b}}$ in Eq. (\ref{FL}) gives an additional compensation. This additional compensation could make it possible
to form a microphase with a smaller domain size. Therefore $q_{ \ast }L$ increases when $r$ becomes larger. In the same way it can be explained why $\chi _{s} L$ is lowered by a stronger Maier-Saupe interaction.

From the analysis and discussion of the phase behaviour in Fig. (1) it has been concluded that the Maier-Saupe interaction counteracts microphase separation. This also appears from the spinodal calculations using Eq. (\ref{spinodalChi}), because $q_{ \ast }L$ is increasing when $r$ increases. However, the opposite effect also takes place, because the spinodal $\chi _{s} L$ is lowered when the Maier-Saupe interaction is increasing. So in that case microphase separation is stimulated. The stimulation takes place when te melt is close to the phase transition, but it is still in the disordered phase. When the spinodal $\chi _{s} L$ is lowered the disordered phase is converted into a microphase structure. However, the counteracting effect occurs when a microphase structure is already present before increasing the Maier-Saupe interaction.

\includegraphics[scale=0.50]{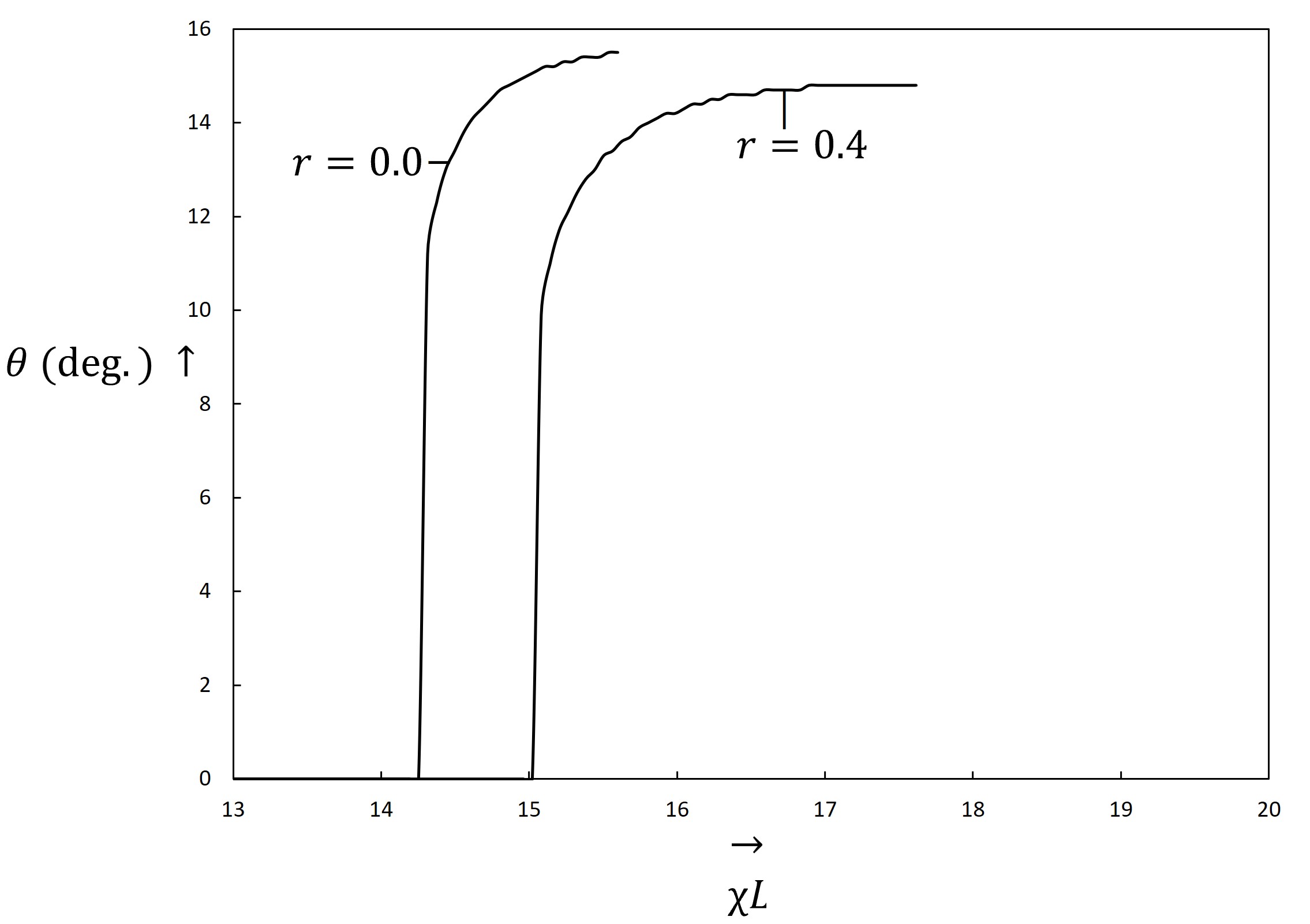} \newline

In Fig. (2) the angle $\theta $ in degrees is calculated as a function of $\chi  L$ at $r =0.0$ and $0.4$ when $f =0.68$. If $r =0.7$ the angle $\theta $ of the smectic-C state cannot be investigated, because the fourth order term in Eq. (\ref{Fx}) is negative
for certain angles $\theta $. At the critical point at $f =0.68$ just above the spinodal the smectic-A state is formed. When $\chi $ becomes larger at a certain point the smectic-A state is converted into the smectic-C state. At the transition we see in Fig. (2)
that the angle $\theta $ changes discontinuously from $\theta  =0$ degrees to a nonzero $\theta $. After the transition when $\chi $ is further increased the angle $\theta $ increases and converges to a constant value. At a larger $\chi $ the function $\theta  (\chi  L)$ cannot be further drawn, because the coefficient $C_{1 1 1 1}^{(4)}$ in Eq. (\ref{Fx}) becomes negative. As mentioned earlier it has been concluded that the
Maier-Saupe interaction counteracts microphase separation. This may also explain the influence of $\omega $ in the smectic structure observed in Fig. (2). When the nematic interation $\omega $ is switched on the transtion from smectic-A to smectic-C takes place at a larger $\chi $. At the same time the angle $\theta $ in the smectic-C state has become smaller.

\includegraphics[scale=0.37]{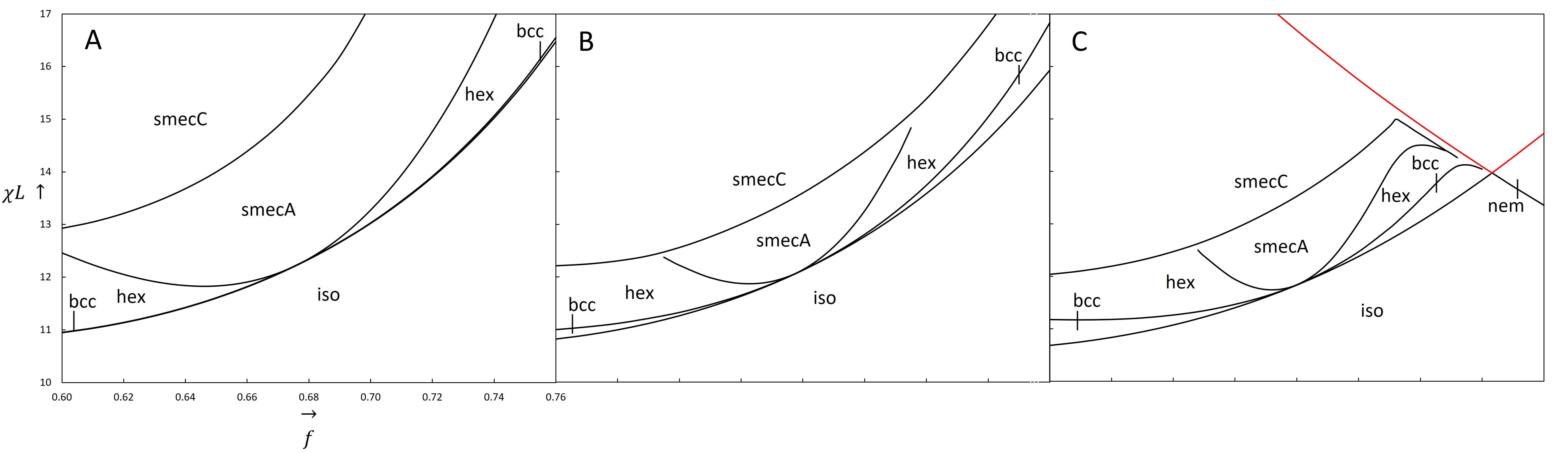} \newline

In Fig. (3) the phase diagram is calculated in which
the persistence length of the A-block $\lambda _{A}$ is again much larger than the persistence length of the B-block, $\lambda _{A}$ $ \gg \lambda _{B}$. In the previous phase diagrams the A-block is made very stiff such that it can be regarded as a rigid rod. In Fig. (3)
the stiffness of the A-block is made smaller, $\lambda _{A}/L_{A} =10^{ -1}$. Here the A-block is called semi-flexible. The stiffness is such that it does not behave as a rigid rod or a Gaussian chain. The B-block
is again totally flexible, $\lambda _{B}/L_{B} =10^{ -4}$. Therefore the orientation $\Upsilon _{B}$ has been neglected in the calculation, $\Upsilon _{B} \approx 0$, so that the Maier-Saupe parameters $\omega _{B B}$ and $\omega _{A B}$\ are excluded. In Fig (3A), (3B) and (3C) the ratio $r =\frac{\omega _{A A}}{\chi } =\frac{\omega }{\chi }$ is taken equal to $0$, $3$ and $5$, respectively. In all three phase diagrams $\chi L$ and $f$ are varied within the same interval. In the red line $\omega  =\omega _{s} (q_{ \ast } =0)$ or $\chi  =\chi _{s} (q_{ \ast } \neq 0)$. If the parameter $r$ is increased we see that the bcc region becomes broader which is also observed in Fig. (1). The smectic-C phase is not
formed at $r =0$ when $f >0.706$. When $r$ increases the smectic-C domain is shifted to a smaller $\chi $ which makes the hexagonal and smectic-A part smaller. Only very close to the critical point the size of the hexagonal part increases
a little bit. Here the influence of $r$ on the phase behaviour is the same as in the previous phase diagrams and can be explained in the same way. However, the influence of
$r$ on the smectic-C domain is different. The angle $\theta $ in the previous phase diagrams is within the interval $8 <\theta  <18$ degrees and depends on the parameters $\chi $, $r$ and $f$. In Fig. (3) in the smectic-C phase the angle $\theta $ is constant, $\theta  =54.7$ degrees. Exactly at $\theta  =54.7$ degrees the director $\underline{\eta }$ is such that $\eta _{x}^{2} -\frac{1}{3} =\cos ^{2} \theta  -\frac{1}{3} =0$. Then in the smectic-C state the tensor component $\Upsilon _{A}^{x x} (\underline{x})$ vanishes, $\Upsilon _{A}^{x x} (\underline{x}) =0$. This means that orientation of A-blocks is random in the $x$-direction. Because $\Upsilon _{A}^{x x} (\underline{x}) =0$ the components $\Upsilon _{A}^{z z} (\underline{x})$ and $\Upsilon _{A}^{y y} (\underline{x})$ are compensating each other, $\Upsilon _{A}^{z z} (\underline{x}) = -\Upsilon _{A}^{y y} (\underline{x})$. So if $\theta  =54.7$ degrees the orientation direction in an A-rich layer is perpendicular to the orientation in a B-rich layer. From numerical calculations
it appears that such a state is not formed. If it appears that in the primary eigenvector $\underline{x}_{1} =(\Psi _{1} ,\Upsilon _{1}^{A} ,\Upsilon _{1}^{B})$ the component $\Upsilon _{1}^{A}$ becomes zero, so that $\underline{x}_{1} =(\Psi _{1} ,0 ,0)$. Only in the secondary eigenvectors a nonzero orientation of A-blocks is possible. The contribution of these vectors is neglibly small
so if $\theta  =54.7$ degrees the smectic-C state can be regarded as a lamellar state with only a nonzero density order-parameter.

This effect does not only
occur in Fig. (3). In other kinds of diblocks it also appears that when $\theta  =54.7$ degrees the primary eigenvector is $\underline{x}_{1} =(\Psi _{1} ,0 ,0)$ for each $\chi $ and $\omega _{\alpha  \beta }$. If $\theta  \neq 54.7$ degrees there is always some orientation $\Upsilon _{1}^{A}$ or $\Upsilon _{1}^{B}$ if the persistence length of the A- or B-block is large enough. The alignment is such that the A-B contact is as small as possible.
Without orientation the A- and B-blocks may have more contact which is enthalpically less favourable. However, the entropy of the melt is higher without
orientation. Maybe the enthalpy gain due to the alligment has a larger effect and compensates for the entropy loss. 

To explain the influence
of $r$ on the phase behaviour the components $\vert \Psi _{1}\vert $ and $\vert \Upsilon _{1}^{A}\vert $ in the primary eigenvector $\underline{x}_{1} =(\Psi _{1} ,\Upsilon _{1}^{A} ,\Upsilon _{1}^{B})$ are again investigated in the different microphase structures. It appears that the density parameter $\vert \Psi _{1}\vert $\ is also ordered according to Eq. (\ref{psiA}),
\begin{equation}\vert \Psi _{1 ,s m e c C}\vert  \geq \vert \Psi _{1 ,s m e c A}\vert  \geq \vert \Psi _{1 ,h e x}\vert  \geq \vert \Psi _{1 ,b c c}\vert  , \label{psiA2}
\end{equation}The orientation strength $\vert \Upsilon _{1}^{A}\vert $ is ordered by,
\begin{equation}\vert \Upsilon _{1 ,s m e c A}^{A}\vert  \geq \vert \Upsilon _{1 ,h e x}^{A}\vert  \geq \vert \Upsilon _{1 ,b c c}^{A}\vert  \geq \vert \Upsilon _{1 ,s m e c C}^{A}\vert  =0 , \label{upsA2}
\end{equation}which is different if we look at Eq. (\ref{upsA}). In Eq. (\ref{upsA})
the smectic-C phase has the strongest orientation, but in Eq. (\ref{upsA2}) $\vert \Upsilon _{1 ,s m e c C}^{A}\vert  =0$ because $\theta  =54.7$ degrees. Therefore it has appeared that the density parameter $\vert \Psi _{1 ,s m e c C}\vert $ does not change very much when $r$ increases. $\vert \Psi _{1 ,s m e c A}\vert $, $\vert \Psi _{1 ,h e x}\vert $ and $\vert \Psi _{1 ,b c c}\vert $ become smaller at a larger $r$. The influence of the Maier-Saupe interaction on $\vert \Upsilon _{1}^{A}\vert $ in the different microphase structures is not very big. From Eq. (\ref{psiA2}) and Fig. (3) it can be concluded that close to the spinodal $\chi _{s} L$ the melt prefers a microphase with a smaller $\vert \Psi _{1}\vert $ when $r$ increases. Here microphase separation is counteracted by the Maier-Saupe interaction. However, for a larger $\chi  L$ the smectic-C phase is formed in which $\vert \Psi _{1}\vert $ is maximal according to Eq. (\ref{psiA2}). In this phase the Maier-Saupe interaction cannot have
much influence on the microphase separation because there is no space dependent orientation, $\vert \Upsilon _{1 ,s m e c C}^{A}\vert  =0$. Only a very weak global orientation $\vert \Upsilon _{1 0}^{A}\vert  =\vert x_{1 0}\vert $ given by Eq. (\ref{ups10}) is possible which increases at a larger $r$. In the other microphases $\vert \Psi _{1}\vert $ is lowered by $r$. Then the smectic-C phase could become more favourable because of the separation enthalpy. This could explain why the smectic-C domain
is shifted to a smaller $\chi  L$ when $r$ becomes larger. So further away from the spinodal $\chi _{s} L$ the Maier-Saupe interaction enhances microphase separation by forming a smectic-C state with a larger density parameter $\vert \Psi _{1}\vert $.

\includegraphics[scale=0.37]{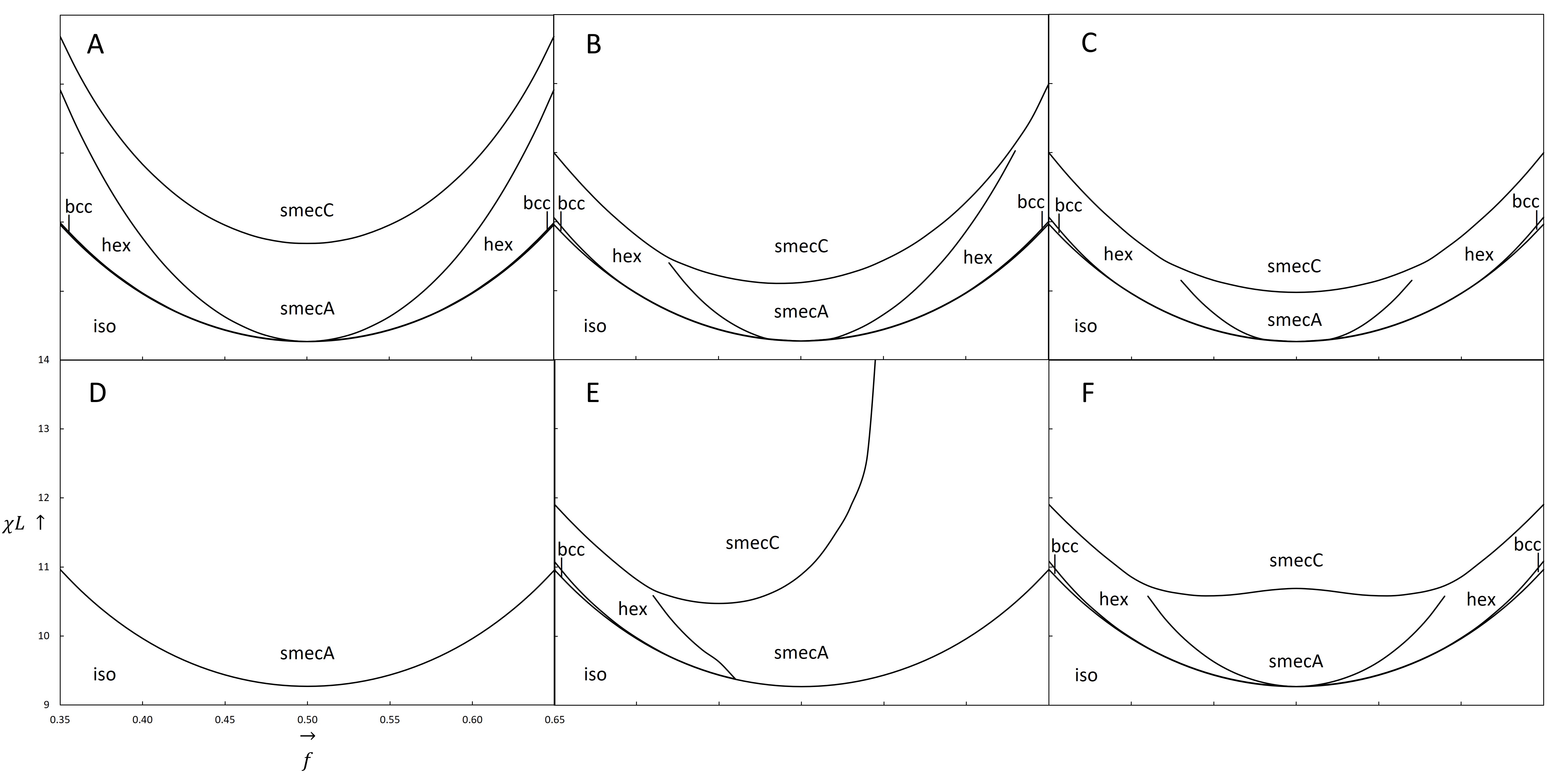} \newline

In Fig. (4) the phase diagram has been determined in which both the A- and B-block are semi-flexible,
$\frac{\lambda _{A}}{L_{A}} =\frac{\lambda _{B}}{L_{B}} =10^{ -1}$. In the previous phase diagrams it was justified to neglect the orientation of the B-block $\Upsilon _{B}$, because the B-block is totally flexible. However in Fig. (4) both the A- and B-block are semi-flexible so that both orientations
$\Upsilon _{A}$ and $\Upsilon _{B}$ have to be taken into account. Not only the Maier-Saupe parameter $\omega _{A A}$ influences the phase behaviour, but also the other parameters $\omega _{A B}$ and $\omega _{B B}$ may change the structure of the melt. In Fig. (4) it has been investigated how the phase behaviour is influenced by the three parameters $\omega _{A A}$, $\omega _{A B}$ and $\omega _{B B}$. The Maier-Saupe paramaters are coupled to the Flory-Huggins interaction $\chi $ by choosing each ratio $r_{\alpha  \beta } =\frac{\omega _{\alpha  \beta }}{\chi }$ constant. The three parameters $r_{\alpha  \beta }$ can have two different values $0$ and $5$ which gives 6 different combinations if we choose $r_{A A} \leq r_{B B}$. Because of symmetry it is not neccessary to consider combinations in which $r_{A A} >r_{B B}$. For each combination a phase diagram is calculated. In Fig. (4A), (4B) and (4C) $r_{A B}=0$ is constant and the other parameters $r_{A A}$ and $r_{B B}$ are varied. In Fig. (4A) and (4C) these parameters are equal and given by $r_{A A}=r_{B B}=0$ and $r_{A A}=r_{B B}=5$, respectively. The unequal combination $r_{A A}=0$ and $r_{B B}=5$ is applied in Fig. (4B). In Fig. (4D), (4E) and (4F) $r_{A B} =5$ and $r_{A A}$ and $r_{B B}$ are varied in the same way as in Fig. (4A), (4B) and (4C). If we compare Fig. (4A) and (4B) we see in Fig. (4B) that the smectic-C
domain is shifted to a smaller $\chi L$. The parameter has also a very small influence on the bcc line and the hexagonal domain which cannot be seen in Fig. (4B). The bcc
line becomes broader and in the neigbourhood of the critical point the size of the hexagonal domain increases. The influence of $r_{B B}$ is larger in the left part of the phase diagram, because here the B-block is longer. In the same way the parameter $r_{A A}$ influences the phase behaviour in the right part of the phase diagram which can be seen in Fig. (4C). In Fig. (4D), (4E) and (4F) in
which $r_{A B} = 5$ we see the same effect if $r_{A A}$ and $r_{B B}$ are changed. The parameter $r_{A B}$ has the contrary effect. If $r_{A B}$ is increased the size of the bcc and hexagonal domain becomes smaller and the smectic-A part becomes larger. The smectic-C part is
shifted to a larger $\chi  L$. The influence of $r_{A B}$ is stronger in the neighbourhood of $f =0.5$, because here more interaction between A- and B-blocks is possible. In Fig. (4D) $r_{A A} =r_{B B} =0$ so that the influence of $r_{A B}$ is very strong. Here the bcc, hexagonal and smectic-C phase have even disappeared and only the smectic-A phase is formed when $\chi  L >\chi _{s} L$. In the other phase diagrams in Fig. (4E) and (4F) the effect of $r_{A B}$ is compensated by the contrary effect of $r_{A A}$ or $r_{B B}$. This can be illustrated by means of Fig. (4A) and (4F) in which the three $r$-parameters are equal, $r_{A A} =r_{A B} =r_{B B}$. In these figures at $f =0.5$ the smectic-C state is formed at exactly the same Flory-Huggins strength $\chi  L =10.69$. Here the influence of $r_{A B}$ is exactly compensated by $r_{A A}$ and $r_{B B}$.

In the smectic-C state in Fig. (4) the orientation $\theta $ is $\theta  =54.7$ degrees which is the same as in Fig. (3). Then in the primary eigenvector $\underline{x}_{1} =(\Psi _{1} ,\Upsilon _{1}^{A} ,\Upsilon _{1}^{B})$ the orientation parameters $\Upsilon _{1}^{A}$ and $\Upsilon _{1}^{B}$ vanish. The influence of $r_{A A}$ and $r_{B B}$ on the phase behaviour is the same as the influence of $r$ observed in Fig. (3) and can be explained in the same way. The parameter $r_{A B}$ has a contrary effect on the microphase structure of the melt which is rather obvious. 

The phase
diagram has also been determined in which both the A- and B-block have a persistence length close to zero, $\frac{\lambda _{A}}{L_{A}} =\frac{\lambda _{B}}{L_{B}} =10^{ -4}$. Here the diblock chain can be regarded as totally flexible. In the general expression of the Landau free energy in Appendix B given by Eq. (\ref{FL}) the terms which contain an oriention tensor $\Upsilon ^{\overline{a}}$ will dissappear if the persistence length of each block approaches zero. Therefore it is justified to neglect the orientation and write
the Landau free energy as a function of only the density order-parameter $\Psi ^{a}$. In this way the phase diagram is calculated which is in agreement with Leibler's result in Ref. \cite{Leibler}.

In Fig. (4) and Ref. \cite{Leibler} the critical point in the phase diagram is found at $f =f_{c} =0.5$, but in Fig. (1) and (3) $f_{c}$ has been shifted to $f_{c} =0.68$. If $f <f_{c}$ the bcc and hexagonal phase are formed by A-rich domains which prefer to be embedded by a B-rich matrix. The size of the embedded A-rich
domains increases when the length $L_{A}$ becomes larger. At $f >f_{c}$ the size is too large and therefore the melt prefers the reverse state in which the B-blocks are embedded by a matrix of A-blocks.
Exactly at the critical point the probability that an A-block is inside the matrix is the same as the probability to be outside the matrix. Then a nonzero
density order-parameter $\Psi $ is not possible. Only the smectic-A or C state can be formed at the critical point. In Fig. (4) and Ref. \cite{Leibler} the A- and B-block
have the same ratio $\lambda _{\alpha }/L_{\alpha }$, so that at $f =0.5$ the A- and B-block are mathematically identical. This explains why the critical point is reached at $f =0.5$. The critical point in Fig. (1) and (3) is shifted to $f =0.68$ at which the stiffer A-block is longer than the flexible B-block. So the stiffer A-block prefers to be embedded by a matrix of flexible
B-blocks in a larger domain of the phase diagram. The stiffness of the A-block makes it difficult to form an A-rich matrix. The number of possible melt
configurations could be bigger if a matrix of flexible B-blocks is formed. Even if the B-block is shorter than the A-block more melt configurations could
still be possible because of the large flexibility of the B-blocks. Only if the B-block becomes too short the melt prefers to form the reverse state.\newline

\includegraphics[scale=0.50]{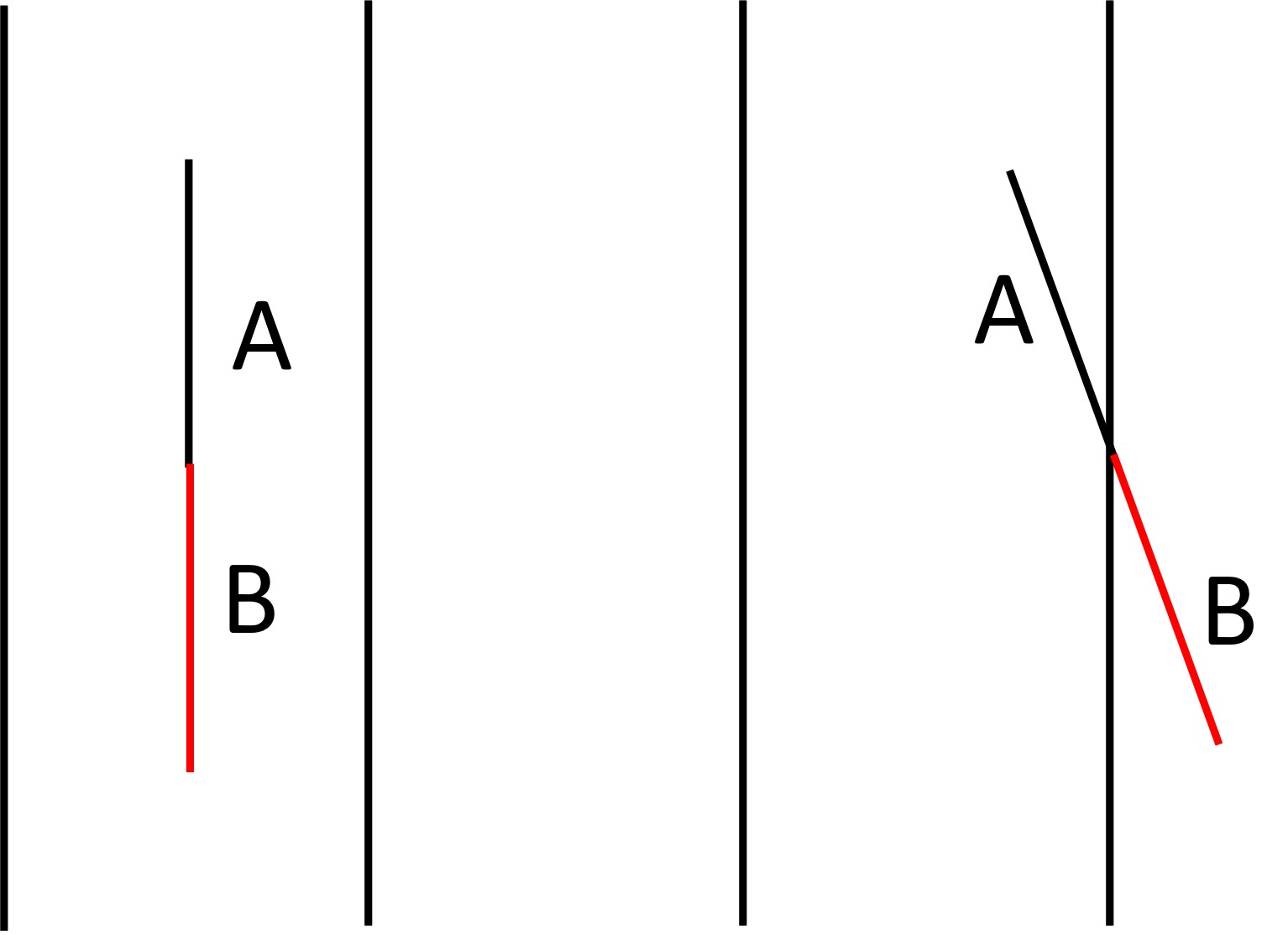} \newline

In each phase diagram in this section the smectic-A state is formed prior to the smectic-C state when $\chi $ is increasing. By means of Fig. (5) it is explained why the A- and B-blocks could be separated easier in the smectic-C state. In Fig.
(5) an A-block in the the middle of an A-rich layer is drawn in the smectic-A and C state. Extensive analysis and visualizations of the orientations of A- and B-blocks show that in the smectic-A phase the A-blocks
are stronger aligned along the wave vector $\underline{q}$ in a B-rich layer. However, in the middle of an A-rich layer there are on average some more A-blocks which are oriented perpendicular to $\underline{q}$. This will also bring some more B-blocks to the middle of the A-rich layer which will make the separation between A- and B-blocks weaker.
So in the smectic-A phase only a weak separation is possible. In the smectic-C phase in Fig. (5) the A-block is rotated which will remove the B-block from
the middle of the A-rich layer. Then a stronger separation becomes possible. Additionally the smectic-A state is rotational symmetric with respect to an arbitrary
axis perpendicular to the layers. The smectic-C state does not have this symmetry and is therefore entropically less favourable. This entropy loss must
also be compensated by a larger $\chi $. This also may explain why the smectic-C phase is always formed at a larger $\chi $.

Close to the order-disorder phase transition just above the spinodal $\chi _{s}$ the bcc phase is always formed prior to the hexagonal phase when $\chi $ is increasing and the hexagonal phase appears prior to a smectic-A or -C state. These sequence of microphase structures in the phase
diagram is strongly determined by total contact area between A-rich and B-rich domains in the whole melt. The contact areas $A_{b c c}$, $A_{h e x}$ and $A_{s m e c}$ of the bcc, hexagonal and smectic phase are related as $A_{b c c} :A_{h e x}$ $ :A_{s m e c} =3 :2 :1$. At a smaller $\chi $ the melt prefers a microphase with a larger contact area so that the mixing entropy is larger. At a larger $\chi $ a state with a smaller contact area is formed which is favourable because of the separation enthalpy.

\section{Concluding remarks}

In this paper in section "The model" and Appendix A we develop a general theory of a polydisperse semi-flexible multi-block copolymer melt in which an expression of the Landau free energy has been derived. This general theory is an extension of the theory in Ref. \cite{Slot} in which the chains are totally flexible. In that case a possible orientation of chains can be neglected so that the Maier-Saupe interaction cannot have any influence on the phase behaviour. Only the Flory-Huggins interaction and density order-parameters are included in the derivation of the Landau free energy. However, in this paper the theory has been extended by adding a certain bending stiffness to each block using the Bawendi-Freed approach. Due to the stiffness orientation of chains may become possible so that the Maier-Saupe interaction and orientation order-parameters have to be included in the derivation of the Landau free energy besides the Flory-Huggins interaction and density order-parameters.      

In Appendix B the general theory has been applied to a less complicated melt of monodisperse semi-flexible diblock copolymers. The derived expression of the Landau free energy of this melt has been written in terms of eigenvalues and eigenvectors of the matrices in the second order term given by Eq. (\ref{matrixvorm}). In this way the minimum of the Landau free
energy with respect to the order-parameters can be determined analytically. Besides a density order-parameter also the orientation order-parameters of the A-
and B-block are taken into account. These orientation order-parameters contain a global and a local contribution. The local contribution is described in the smectic-A phase and in more complicated structures such as the hexagonal phase. This description of the orientation order-parameters
is applied in the expression of the Landau free energy. By means of the Landau theory the phase structure can be predicted as a function of the composition,
persistence length and the strength of the Flory-Huggins and Maier-Saupe interaction. In several phase diagrams the bcc, hexagonal, smectic-A, smectic-C
and nematic phase are observed. First a diblock melt has been investigated in which the A-block is a rigid rod and the B-block is a totally flexible. In
such a rod-coil diblock melt it has appeared that the Maier-Saupe interaction $\omega _{A A}$ between stiff A-blocks weakens microphase separation. At a larger $\chi  L$ a smectic-C phase is found in which $8 <\theta  <18$ degrees. If the stiff A-block is replaced by a semi-flexible block a different phase behaviour is observed.
Here the Maier-Saupe interaction $\omega _{A A}$ weakens microphase separation only close to the spinodal $\chi _{s} L$. At a larger $\chi  L$ a smectic-C phase is found in which $\theta $ is always $\theta  =54.7$ degrees. In this smectic-C phase the separation between A- and B-blocks is stronger than in other microphase structures which are
found in the phase diagram. However, the orientation order-parameter of both the A- and B-block is negligible small. So this smectic-C phase can be regarded
as a lamellar phase with only a nonzero density order-parameter. This smectic-C phase is shifted to a smaller $\chi  L$ when $\omega _{A A}$ increases. Then microphase separation is enhanced by $\omega _{A A}$ if the melt is converted into the smectic-C phase with $\theta  =54.7$ degrees. The phase diagram is also calculated if both the A- and B-block are semi-flexible.
Then not only the Maier-Saupe interaction $\omega _{A A}$ between A-blocks, but also the parameters $\omega _{A B}$ and $\omega _{B B}$ may influence the phase behaviour. The Maier-Saupe interaction $\omega _{A A}$ and $\omega _{B B}$ weakens microphase separation close to the spinodal $\chi _{s} L$. At a larger $\chi  L$ a smectic-C with again $\theta  =54.7$ degrees is found which is shifted to a smaller $\chi  L$ when $\omega _{A A}$ or $\omega _{B B}$ increases. This smectic-C phase can also be regarded as a lamellar phase without any alignment of A- or B-blocks. The effect of the
parameter $\omega _{A B}$ on the phase behaviour appeared to be contrary to the effect of $\omega _{A A}$ and $\omega _{B B}$. 

This paper contains three new aspects which have not yet been applied or investigated in other papers before. First the Bawendi-Freed approach is applied to model semi-flexibilty. This approach makes the further derivation of the Landau free energy in the weak segregation regime possible. Secondly the complete phase diagram has been calculated for a melt of monodisperse semi-flexible diblock copolymers. Other papers are restricted to spinodal calculations or the complete phase diagram has been calculated for more simplified systems. Thirdly a space dependent orientation order-parameter is described in microphase structures such as the hexagonal phase. In other papers this was not necessary, because these were restricted to spinodal calculations or the space dependent orientation order-parameter was not included in the calcalution of the complete phase diagram.   

\section*{Appendix A}
\markboth{Appendix A}{}\renewcommand{\theequation}{A.\arabic{equation}}%
\setcounter{equation}{0}

In this appendix the Landau free energy given by Eq. (\ref{vrijeenergie}) is
derived. The starting point of the derivation is the partition function $Z$,
i.e. the sum of the \textit{Boltzmann weights} over all allowed states of
the system. The set of all allowed states furnishes the so-called \textit{%
state-space} or \textit{configuration-space }$\Gamma $ of the system, which
in this case is given by, 
\begin{equation}
\Gamma \equiv \{\{\underline{R}_{m}^{s},\underline{u}_{m}^{s}\}_{sm}\mid 
\underline{u}_{m}^{s}\equiv \underline{\dot{R}}_{m}^{s},\;\forall \;m,s\quad
\&\quad \hat{\rho}(\underline{x})=1,\;\forall \underline{x}\in V\}.
\end{equation}%
As we are ultimately only interested in differences in free energy between
possible inhomogeneous and/or anisotropic phases of the system, all
combinatorial terms will be left out of this partition function since they
only lead to constant terms in the free energy. With this in mind $Z$ can be
written as, 
\begin{equation}
Z\equiv \prod_{sm}\int d^{3}\widetilde{U}_{m}^{s}\int d^{3}U_{m}^{s}G(\{%
\underline{U}_{m}^{s},L_{s}\}\mid \{\widetilde{\underline{U}}_{m}^{s},0\}),
\label{21}
\end{equation}%
where the \textit{orientational Green's function} $G(\{\underline{U}%
_{m}^{s},L_{s}\}\mid \{\widetilde{\underline{U}}_{m}^{s},0\})$ is defined
by, 
\begin{equation}
G\equiv \prod_{sm}\int D\underline{R}_{m}^{s}\,\int_{(0,\widetilde{%
\underline{U}}_{m}^{s})}^{(L_{s},\underline{U}_{m}^{s})}D\underline{u}%
_{m}^{s}\,\delta \left[ \underline{R}_{m}^{s}-\int dl\,\underline{u}%
_{m}^{s}(l)\right] \,\delta \left[ 1-\hat{\rho}\,\right] \,e^{-(\hat{H}_{0}+%
\hat{W})}  \label{22}
\end{equation}%
which gives the probability that each chain has a certain initial and final
orientation. In this coarse grained description incompressibility, which is
due to interactions at the molecular level, has to be explicitly accounted
for via the delta function $\delta \left[ 1-\hat{\rho}\,\right] $. This
partition function will be transformed in four steps into a form which is
more amendable for further analysis. The first step involves a formal shift
of the state-variable dependence of $e^{-\hat{W}}$ in (\ref{22}). This is
done by introducing the following two \textit{functional decompositions of
the identity} into $G$,

\begin{equation}
\prod_{\mu }\text{\/}^{\prime }\int D\psi ^{\mu }\,\delta \left[ \psi ^{\mu
}-\hat{\psi}^{\mu }\right] =1  \label{23}
\end{equation}%
and 
\begin{equation}
\prod_{\nu }\text{\/}\int D\underline{\underline{Q}}^{\nu }\,\,\delta \left[ 
\underline{\underline{Q}}^{\nu }-\underline{\underline{\hat{Q}}}^{\nu }%
\right] =\underline{\underline{I}}  \label{24}
\end{equation}%
which yield, 
\begin{eqnarray}
G &=&\prod_{\mu }\text{\/}^{\prime }\int D\psi ^{\mu }\prod_{\nu }\text{\/}%
\int D\underline{\underline{Q}}^{\nu }\,e^{-W}\times  \notag \\
&&\times \prod_{sm}\int D\underline{R}_{m}^{s}\,\int_{(0,\widetilde{%
\underline{U}}_{m}^{s})}^{(L_{s},\underline{U}_{m}^{s})}D\underline{u}%
_{m}^{s}\,e^{-\hat{H}_{0}}\,\delta \left[ \underline{R}_{m}^{s}-\int dl\,%
\underline{u}_{m}^{s}(l)\right] \,\delta \left[ 1-\hat{\rho}\,\right] \times
\notag \\
&&\times \prod_{\lambda }\text{\/}^{\prime }\,\prod_{\eta }\text{\/}\delta %
\left[ \psi ^{\lambda }-\hat{\psi}^{\lambda }\right] \,\delta \left[ 
\underline{\underline{Q}}^{\eta }-\underline{\underline{\hat{Q}}}^{\eta }%
\right] .  \label{25}
\end{eqnarray}%
The elements $Q_{ij}^{\nu }$ and $Q_{ji}^{\nu }$ of the tensor $\underline{%
\underline{Q}}^{\nu }$ in which $i\neq j$ are identical, but in Eq. (\ref{25}%
) they are treated as independent parameters. At a certain point this would
obstruct the further derivation of the Landau free energy. This happens when
we solve Eq. (\ref{74}) and (\ref{75}) iteratively, because then the matrix $%
A^{\overline{a}\overline{b}}$ would not be invertible. In this matrix the
rows and columns in which $ij=xy$, $yz$ and $xz$ would be identical to the
rows and collumns with reversed indices $ij=yx$, $zy$ and $zx$. Such a
matrix is not invertible. Therefore in the functional integral in Eq. (\ref%
{25}) only the elements $Q_{ij}^{\nu }$ with unique pairs $xx$, $yy$, $zz$, $%
xy$, $yz$ and $xz$ may occur if we want to derive the desired expression of
the Landau free energy. In the rest of the derivation we will ignore the
other pairs $yx$, $zy$ and $zx$. In the interaction energy $W$ given by Eq. (%
\ref{14}) the terms containing elements $Q_{ij}^{\nu }$ with $i\neq j$ are
counted twice.

The second step involves substitution of the following \textit{functional
spectral representations }for the last $2M$ ''delta-functions''\ in the
above expression, i.e., 
\begin{eqnarray}
\delta \left[ 1-\hat{\rho}\,\right] &\equiv &\int
Dh^{0}\,e^{i\int_{V}d^{3}x\,h^{0}(\underline{x})\,\{1-\hat{\rho}(\underline{x%
})\}}  \label{26} \\
\delta \left[ \psi ^{\lambda }-\hat{\psi}^{\lambda }\right] &\equiv &\int
Dh^{\lambda }\,e^{i\int_{V}d^{3}x\,h^{\lambda }(\underline{x})\,\{\psi
^{\lambda }(\underline{x})-\hat{\psi}^{\lambda }(\underline{x})\}}\quad
(\lambda =1,...,M-1)  \notag \\
&&  \label{27} \\
\delta \left[ \underline{\underline{Q}}^{\eta }-\underline{\underline{\hat{Q}%
}}^{\eta }\right] &\equiv &\int D\underline{\underline{K}}^{\eta
}\,e^{i\int_{V}d^{3}x\,\underline{\underline{K}}^{\eta }(\underline{x}%
)\,:\,\{\underline{\underline{Q}}^{\eta }(\underline{x})-\underline{%
\underline{\hat{Q}}}^{\eta }(\underline{x})\}}\quad (\eta =1,...,M)  \notag
\\
&&\quad  \label{28}
\end{eqnarray}%
resulting in, 
\begin{eqnarray}
G &=&\prod_{\mu }\text{\/}^{\prime }\int D\psi ^{\mu }\prod_{\nu }\text{\/}%
\int D\underline{\underline{Q}}^{\nu }\,e^{-W}\times  \notag \\
&&\times \int Dh^{0}\prod_{\lambda }\text{\/}^{\prime }\int Dh^{\lambda
}\prod_{\eta }\text{\/}\int D\underline{\underline{K}}^{\eta
}\,e^{i\int_{V}d^{3}x\,\{h^{0}(\underline{x})\,+\sum\limits_{\alpha }\text{\/%
}^{^{\prime }}\,h^{\alpha }(\underline{x})\,\psi ^{\alpha }(\underline{x}%
)+\sum\limits_{\beta }\underline{\underline{K}}^{\beta }(\underline{x}%
)\,\,:\,\underline{\underline{Q}}^{\beta }(\underline{x})\}}\times  \notag \\
&&\times \prod_{sm}\int D\underline{R}_{m}^{s}\,\int_{(0,\widetilde{%
\underline{U}}_{m}^{s})}^{(L_{s},\underline{U}_{m}^{s})}D\underline{u}%
_{m}^{s}\,e^{-\hat{H}_{0}}\,\delta \left[ \underline{R}_{m}^{s}-\int dl\,%
\underline{u}_{m}^{s}(l)\right] \,\times  \notag \\
&&\times e^{-\,\,i\int_{V}d^{3}x\,\{h^{0}(\underline{x})\,\,\hat{\rho}(%
\underline{x})\,+\sum\limits_{\alpha }\text{\/}^{^{\prime }}\,h^{\alpha }(%
\underline{x})\,\hat{\psi}^{\alpha }(\underline{x})+\sum\limits_{\beta }%
\underline{\underline{K}}^{\beta }(\underline{x})\,\,:\,\underline{%
\underline{\hat{Q}}}^{\beta }(\underline{x})\}}\,.  \label{29}
\end{eqnarray}%
In the third step the auxiliary integration fields $h^{0}(\underline{x}%
),h^{1}(\underline{x}),...,h^{M-1}(\underline{x})$ are transformed to new
fields $J^{1}(\underline{x}),...,J^{M}(\underline{x})$, defined in the
following way, 
\begin{eqnarray}
J^{\alpha }(\underline{x}) &\equiv &h^{\alpha }(\underline{x})+h^{0}(%
\underline{x})\quad (\alpha =1,...,M-1)  \notag \\
J^{M}(\underline{x}) &\equiv &h^{0}(\underline{x}).  \label{30}
\end{eqnarray}%
Using this isometric transformation and (\ref{2}), (\ref{8}), (\ref{9}) and (%
\ref{10}) it is easily verified that the following identities hold, 
\begin{subequations}
\begin{eqnarray}
h^{0}(\underline{x})\,\,\hat{\rho}(\underline{x})\,+\sum\limits_{\alpha }%
\text{\/}^{^{\prime }}\,h^{\alpha }(\underline{x})\,\hat{\psi}^{\alpha }(%
\underline{x}) &=&J^{M}(\underline{x})+\sum\limits_{\alpha }\text{\/}%
^{^{\prime }}\,J^{\alpha }(\underline{x})\,\hat{\rho}^{\alpha }(\underline{x}%
)-\sum_{\alpha }f^{\alpha }\,J^{\alpha }(\underline{x})  \notag \\
&& \\
h^{0}(\underline{x})\,+\sum\limits_{\alpha }\text{\/}^{^{\prime
}}\,h^{\alpha }(\underline{x})\,\psi ^{\alpha }(\underline{x}) &=&J^{M}(%
\underline{x})+\sum\limits_{\alpha }\text{\/}J^{\alpha }(\underline{x}%
)\,\psi ^{\alpha }(\underline{x})
\end{eqnarray}%
and so by combining (\ref{29}) with (\ref{21}) $Z$ can be written as, 
\end{subequations}
\begin{eqnarray}
Z &=&\prod_{\mu }\text{\/}^{\prime }\int D\psi ^{\mu }\prod_{\nu }\text{\/}%
\int D\underline{\underline{Q}}^{\nu }\,e^{-W}\times  \notag \\
&&\times \prod_{\lambda }\text{\/}\int DJ^{\lambda }\prod_{\eta }\text{\/}%
\int D\underline{\underline{K}}^{\eta
}\,e^{i\int_{V}d^{3}x\,\{\sum\limits_{\alpha }\text{\/}J^{\alpha }(%
\underline{x})\,[\psi ^{\alpha }(\underline{x})+f^{\alpha
}]+\sum\limits_{\beta }\underline{\underline{K}}^{\beta }(\underline{x}%
)\,\,:[\,\underline{\underline{Q}}^{\beta }(\underline{x})+\frac{1}{3}\rho
^{\beta }(\underline{x})\,\underline{\underline{I}}]\}}\times  \notag \\
&&\times \langle e^{-\,\,i\int_{V}d^{3}x\,\{\sum\limits_{\alpha }\text{\/}%
J^{\alpha }(\underline{x})\,\hat{\rho}^{\alpha }(\underline{x}%
)+\sum\limits_{\beta }\underline{\underline{K}}^{\beta }(\underline{x}%
)\,\,:\,\underline{\underline{\widehat{Q}}}^{\beta }(\underline{x}%
)\}}\,\rangle _{0}.  \label{32}
\end{eqnarray}%
In this expression $\langle ...\rangle _{0}$ denotes an average with respect
to the \textit{unperturbed ensemble of chain conformations }defined by $%
\hat{H}_{0}$, i.e., 
\begin{eqnarray}
\langle \,\hat{A}\,\rangle _{0} &\equiv &\prod_{sm}\int d^{3}\widetilde{U}%
_{m}^{s}\int d^{3}U_{m}^{s}\int D\underline{R}_{m}^{s}\,\int_{(0,\widetilde{%
\underline{U}}_{m}^{s})}^{(L_{s},\underline{U}_{m}^{s})}D\underline{u}%
_{m}^{s}\,e^{-\hat{H}_{0}}\,\times  \notag \\
&&\times \delta \left[ \underline{R}_{m}^{s}-\int dl\,\underline{u}%
_{m}^{s}(l)\right] \,\,\hat{A},  \label{33}
\end{eqnarray}%
where $\hat{H}_{0}$ is the \textit{unperturbed Hamiltonian} according to the Bawendi-Freed approach which has earlier been defined in section "The model" in Eq. (\ref{18}). In Eq. (\ref{33}) the functional integrations over $\{\underline{R}_{m}^{s}\}$ and $\{%
\underline{u}_{m}^{s}\}$ are defined in such a way that $\langle 1\,\rangle
_{0}\equiv 1$. The last step in the transformation of $Z$ boils down to
rewriting the integrand of (\ref{32}) using the fields $\tilde{J}^{1}(%
\underline{x}),...,\tilde{J}^{M}(\underline{x})$ in the spirit of \cite%
{Fredrickson1}, 
\begin{equation}
\tilde{J}^{\alpha }(\underline{x})\equiv J^{\alpha }(\underline{x})-\frac{1}{%
V}\int_{V}d^{3}y\,J^{\alpha }(\underline{y})\quad (\alpha =1,...,M).
\label{JKthilde}
\end{equation}%
It is easy to see that the use of these new fields in conjunction with (\ref%
{4}) will eliminate the terms in (\ref{32}) involving $f^{\alpha }$. Thus we
finally end up with, 
\begin{eqnarray}
Z &=&\prod_{\mu }\text{\/}^{\prime }\int D\psi ^{\mu }\prod_{\nu }\text{\/}%
\int D\underline{\underline{Q}}^{\nu }\,e^{-W}\times  \notag \\
&&\times \prod_{\lambda }\text{\/}\int DJ^{\lambda }\prod_{\eta }\text{\/}%
\int D\underline{\underline{K}}^{\eta
}\,e^{i\int_{V}d^{3}x\,\{\sum\limits_{\alpha }\text{\/}\tilde{J}^{\alpha }(%
\underline{x})\,\psi ^{\alpha }(\underline{x})+\sum\limits_{\beta }%
\underline{\underline{K}}^{\beta }(\underline{x})\,\,:\,\underline{%
\underline{Q}}^{\beta }(\underline{x})\}+\Lambda }  \label{35A}
\end{eqnarray}%
with $\Lambda $ defined by, 
\begin{equation}
\Lambda \equiv \ln \,\langle e^{-\,\,i\int_{V}d^{3}x\,\{\sum\limits_{\alpha }%
\text{\/}\tilde{J}^{\alpha }(\underline{x})\,\hat{\rho}^{\alpha }(\underline{%
x})+\sum\limits_{\beta }\underline{\underline{K}}^{\beta }(\underline{x}%
)\,\,:\,\underline{\underline{\widehat{Q}}}^{\beta }(\underline{x}%
)\}}\,\rangle _{0}.  \label{36}
\end{equation}%
By Fourier-transforming all the integrals involving the $\psi $, $\underline{\underline{Q}}$, $J $ and $\underline{\underline{K}}$ fields and making use of the fact that
according to the definition of $\tilde{\chi}_{\alpha \beta }$ (see Eq. (\ref{16}%
)) $\tilde{\chi}_{MM}\equiv 0$, the partition function $Z$ (\ref{35A}) can be
written as,
\begin{equation}
Z\equiv \prod_{c}\text{\/}^{\prime }\prod_{\overline{d}}\text{\/}\int D\Psi
^{c}\int D\Upsilon ^{\overline{d}}\,e^{V\,\{\tilde{\chi}_{ab}\Psi ^{a}\Psi
^{b}+\frac{1}{2}\omega _{\overline{a}\overline{b}}\Upsilon ^{\overline{a}%
}\Upsilon ^{\overline{b}}\}}\,\widetilde{Z}[\underline{\Psi },\overline{%
\Upsilon }]  \label{71A}
\end{equation}%
with $\tilde{\chi}_{ab}\equiv \tilde{\chi}_{\alpha \beta }\,\delta (%
\underline{q}_{1}+\underline{q}_{2})$, $\omega _{\overline{a}\overline{b}%
}\equiv \omega _{\alpha \beta }\,|2\delta _{ii^{\prime }}\delta _{jj^{\prime
}}-\delta _{ij}\delta _{i^{\prime }j^{\prime }}|\,\delta (\underline{q}_{1}+%
\underline{q}_{2})$, $\Psi ^{a}$ $\equiv \frac{\psi ^{\alpha }(-\underline{q}%
)}{V}$, $\Upsilon ^{\overline{a}}\equiv \frac{Q_{ij}^{\alpha }(-\underline{q}%
)}{V}$ and 
\begin{equation}
\widetilde{Z}[\underline{\Psi },\overline{\Upsilon }]\equiv \prod_{g}\prod_{%
\overline{h}}\int Dv_{g}\int Dw_{\overline{h}}\,\,e^{V\,\{i\,[v_{a}\Psi
^{a}+w_{\overline{a}}\Upsilon ^{\overline{a}}]+\frac{\Lambda \lbrack 
\underline{v},\overline{w}]}{V}\,\}},  \label{72A}
\end{equation}%
where $v_{a}\equiv \frac{J^{\alpha }(\underline{q})}{V}$, $w_{\overline{a}}\equiv \frac{K_{ij}^{\alpha }(\underline{q})}{V}$, $\underline{\Psi }\equiv \{\Psi ^{a}\}_{a}$ and $\overline{\Upsilon }%
\equiv \{\Upsilon ^{\overline{a}}\}_{\overline{a}}$.

In (\ref{71A}) and (\ref{72A}) we use the \textit{composite labels} $%
a\equiv (\underline{q}_{1}\neq \underline{0},\alpha )$, $b\equiv (\underline{%
q}_{2}\neq \underline{0},\beta )$ etc. and $\overline{a}\equiv (\underline{q}%
_{1},ij,\alpha )$, $\overline{b}\equiv (\underline{q}_{2},i^{\prime
}j^{\prime },\beta )$ etc. in which the
pairs $ij$ and $i^{\prime }j^{\prime }$ are one of the six unique pairs $xx$%
, $yy$, $zz$, $xy$, $xz$ and $yz$. Furthermore we see that each composite label $a$, $b$, $\overline{a}$ or $\overline{b}$ in a subscript matches with a composite label in a superscript. In such a match the Einstein summation convention is applied. More details about the derivation of Eq. (\ref{71A}) and (\ref{72A}) from Eq. (\ref{35A}) can be found in Supplemental material III.

For large values of the
system's volume $V$, $\widetilde{Z}[\underline{\Psi },\overline{\Upsilon }]$
can be evaluated with the well-known \textit{saddle-point method}, i.e.
approximating $\widetilde{Z}[\underline{\Psi },\overline{\Upsilon }]$ by, 
\begin{equation}
\widetilde{Z}[\underline{\Psi },\overline{\Upsilon }]\simeq e^{V\,\Phi
\lbrack \underline{\Psi },\overline{\Upsilon }]},  \label{73A}
\end{equation}%
where $\Phi \lbrack \underline{\Psi },\overline{\Upsilon }]$ is the
stationary value of $i\,[v_{a}\Psi ^{a}+w_{\overline{a}}\Upsilon ^{\overline{%
a}}]+\frac{\Lambda \lbrack \underline{v},\overline{w}]}{V}$ with respect to
the set of $v$'s and $w$'s for which its absolute value is the smallest.
This stationary point is a solution of the following set of equations, 
\begin{eqnarray}
i\,\Psi ^{a} &=&A^{ab}\,v_{b}+A^{a\overline{b}}\,w_{\overline{b}}-\frac{i}{2}%
\,B^{abc}\,v_{b}v_{c}-i\,B^{a\overline{b}c}\,w_{\overline{b}}v_{c}-\frac{i}{2%
}\,B^{a\overline{b}\overline{c}}\,w_{\overline{b}}w_{\overline{c}}+  \notag
\\
&&-\frac{1}{6}\,C^{abcd}\,v_{b}v_{c}v_{d}-\frac{1}{2}\,C^{a\overline{b}%
cd}\,w_{\overline{b}}v_{c}v_{d}-\frac{1}{2}\,C^{a\overline{b}\overline{c}%
d}\,w_{\overline{b}}w_{\overline{c}}v_{d}-\frac{1}{6}\,C^{a\overline{b}%
\overline{c}\overline{d}}\,w_{\overline{b}}w_{\overline{c}}w_{\overline{d}} 
\notag \\
&&,\forall a  \label{74A}
\end{eqnarray}%
and 
\begin{eqnarray}
i\,\Upsilon ^{\overline{a}} &=&A^{\overline{a}\overline{b}}\,w_{\overline{b}%
}+A^{\overline{a}b}\,v_{b}-\frac{i}{2}\,B^{\overline{a}bc}\,v_{b}v_{c}-i\,B^{%
\overline{a}\overline{b}c}\,w_{\overline{b}}v_{c}-\frac{i}{2}\,B^{\overline{a%
}\overline{b}\overline{c}}\,w_{\overline{b}}w_{\overline{c}}+  \notag \\
&&-\frac{1}{6}\,C^{\overline{a}bcd}\,v_{b}v_{c}v_{d}-\frac{1}{2}\,C^{%
\overline{a}\overline{b}cd}\,w_{\overline{b}}v_{c}v_{d}-\frac{1}{2}\,C^{%
\overline{a}\overline{b}\overline{c}d}\,w_{\overline{b}}w_{\overline{c}%
}v_{d}-\frac{1}{6}\,C^{\overline{a}\overline{b}\overline{c}\overline{d}}\,w_{%
\overline{b}}w_{\overline{c}}w_{\overline{d}}  \notag \\
&&,\forall \overline{a}.  \label{75A}
\end{eqnarray}%

In Eq. (\ref{74A}) en (\ref{75A}) the $A$'s, $B$'s and $C$'s are second, third and fourth order single-chain correlation functions, respectively. These correlation functions are introduced in Supplemental material III and calculated in Apppendix B.   
As we ultimately want to arrive at a Landau free energy as an expansion
up-to fourth order in the $\Psi ^{a}$ - and the $\Upsilon ^{\overline{a}}$
fields, we only need to solve these last two vector-equations iteratively
for $v_{a}$ and $w_{\overline{a}}$ up-to third order in the $\Psi $'s and
the $\Upsilon $'s. This iterative solution can be found in Supplemental material II and is subsituted into $i\,[v_{a}\Psi ^{a}+w_{\overline{a}}\Upsilon ^{%
\overline{a}}]+\frac{\Lambda \lbrack \underline{v},\overline{w}]}{V}$ such that we
obtain $\Phi \lbrack \underline{\Psi },\overline{\Upsilon }]$ and hence the
partition function, 
\begin{equation}
Z\simeq \prod_{c}\text{\/}^{\prime }\prod_{\overline{d}}\text{\/}\int D\Psi
^{c}\int D\Upsilon ^{\overline{d}}\,e^{V\,\{\tilde{\chi}_{ab}\Psi ^{a}\Psi
^{b}+\frac{1}{2}\omega _{\overline{a}\overline{b}}\Upsilon ^{\overline{a}%
}\Upsilon ^{\overline{b}}+\Phi \lbrack \underline{\Psi },\overline{\Upsilon }%
]\}}.\,  \label{87A}
\end{equation}%
The Landau free energy, that is the free energy of the system within the 
\textit{mean field approximation}, can be obtained by again applying the
saddle-point method, but now to approximately evaluate this last set of
functional integrals. If we write the result as, 
\begin{equation}
Z\simeq e^{-F_{L}},  \label{88}
\end{equation}%
then this Landau free energy $F_{L}$ (in units of $k_{B}T$) is given by, 
\begin{gather}
\frac{F_{L}}{V}=\underset{\underline{\Psi },\overline{\Upsilon }}{\min }%
\{(\Gamma _{ab}^{(2)}-\widetilde{\chi }_{ab})\Psi ^{a}\Psi ^{b}+2\Gamma _{a%
\overline{b}}^{(2)}\Psi ^{a}\Upsilon ^{\overline{b}}+  \notag \\
(\Gamma _{\overline{a}\overline{b}}^{(2)}-\frac{1}{2}\omega _{\overline{a}%
\overline{b}})\Upsilon ^{\overline{a}}\Upsilon ^{\overline{b}}-\frac{1}{3}%
\omega _{ab}\Upsilon ^{a,ij}\delta _{ij}(\Psi ^{b}+f^{b})+  \notag \\
\Gamma _{abc}^{(3)}\Psi ^{a}\Psi ^{b}\Psi ^{c}+3\Gamma _{ab\overline{c}%
}^{(3)}\Psi ^{a}\Psi ^{b}\Upsilon ^{\overline{c}}+3\Gamma _{a\overline{b}%
\overline{c}}^{(3)}\Psi ^{a}\Upsilon ^{\overline{b}}\Upsilon ^{\overline{c}%
}+\Gamma _{\overline{a}\overline{b}\overline{c}}^{(3)}\Upsilon ^{\overline{a}%
}\Upsilon ^{\overline{b}}\Upsilon ^{\overline{c}}+  \notag \\
\Gamma _{abcd}^{(4)}\Psi ^{a}\Psi ^{b}\Psi ^{c}\Psi ^{d}+4\Gamma _{abc%
\overline{d}}^{(4)}\Psi ^{a}\Psi ^{b}\Psi ^{c}\Upsilon ^{\overline{d}%
}+6\Gamma _{ab\overline{c}\overline{d}}^{(4)}\Psi ^{a}\Psi ^{b}\Upsilon ^{%
\overline{c}}\Upsilon ^{\overline{d}}+  \notag \\
4\Gamma _{a\overline{b}\overline{c}\overline{d}}^{(4)}\Psi ^{a}\Upsilon ^{%
\overline{b}}\Upsilon ^{\overline{c}}\Upsilon ^{\overline{d}}+\Gamma _{%
\overline{a}\overline{b}\overline{c}\overline{d}}^{(4)}\Upsilon ^{\overline{a%
}}\Upsilon ^{\overline{b}}\Upsilon ^{\overline{c}}\Upsilon ^{\overline{d}}\}.
\label{89A}
\end{gather}%
The coefficient functions ($\Gamma $'s) in this expression are called \textit{vertices} which follow from applying the iterative solution of Eq. (\ref{74A}) en (\ref{75A}) to the partition function (\ref{87A}), see also Supplemental material II for more details. In this way the vertices only depend on the single-chain correlation functions as mentioned earlier.

\section*{Appendix B}
\markboth{Appendix B}{}\renewcommand{\theequation}{B.\arabic{equation}}%
\setcounter{equation}{0}
For a general melt of polydisperse multiblock copolymers the expression of the Landau free energy has been derived in appendix A.
In this appendix the expression for this general melt is applied to a melt of monodisperse diblocks. The minimum with respect to the order parameters is
determined analytically. In the next appendix it is explained how the orientation of A- and B-blocks can be described in several microphase structures. 

Below the spinodal $\chi  <\chi _{s}$ the melt is in the isotropic state. When $\chi $ is increased a phase transition will take place at the the spinodal. A certain microphase is formed when $\chi $ is further increased. In a microphase structure besides a density fluctuation there is also a certain amount of nematic ordening possible
because of the bending stiffness of the chains. In the first harmonics approximation the Fourier tranformed density order-parameter $\Psi ^{a}$ and the orientation tensor $\Upsilon ^{\overline{a}} =$ $\frac{Q_{\alpha }^{\mu  \nu } ( -\underline{q})}{V}$ have the following form according to Ref. \cite{Singh},
\begin{gather}\Psi ^{a} =\Psi ^{A} (\underline{q}) = -\Psi ^{B} (\underline{q}) =\Psi  (\underline{q}) =\Psi  \sum _{\underline{q}^{ \prime } \in H}\exp  (i \varphi _{\underline{q}^{ \prime }}) \delta  (\underline{q} -\underline{q}^{ \prime }) \label{psi1} \\
\text{and} \nonumber  \\
\Upsilon ^{\overline{a}} =\Upsilon _{\alpha }^{\mu  \nu } (\underline{q}) =\Upsilon _{\alpha } (\eta _{\alpha }^{\mu } \eta _{\alpha }^{\nu } -\frac{1}{3} \delta ^{\mu  \nu }) \sum _{\underline{q}^{ \prime } \in H}\exp  (i \varphi _{\underline{q}^{ \prime }}) \delta  (\underline{q} -\underline{q}^{ \prime }) + \nonumber  \\
\Upsilon _{\alpha }^{0} (\eta _{\alpha }^{\mu } \eta _{\alpha }^{\nu } -\frac{1}{3} \delta ^{\mu  \nu }) \delta  (\underline{q}) . \label{ups}\end{gather}The summation over the wave vectors $\underline{q}^{ \prime }$ is limited over a set $H=\{\pm\underline{q}_1, \pm\underline{q}_2, ...\}$ in which all wave vectors have the same fixed magnitude $q_*$. This set $H$ defines the symmetry properties and wavelength of a phase structure. The magnitude $q_*$ follows from the spinodal expression given by Eq. (\ref{spinodalChi}) which is minimized with respect to the magintude $q_*$. Besides a local orientation a possible contribution of a global orientation is also taken into account in Eq. (\ref{ups}).
In $\Upsilon ^{\overline{a}}$ the vector $\eta _{\alpha }^{\mu _{1}}$ is a unit vector which is the director of the orientation of block $\alpha $. If the orientation of the A- and B-block are different each semi-flexible chain must be bent over a certain angle. In the derivation
of the single-chain correlation function in Supplemental material I the tangent vectors of the A- and B-block at the
connection point have been taken equal. So the A- and B-block cannot be rotated freely with respect to each other. Therefore it is energetically not favourable
that there are two different orientations. It is then justified to assume that in Eq.(\ref{ups}) the directors $\eta _{A}^{\mu }$ and $\eta _{B}^{\mu }$ are equal, $\eta _{A}^{\mu } =\eta _{B}^{\mu } =\eta ^{\mu }$. 

In the nematic and smectic state the first harmonics approximation according to Eq. (\ref{ups})
can be applied, but it is not possible in the hexagonal and bcc structure. The orientation of nematic and smectic state is described by a constant director
$\underline{\eta }_{\alpha } (\underline{x}) =\underline{\eta }$ in Eq. (\ref{macr}), but in the hexagonal and bcc state it must be space dependent. In the next appendix
this space dependent director $\underline{\eta }_{\alpha } (\underline{x})$ is chosen such that the orientation tensors $\underline{\underline{Q}}_{h e x}^{\alpha } (\underline{q})$ and $\underline{\underline{Q}}_{b c c}^{\alpha } (\underline{q})$\ can be written as a lineair combination of smectic-A states with different directors $\underline{\eta }_{m}$,
\begin{gather}\underline{\underline{Q}}_{h e x}^{\alpha } (\underline{q}) =\frac{1}{3} \overset{3}{\sum _{m =1}} \underline{\underline{Q}}_{s m e c A}^{\alpha } (\underline{q} ,\underline{\eta }_{m}) \\
\text{and} \nonumber  \\
\underline{\underline{Q}}_{b c c}^{\alpha } (\underline{q}) =\frac{1}{6} \overset{6}{\sum _{m =1}} \underline{\underline{Q}}_{s m e c A}^{\alpha } (\underline{q} ,\underline{\eta }_{m})\text{.}\end{gather} In this way the orientation tensor is invariant under the symmetry operations belonging to the corresponding microphase structure.

In the Landau free energy,
\begin{gather}\frac{F_{L}}{V} =\min_{\underline{\Psi } ,\overline{\Upsilon }}\{(\Gamma _{a b}^{(2)} -\widetilde{\chi }_{a b}) \Psi ^{a} \Psi ^{b} +2 \Gamma _{a \overline{b}}^{(2)} \Psi ^{a} \Upsilon ^{\overline{b}} + \nonumber  \\
(\Gamma _{\overline{a} \overline{b}}^{(2)} -\frac{1}{2} \omega _{\overline{a} \overline{b}}) \Upsilon ^{\overline{a}} \Upsilon ^{\overline{b}} -\frac{1}{3} \omega _{a b} \Upsilon ^{a ,i j} \delta _{i j} (\Psi ^{b} +f^{b}) + \nonumber  \\
\Gamma _{a b c}^{(3)} \Psi ^{a} \Psi ^{b} \Psi ^{c} +3 \Gamma _{a b \overline{c}}^{(3)} \Psi ^{a} \Psi ^{b} \Upsilon ^{\overline{c}} +3 \Gamma _{a \overline{b} \overline{c}}^{(3)} \Psi ^{a} \Upsilon ^{\overline{b}} \Upsilon ^{\overline{c}} +\Gamma _{\overline{a} \overline{b} \overline{c}}^{(3)} \Upsilon ^{\overline{a}} \Upsilon ^{\overline{b}} \Upsilon ^{\overline{c}} + \nonumber  \\
\Gamma _{a b c d}^{(4)} \Psi ^{a} \Psi ^{b} \Psi ^{c} \Psi ^{d} +4 \Gamma _{a b c \overline{d}}^{(4)} \Psi ^{a} \Psi ^{b} \Psi ^{c} \Upsilon ^{\overline{d}} +6 \Gamma _{a b \overline{c} \overline{d}}^{(4)} \Psi ^{a} \Psi ^{b} \Upsilon ^{\overline{c}} \Upsilon ^{\overline{d}} + \nonumber  \\
4 \Gamma _{a \overline{b} \overline{c} \overline{d}}^{(4)} \Psi ^{a} \Upsilon ^{\overline{b}} \Upsilon ^{\overline{c}} \Upsilon ^{\overline{d}} +\Gamma _{\overline{a} \overline{b} \overline{c} \overline{d}}^{(4)} \Upsilon ^{\overline{a}} \Upsilon ^{\overline{b}} \Upsilon ^{\overline{c}} \Upsilon ^{\overline{d}}\} , \label{FL}\end{gather}the first harmonics approximation according to Eq. (\ref{psi1}) and (\ref{ups})
is applied to the order-parameters. The free energy is written as a power series expansion of the vector $(\Psi  ,\Upsilon _{A} ,\Upsilon _{B} ,\Upsilon _{A}^{0} ,\Upsilon _{B}^{0})$. In the second order term $\frac{F_{L}^{(2)}}{V}$ there is no coupling between the global and local order-parameters so that it can be written in the following form,
\begin{align}\frac{F_{L}^{(2)}}{V} &  = & \left [\begin{array}{ccc}\Psi  & \Upsilon _{A} & \Upsilon _{B}\end{array}\right ] \left [\begin{array}{ccc}\widetilde{\Gamma }^{(2)} -\chi  & \widetilde{\Gamma }_{A}^{(2)} & \widetilde{\Gamma }_{B}^{(2)} \\
\widetilde{\Gamma }_{A}^{(2)} & \widetilde{\Gamma }_{A A}^{(2)} -\frac{1}{3} \omega _{A A} & \widetilde{\Gamma }_{A B}^{(2)} -\frac{1}{3} \omega _{A B} \\
\widetilde{\Gamma }_{B}^{(2)} & \widetilde{\Gamma }_{A B}^{(2)} -\frac{1}{3} \omega _{A B} & \widetilde{\Gamma }_{B B}^{(2)} -\frac{1}{3} \omega _{B B}\end{array}\right ] \left [\begin{array}{c}\Psi  \\
\Upsilon _{A} \\
\Upsilon _{B}\end{array}\right ] + \nonumber  \\
 &  & \left [\begin{array}{cc}\Upsilon _{A}^{0} & \Upsilon _{B}^{0}\end{array}\right ] \left [\begin{array}{cc}\widetilde{\Gamma }_{A A ,00}^{(2)} -\frac{1}{3} \omega _{A A} & \widetilde{\Gamma }_{A B ,00}^{(2)} -\frac{1}{3} \omega _{A B} \\
\widetilde{\Gamma }_{A B ,00}^{(2)} -\frac{1}{3} \omega _{A B} & \widetilde{\Gamma }_{B B ,00}^{(2)} -\frac{1}{3} \omega _{B B}\end{array}\right ] \left [\begin{array}{c}\Upsilon _{A}^{0} \\
\Upsilon _{B}^{0}\end{array}\right ] , \label{matrixvorm}\end{align}in which the notations
\begin{gather}\widetilde{\Gamma }^{(2)} =\Gamma _{a b}^{(2)} S^{a} S^{b} ,\text{}\widetilde{\Gamma }_{\alpha }^{(2)} =\Gamma _{a \overline{b}}^{(2)} S^{a} d_{\alpha }^{\overline{b}} \nonumber  \\
\widetilde{\Gamma }_{\alpha  \beta }^{(2)} =\Gamma _{\overline{a} \overline{b}}^{(2)} d_{\alpha }^{\overline{a}} d_{\beta }^{\overline{b}}\text{  and }\widetilde{\Gamma }_{\alpha  \beta  ,00}^{(2)} =\Gamma _{\overline{a} \overline{b}}^{(2)} d_{\alpha  ,0}^{\overline{a}} d_{\beta  ,0}^{\overline{b}}\text{,\ }\end{gather}
\begin{equation}S^{a} \equiv (1 -2 \delta _{\alpha  B}) \exp  (i \varphi _{\underline{q}}) = \pm \exp  (i \varphi _{\underline{q}})\text{  and }a =\underline{q} ,\alpha 
\end{equation}and
\begin{equation}d_{\alpha }^{\overline{b}} =d_{\alpha  ,0}^{\overline{b}} \equiv (\eta _{\beta }^{\mu _{2}} \eta _{\beta }^{\nu _{2}} -\frac{1}{3} \delta ^{\mu _{2} \nu _{2}}) \exp  (i \varphi _{\underline{q}}) \delta _{\alpha  \beta }\text{  and }\overline{b} =\underline{q} ,\beta  ,\mu _{2} \nu _{2}
\end{equation}are applied. 

The 3$ \times $3 matrix in the second order term is symmetric, so it has real eigenvalues. In general the eigenvalues $\lambda _{1}$, $\lambda _{2}$ and $\lambda _{3}$ of this matrix are different. $\lambda _{1}$ is the smallest eigenvalue which becomes negative if $\chi  >\chi _{s} (q_{ \ast })$ or $\omega _{\alpha  \beta } >\omega _{\alpha  \beta  ,s} =\omega _{\alpha  \beta  ,s} (q_{ \ast })$ in which the wavenumber $q_{ \ast }$ is nonzero. The wavenumber $q_{ \ast }$ in the 2$ \times $2 matrix in Eq. (\ref{matrixvorm}) is zero and the two eigenvalues of this matrix are denoted
as $\lambda _{1 0}$ and $\lambda _{2 0}$. The smallest eigenvalue $\lambda _{1 0}$ becomes negative if $\omega _{\alpha  \beta } >\omega _{\alpha  \beta  ,s} =\omega _{\alpha  \beta  ,s} (q_{ \ast } =0)$. Exactly at $\chi  =\chi _{s}$ the eigenvalue $\lambda _{1}$ is zero so that the determinant of 3$ \times $3 matrix in Eq. (\ref{matrixvorm}) is zero. From this determinant the spinodal $\chi _{s}$ can be easily derived which is,
\begin{equation}\chi _{s} =\min_{q_{ \ast }}\{\widetilde{\Gamma }^{(2)} -\frac{(\widetilde{\Gamma }_{B}^{(2)})^{2} (\widetilde{\Gamma }_{A A}^{(2)} -\frac{1}{3} \omega _{A A}) +(\widetilde{\Gamma }_{A}^{(2)})^{2} (\widetilde{\Gamma }_{B B}^{(2)} -\frac{1}{3} \omega _{B B}) -2 \widetilde{\Gamma }_{A}^{(2)} \widetilde{\Gamma }_{B}^{(2)} (\widetilde{\Gamma }_{A B}^{(2)} -\frac{1}{3} \omega _{A B})}{(\widetilde{\Gamma }_{A A}^{(2)} -\frac{1}{3} \omega _{A A}) (\widetilde{\Gamma }_{B B}^{(2)} -\frac{1}{3} \omega _{B B}) -(\widetilde{\Gamma }_{A B}^{(2)} -\frac{1}{3} \omega _{A B})^{2}}\}\text{.}\label{spinodalChi}
\end{equation}This expression of $\chi _{s}$ can also be used to calculate the spinodals $\omega _{\alpha  \beta  ,s} (q_{ \ast })$ in which the wavenumber $q_{ \ast }$ is the same as in $\chi _{s} =\chi _{s} (q_{ \ast })$. In the same way the 2$ \times $2 matrix in Eq. (\ref{matrixvorm}) is used to determine the spinodals $\omega _{\alpha  \beta  ,s} =\omega _{\alpha  \beta  ,s} (q_{ \ast } =0)$. 

In this appendix we consider the free energy of a microphase in which $\lambda _{1}$ is negative and the other eigenvalues $\lambda _{2}$, $\lambda _{3}$, $\lambda _{1 0}$ and $\lambda _{2 0}$ are still positive. Then the primary eigenvector $\underline{x}_{1}$ which belongs to $\lambda _{1}$ is nonzero at the minimum of the free energy. The secondary eigenvectors $\underline{x}_{2}$, $\underline{x}_{3}$, $\underline{x}_{1 0}$ and $\underline{x}_{2 0}$ belonging to $\lambda _{2}$, $\lambda _{3}$, $\lambda _{1 0}$ and $\lambda _{2 0}$, respectively, may also become nonzero. These vectors are induced by third and higher order terms in the free energy in which $\underline{x}_{1}$ is coupled to the secondary eigenvectors. The directions of the eigenvectors can be determined by the 3$ \times $3 matrix and 2$ \times $2 matrix in Eq. (\ref{matrixvorm}). If we know the directions of $\underline{x}_{1}$, $\underline{x}_{2}$, $\underline{x}_{3}$, $\underline{x}_{1 0}$ and $\underline{x}_{2 0}$ it is only necassary to write the Landau free energy as a power series expansion in $x_{1} = \pm \vert \underline{x}_{1}\vert $, $x_{2} = \pm \vert \underline{x}_{2}\vert $, $x_{3} = \pm \vert \underline{x}_{3}\vert $, $x_{1 0} = \pm \vert \underline{x}_{1 0}\vert $ and $x_{2 0} = \pm \vert \underline{x}_{2 0}\vert $,
\begin{gather}\frac{F_{L}}{V} =\min_{\{x_{1} ,x_{2} ,x_{3} ,x_{1 0} ,x_{2 0}\}}\{\lambda _{1} x_{1}^{2} +\lambda _{2} x_{2}^{2} +\lambda _{3} x_{3}^{2} +\lambda _{1 0} x_{1 0}^{2} +\lambda _{2 0} x_{2 0}^{2} + \nonumber  \\
C_{1 1 1}^{(3)} x_{1}^{3} +C_{1 1 2}^{(3)} x_{1}^{2} x_{2} +\ldots  C_{1 1 1 1}^{(4)} x_{1}^{4} +\ldots \} . \label{Fx}\end{gather}In the power series expansion fifth and higher order terms are not taken into account. The normalized eigenvectors $\underline{\widehat{x}}_{n} =(\widehat{\Psi }_{n} ,\widehat{\Upsilon }_{n}^{A} ,\widehat{\Upsilon }_{n}^{B})$ with $n =1$, $2$, $3$, $10$ or $20$ are used to determine the $C -$coefficients in Eq. (\ref{Fx}). For example $C_{1 1 2}^{(3)}$ and $C_{11 ,20}^{(3)}$ are given by
\begin{subequations}
\begin{gather}C_{1 1 2}^{(3)} =\text{}\widetilde{\Gamma }_{\alpha  \beta  \gamma }^{(3)} \widehat{\Upsilon }_{1}^{\alpha } \widehat{\Upsilon }_{1}^{\beta } \widehat{\Upsilon }_{2}^{\gamma } +\widetilde{\Gamma }_{\alpha  \beta }^{(3)} \widehat{\Upsilon }_{1}^{\alpha } \widehat{\Upsilon }_{1}^{\beta } \widehat{\Psi }_{2} +2 \widetilde{\Gamma }_{\alpha  \beta }^{(3)} \widehat{\Upsilon }_{1}^{\alpha } \widehat{\Upsilon }_{2}^{\beta } \widehat{\Psi }_{1} + \nonumber  \\
\widetilde{\Gamma }_{\alpha }^{(3)} \widehat{\Upsilon }_{2}^{\alpha } \widehat{\Psi }_{1}^{2} +2 \widetilde{\Gamma }_{\alpha }^{(3)} \widehat{\Upsilon }_{1}^{\alpha } \widehat{\Psi }_{1} \widehat{\Psi }_{2} +\widetilde{\Gamma }^{(3)} \widehat{\Psi }_{1}^{2} \widehat{\Psi }_{2} \\
\text{and} \nonumber  \\
C_{11 ,20}^{(3)} =\text{}\widetilde{\Gamma }_{\alpha  \beta  \gamma }^{(3)} \widehat{\Upsilon }_{1}^{\alpha } \widehat{\Upsilon }_{1}^{\beta } \widehat{\Upsilon }_{2 0}^{\gamma } +2 \widetilde{\Gamma }_{\alpha  \beta }^{(3)} \widehat{\Upsilon }_{1}^{\alpha } \widehat{\Upsilon }_{2 0}^{\beta } \widehat{\Psi }_{1} +\widetilde{\Gamma }_{\alpha }^{(3)} \widehat{\Upsilon }_{2 0}^{\alpha } \widehat{\Psi }_{1}^{2}\text{.}\end{gather}

The third order $\widetilde{\Gamma }$'s in this example are expressed in a similar way as the second order $\widetilde{\Gamma }$'s in Eq. (\ref{matrixvorm}). The free energy has to be minimized with respect to $x_{1}$, $x_{2}$, $x_{3}$, $x_{1 0}$ and $x_{2 0}$. First the free energy is minimized with respect to the secondary parameters $x_{2}$, $x_{3}$, $x_{1 0}$ and $x_{2 0}$ at an arbitrary value of the primary parameter $x_{1}$. In this way the secondary parameters can be expressed as a power series expansion in $x_{1}$. Substituting these expressions in Eq. (\ref{Fx}) yields the free energy as a function of only one
parameter $x_{1}$. The first order partial derivative of $F_{L}$ with respect to $x_{2}$, $x_{3}$, $x_{1 0}$ and $x_{2 0}$ must be equal to zero which yields four equations with four unknows,
\end{subequations}
\begin{equation}\frac{1}{V} \frac{ \partial F_{L}}{ \partial x_{n}} =2 \lambda _{n} x_{n} +C_{1 1 n}^{(3)} x_{1}^{2} +\ldots  =0\text{  with }n =2\text{, }3\text{, }10\text{ or }20\text{.}
\end{equation}In the solutions the third and higher order terms in $x_{1}$ can be neglected, because when these are substituted in $F_{L}$ this yields fifth and higher order terms. These terms are not taken into account. The solutions $x_{n}$ with $n =2$, $3$, $10$ or $20$ does not contain lineair terms and are given by,
\begin{subequations}
\begin{equation}x_{n} =\frac{ -C_{1 1 n}^{(3)} x_{1}^{2}}{2 \lambda _{n}} +O (x_{1}^{3}) . \label{xn}
\end{equation}When $x_{2}$, $x_{3}$, $x_{1 0}$ and $x_{2 0}$ are substituted, $\frac{F_{L}}{V}$ becomes,
\end{subequations}
\begin{gather}\frac{F_{L}}{V} =\min_{x_{1}}\{\lambda _{1} x_{1}^{2} +C_{1 1 1}^{(3)} x_{1}^{3} +(C_{1 1 1 1}^{(4)} -\sum _{n \neq 1}\frac{(C_{1 1 n}^{(3)})^{2}}{4 \lambda _{n}}) x_{1}^{4}\} = \nonumber  \\
\min_{x_{1}}\{\lambda _{1} x_{1}^{2} +C_{1 1 1}^{(3)} x_{1}^{3} +\widetilde{C}_{1 1 1 1}^{(4)} x_{1}^{4}\} . \label{Fx1}\end{gather}At the minimum $x_{1}$ is
\begin{equation}x_{1} =\frac{ -3 C_{1 1 1}^{(3)} \pm \sqrt{9 (C_{1 1 1}^{(3)})^{2} -32 \lambda _{1} \widetilde{C}_{1 1 1 1}^{(4)}}}{8 \widetilde{C}_{1 1 1 1}^{(4)}}\text{.}
\end{equation}

\section*{Appendix C}
\markboth{Appendix C}{}\renewcommand{\theequation}{C.\arabic{equation}}%
\setcounter{equation}{0}

In the previous appendix besides a global orientation of A- and B-blocks a possible space dependent orientation is also taken into account in
the Landau free energy. Such a local orientation becomes possible in a microphase. The first harmonics approximation according to Ref. \cite{Singh}
given by Eq. (\ref{ups}) can only be applied to a nematic or smectic state. In this section it will be explained how an
orientation tensor can be constructed such that it can also be applied to the hexagonal and bcc state. 

The macroscopic orientation
tensor $\underline{\underline{Q}}^{\alpha } (\underline{x})$ in real space is written in the following form,
\begin{equation}\underline{\underline{Q}}^{\alpha } (\underline{x}) = \langle \underline{\underline{\widehat{Q}}}^{\alpha } (\underline{x}) \rangle  =Q^{\alpha } (\underline{x}) (\underline{\eta }^{\alpha } (\underline{x}) \underline{\eta }^{\alpha } (\underline{x}) -\frac{1}{3} \underline{\underline{I}}) . \label{macr}
\end{equation}In this form $\underline{\underline{\widehat{Q}}}^{\alpha } (\underline{x})$ is the microscopic orientation tensor defined by Eq. (\ref{9})
and $ \langle \underline{\underline{\widehat{Q}}}^{\alpha } (\underline{x}) \rangle $ is the corresponding coarse grained order-parameter field. The unit vector $\underline{\eta }^{\alpha } (\underline{x})$ is the local director of monomers of kind $\alpha $. The scalar
$Q^{\alpha } (\underline{x})$ can be regarded as the strength of the alignment of the $\alpha $-monomers along the director $\underline{\eta }^{\alpha } (\underline{x})$. If $Q^{\alpha } (\underline{x}) =0$ the alignment is completely arbitrary on average. A positive strength $Q^{\alpha } (\underline{x})$ means that the monomers prefers to be aligned parallel to $\underline{\eta }^{\alpha } (\underline{x})$ and if $Q^{\alpha } (\underline{x})$ is negative the monomers are stronger directed perpendicular to $\underline{\eta }^{\alpha } (\underline{x})$. 

The orientation in the nematic and smectic state can be described by a space independent director $\underline{\eta }^{\alpha } (\underline{x}) =\underline{\eta }^{\alpha }$ in Eq. (\ref{macr}). Then the form of Eq. (\ref{macr}) is in agreement
with Eq. (\ref{ups}) which is transformed back in real space. However, in the hexagonal and bcc state Eq. (\ref{ups})
cannot be applied, because there is more than one direction at which the A- and B-blocks prefer to be oriented. Therefore the orientation in these structures
cannot be described by means of one constant director $\underline{\eta }^{\alpha }$. A simple way to express the director $\underline{\eta }^{\alpha } (\underline{x})$ of the hexagonal state is to divide the melt into domains at which $\underline{\eta }^{\alpha } (\underline{x})$ is locally constant. Such director field has been drawn in Fig. (6). 

\includegraphics[scale=0.50]{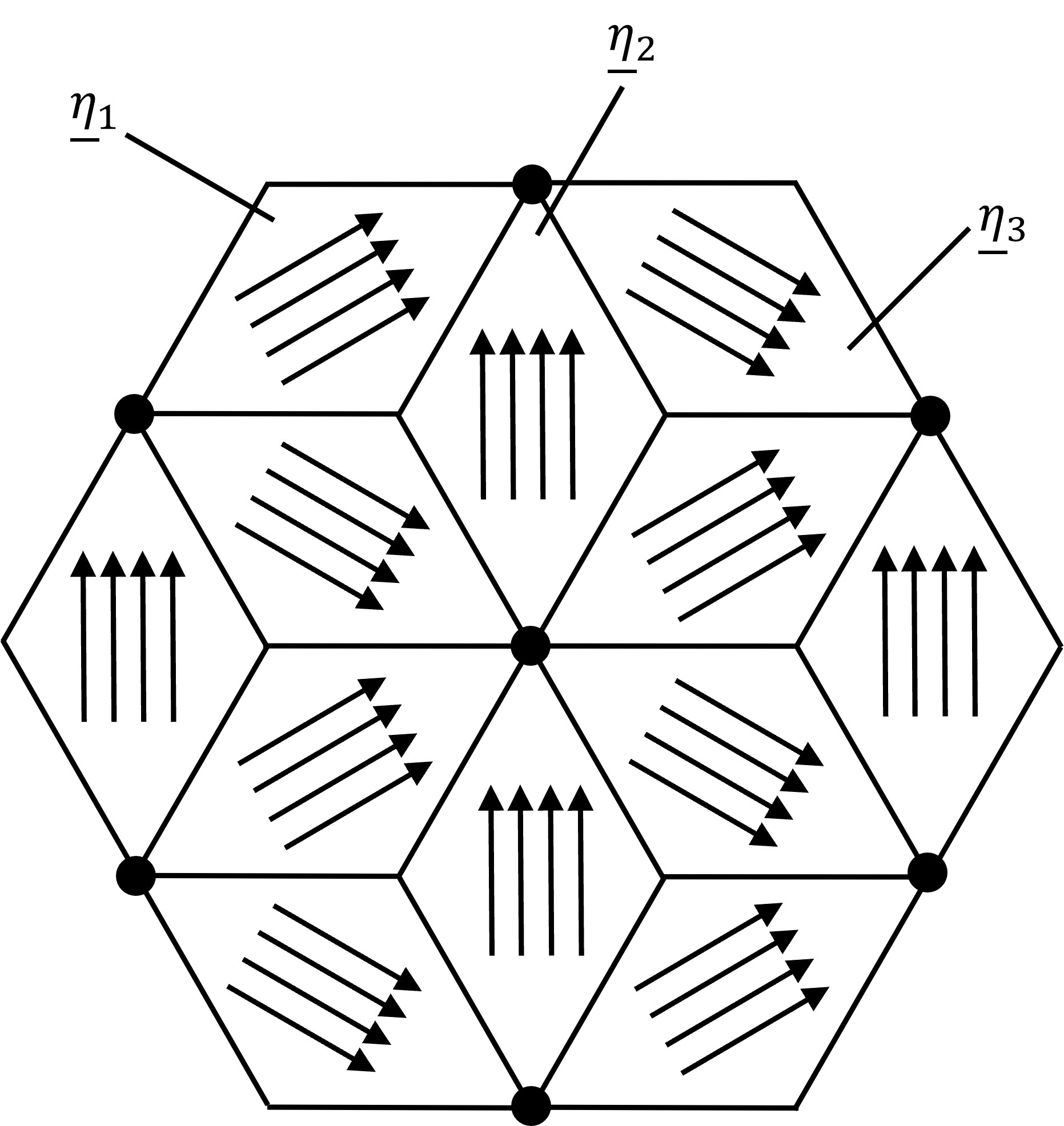} \newline 

In this figure the dots are the points at which the density of
$\alpha $-monomers is maximal. Each line connects a maximum with one of the six nearest points at which the density of $\alpha $-monomers is minimal. Fig. (6) can be applied to both the hexagonal state with $\alpha $-rich cylinders and the reverse state with an $\alpha $-rich matrix. The local directors $\underline{\eta }_{1}$, $\underline{\eta }_{2}$ and $\underline{\eta }_{3}$ are choosen parallel to the set of wave vectors $H_{h e x} =\{ \pm \underline{q}_{1} , \pm \underline{q}_{2} , \pm \underline{q}_{3}\}$ in $\Psi  (\underline{q})$ given by Eq. (\ref{psi1}). The domains at which $\underline{\eta }^{\alpha } (\underline{x}) =\underline{\eta }_{1}$, $\underline{\eta }_{2}$ and $\underline{\eta }_{3}$ are denoted as $V_{1 n}$, $V_{2 n}$ and $V_{3 n}$, respectively. In this notation $n$ is a certain integer. So $\underline{\eta }^{\alpha } (\underline{x})$ can be expressed as,
\begin{equation}\underline{\eta }^{\alpha } (\underline{x}) =\begin{array}{c}\underline{\eta }_{1}\text{ if\ }\underline{x} \in V_{1 n} \\
\underline{\eta }_{2}\text{ if\ }\underline{x} \in V_{2 n} \\
\underline{\eta }_{3}\text{ if\ }\underline{x} \in V_{3 n}\end{array} . \label{dir}
\end{equation}After inserting this director field $\underline{\eta }^{\alpha } (\underline{x})$ in Eq. (\ref{macr}) it can be proven that the Fourier transform of $\underline{\underline{Q}}^{\alpha } (\underline{x}) =\underline{\underline{Q}}_{h e x}^{\alpha } (\underline{x})$ has the form,
\begin{equation}\underline{\underline{Q}}_{h e x}^{\alpha } (\underline{q}) =\frac{1}{3} \overset{3}{\sum _{m =1}} \underline{\underline{Q}}_{s m e c A}^{\alpha } (\underline{q} ,\underline{\eta }_{m})\text{.}
\end{equation}So the orientation tensor of the hexagonal phase is the average of three smectic-A phases. In the same way it can be derived
that the orientation tensor $\underline{\underline{Q}}_{b c c}^{\alpha } (\underline{q})$ of the bcc phase can be expressed as the average of six smectic-A phases,
\begin{equation}\underline{\underline{Q}}_{b c c}^{\alpha } (\underline{q}) =\frac{1}{6} \overset{6}{\sum _{m =1}} \underline{\underline{Q}}_{s m e c A}^{\alpha } (\underline{q} ,\underline{\eta }_{m})\text{.}
\end{equation}The phase factors $\exp  (i \varphi _{\underline{q}^{ \prime }})$ in Eq. (\ref{ups}) and local directors $\underline{\eta }_{m}$ in the orientation tensors $\underline{\underline{Q}}_{h e x}^{\alpha } (\underline{q})$ and $\underline{\underline{Q}}_{b c c}^{\alpha } (\underline{q})$ are chosen such that the symmetry properties of the microphase structure are maintained.

\bigskip 

\newpage
\section*{Supplemental material I}
\markboth{Supplemental material I}{}\renewcommand{\theequation}{I.\arabic{equation}}%
\setcounter{equation}{0}

In this supplemental material the second order single-chain correlation functions as introduced in Supplemental material III are calculated. The third and fourth order single-chain correlation fuctions can be derived in a similar way. All second
order correlation functions can be derived from the following generalised form,  
\begin{eqnarray}
{\small A}^{\alpha \beta }(\underline{q}_{1},\underline{q}_{2}) &\equiv &%
\underset{l_{1},l_{2}}{\sum }{\small \sigma }_{l_{1}}^{\alpha }{\small %
\sigma }_{l_{2}}^{\beta }\left\langle \exp \left( -i\underline{q}_{1}\cdot 
\underline{R}(l_{1})-i\underline{q}_{2}\cdot \underline{R}(l_{2})+\overset{L}%
{\underset{0}{\int }}dl\underline{\eta }(l)\cdot \underline{\overset{.}{R}}%
(l)\right) \right\rangle _{0}{\small =}  \notag \\
&&\underset{l_{1},l_{2}}{\sum }{\small \sigma }_{l_{1}}^{\alpha }{\small %
\sigma }_{l_{2}}^{\beta }\int {\small d}\underline{\widetilde{U}}\int 
{\small d}\underline{U}\int {\small D}\underline{R}\int D\underline{u}\text{ 
}\delta (\underline{u}(l)-\underline{\overset{.}{R}}(l))\times  \notag \\
&&{\small \delta (}\underline{\overset{.}{R}}{\small (0)-}\underline{%
\widetilde{U}}{\small )\delta ((}\underline{\overset{.}{R}}({\small L)-}%
\underline{U}{\small )}\exp \left( -H_{0}\right) \times  \notag \\
&&\exp \left( -i\underline{q}_{1}\cdot \underline{R}(l_{1})-i\underline{q}%
_{2}\cdot \underline{R}(l_{2})+\overset{L}{\underset{0}{\int }}dl\underline{%
\eta }(l)\cdot \underline{u}L)\right) {\small .}  \label{corr}
\end{eqnarray}%
$\underline{R}(l)$ is the position vector of a certain segment labelled by $%
l $. $\underline{\widetilde{U}}$ \ and $\underline{U}$ are the the tangent
vectors of the first and last segment, respectively. $\underline{\eta }(l)$\bigskip\ is an arbitrary path
which is used to calculate the tensorial correlation functions $\left\langle 
\widehat{S}_{\mu _{1}\nu _{1}}^{\alpha }(\underline{q}_{1})\rho ^{\beta }(%
\underline{q}_{2})\right\rangle _{0}$\ and\ $\left\langle \widehat{S}_{\mu
_{1}\nu _{1}}^{\alpha }(\underline{q}_{1})\widehat{S}_{\mu _{2}\nu
_{2}}^{\beta }(\underline{q}_{2})\right\rangle _{0}$, see also Eq. (\ref{corrb}) and (\ref{corrc}). $H_{0}$\ is the
Hamiltonian of an unperturbed semi-flexible chain. For a semi-flexible
homopolymer $H_{0}$ is equal to 
\begin{equation}
H_{0}=\frac{3}{4}\widetilde{U}^{2}+\frac{3}{4}U^{2}+\frac{3}{4}\overset{L}{%
\underset{0}{\int }}dl[\lambda \overset{..}{R}^{2}(l)+\frac{1}{\lambda }%
\overset{.}{R}^{2}(l)],  \label{H0}
\end{equation}%
which is the Hamiltonian corresponding to the Bawendi-Freed model \cite%
{Bawendi}. In the Bawendi-Freed model \cite{Bawendi} the segments are
connected to each other by springs with a spring constant equal to $\frac{3}{%
2\lambda }$. On a coarse grained level $\underline{u}(l)$ is the tangent
vector, but on the microscopic level $\underline{u}(l)$ is the connection
vector between two segments. So the fourth term of $H_{0}$ can be regarded
as the total spring energy. If we consider $\frac{3\lambda }{2}$ as the mass
of a segment, then the third term becomes the total kinetic energy. In this
way $H_{0}$ can be regarded as the Hamiltonian of a chain of coupled harmonic oscillators. $\underline{u}(l)$ is on average equal to a unit
vector. Another possible model to describe a semi-flexible chain is the Sait%
\^{o} model \cite{Saito}. In that model the segments are not connected by
springs. The connection vector $\underline{u}(l)$ between two segments has a
constant unit length so that only the third term of $H_{0}$ remains. So in
Eq. (\ref{corr}) the delta function $\delta (u^{2}-1)$ has to be added in
the functional integration over $\underline{u}$. This too strict condition makes the
calculation of the correlation function analytically impossible. Therefore
we choose the Bawendi-Freed model \cite{Bawendi} in which $\underline{u}(l)$
is only on average equal to a unit vector. The first and second term in $%
H_{0}$ are local potential energies at the ends of the homopolymer. These
terms are added to the free Hamiltonian so that $\left\langle
u(l)^{2}\right\rangle _{0}=1$ everywhere along the chain. If these terms are
omitted, then $\left\langle u(l)^{2}\right\rangle _{0}\neq 1$ close to the
ends of the homopolymer. The local potential energies are necessary to
describe a homogeneous chain according to \cite{Lagowski}. Eq. (\ref{H0}) is
the free Hamiltonian of a semi-flexible homopolymer with a constant
persistent length $\lambda $. It can be proven that in Eq. (\ref{H0}) it is
allowed to replace the constant $\lambda $ by an $l$-dependent persistent
length $\lambda (l)$. Then Eq. (\ref{H0}) can also be applied to a
multiblock copolymer.

The continuous parameter $l$ in Eq. (\ref{H0}) can not only be regarded as a parameter to label segments, but also as a measure of distance between a certain segment and one of the chain ends measured along the contour. As mentioned earlier the segments are connected by springs so that the parameter $l$ is not the actual contour length, but the average contour length. If $l$ would be the actual contour length, then $l$ cannot be used as a parameter to identify each segment unless the springs are replaced by rigid rods with a fixed unit length according to the Sait%
\^{o} model \cite{Saito}.

The tensorial correlation functions $\left\langle 
\widehat{S}_{\mu _{1}\nu _{1}}^{\alpha }(\underline{q}_{1})\rho ^{\beta }(%
\underline{q}_{2})\right\rangle _{0}$\ and\ $\left\langle \widehat{S}_{\mu
_{1}\nu _{1}}^{\alpha }(\underline{q}_{1})\widehat{S}_{\mu _{2}\nu
_{2}}^{\beta }(\underline{q}_{2})\right\rangle _{0}$ will be derived using Eq. (\ref{corrb}) and (\ref{corrc}). However, in Eq. (\ref%
{49}) and (\ref{50}) in Supplemental material III the tensor $\underline{\underline{\hat{Q}}}%
\,^{\alpha }(\underline{x})$ is applied instead of $\underline{\underline{%
\hat{S}}}^{\alpha }(\underline{x})$ which is defined by, 
\begin{equation}
\underline{\underline{\hat{Q}}}\,^{\alpha }(\underline{x})\equiv \underline{%
\underline{\hat{S}}}^{\alpha }(\underline{x})-\frac{1}{3}\hat{\rho}^{\alpha
}(\underline{x})\underline{\underline{\,I}}\qquad (\alpha =1,...,M)\text{.}
\end{equation}%
By means of this definition $\left\langle \widehat{Q}_{\mu _{1}\nu
_{1}}^{\alpha }(\underline{q}_{1})\rho ^{\beta }(\underline{q}%
_{2})\right\rangle _{0}$ and $\left\langle \widehat{Q}_{\mu _{1}\nu
_{1}}^{\alpha }(\underline{q}_{1})\widehat{Q}_{\mu _{2}\nu _{2}}^{\beta }(%
\underline{q}_{2})\right\rangle _{0}$ can be expressed as, 
\begin{subequations}
\begin{gather}
\left\langle \widehat{Q}_{\mu _{1}\nu _{1}}^{\alpha }(\underline{q}_{1})\rho
^{\beta }(\underline{q}_{2})\right\rangle _{0}=\left\langle \widehat{S}_{\mu
_{1}\nu _{1}}^{\alpha }(\underline{q}_{1})\rho ^{\beta }(\underline{q}%
_{2})\right\rangle _{0}-\frac{1}{3}\delta _{\mu _{1}\nu _{1}}\left\langle
\rho ^{\alpha }(\underline{q}_{1})\rho ^{\beta }(\underline{q}%
_{2})\right\rangle _{0} \label{corrbQ} \\
\text{and}  \notag \\
\left\langle \widehat{Q}_{\mu _{1}\nu _{1}}^{\alpha }(\underline{q}_{1})%
\widehat{Q}_{\mu _{2}\nu _{2}}^{\beta }(\underline{q}_{2})\right\rangle
_{0}=\left\langle \widehat{S}_{\mu _{1}\nu _{1}}^{\alpha }(\underline{q}_{1})%
\widehat{S}_{\mu _{2}\nu _{2}}^{\beta }(\underline{q}_{2})\right\rangle _{0}-%
\frac{1}{3}\delta _{\mu _{1}\nu _{1}}\left\langle \rho ^{\alpha }(\underline{%
q}_{1})\widehat{S}_{\mu _{2}\nu _{2}}^{\beta }(\underline{q}%
_{2})\right\rangle _{0}  \notag \\
-\frac{1}{3}\delta _{\mu _{2}\nu _{2}}\left\langle \widehat{S}_{\mu _{1}\nu
_{1}}^{\alpha }(\underline{q}_{1})\rho ^{\beta }(\underline{q}%
_{2})\right\rangle _{0}+\frac{1}{9}\delta _{\mu _{1}\nu _{1}}\delta _{\mu
_{2}\nu _{2}}\left\langle \rho ^{\alpha }(\underline{q}_{1})\rho ^{\beta }(%
\underline{q}_{2})\right\rangle _{0}\text{.}\label{corrcQ}
\end{gather}%
Third and fourth order tensorial correlation functions are determined in a
similar way.

The physical meaning of these single-chain correlation functions can be
explained by Fourier transforming them back into real space. $\left\langle
\rho ^{\alpha }(\underline{x}_{1})\rho ^{\beta }(\underline{x}%
_{2})\right\rangle _{0}$ is the probability that $x_{1}$ is in the $\alpha $%
-block and $x_{2}$ is in the $\beta $-block of a certain chain. The average
is taken over all possible chain configurations which is denoted by $%
\left\langle {}\right\rangle _{0}$. $\left\langle \widehat{S}_{\mu _{1}\nu
_{1}}^{\alpha }(\underline{x}_{1})\rho ^{\beta }(\underline{x}%
_{2})\right\rangle _{0}$ and $\left\langle \widehat{S}_{\mu _{1}\nu
_{1}}^{\alpha }(\underline{x}_{1})\widehat{S}_{\mu _{2}\nu _{2}}^{\beta }(%
\underline{x}_{2})\right\rangle _{0}$ can be regarded as the expectation
value of $u^{\mu _{1}}u^{\nu _{1}}$ and $u^{\mu _{1}}u^{\nu _{1}}u^{\mu
_{2}}u^{\nu _{2}}$, respectively. In both cases $x_{1}$ must be in the $%
\alpha $-block and $x_{2}$ in the $\beta $-block. The meaning of the higher
order correlation functions can be explained in the same way.

According to Eq. (\ref{48}), (\ref{49}) and (\ref{50}) in Supplemental material III and Eq. (\ref{corr}) the second order single-chain correlation functions are,

\end{subequations}
\begin{subequations}
\begin{eqnarray}
\left\langle \rho ^{\alpha }(\underline{q}_{1})\rho ^{\beta }(\underline{q}%
_{2})\right\rangle _{0} &=&\underset{\underline{\eta }\rightarrow \underline{%
0}}{\lim }{\small A}^{\alpha \beta }(\underline{q}_{1},\underline{q}_{2}),\ 
\label{corra} \\
\left\langle \widehat{S}_{\mu _{1}\nu _{1}}^{\alpha }(\underline{q}_{1})\rho
^{\beta }(\underline{q}_{2})\right\rangle _{0} &=&\underset{\underline{\eta }%
\rightarrow \underline{0}}{\lim }\frac{\delta ^{2}}{\delta \eta _{\mu
_{1}}(l_{1})\delta \eta _{\nu _{1}}(l_{1})}{\small A}^{\alpha \beta }(%
\underline{q}_{1},\underline{q}_{2})  \label{corrb} \\
&&\text{and}  \notag \\
\left\langle \widehat{S}_{\mu _{1}\nu _{1}}^{\alpha }(\underline{q}_{1})%
\widehat{S}_{\mu _{2}\nu _{2}}^{\beta }(\underline{q}_{2})\right\rangle _{0}
&=&\underset{\underline{\eta }\rightarrow \underline{0}}{\lim }\frac{\delta
^{4}}{\delta \eta _{\mu _{1}}(l_{1})\delta \eta _{\nu _{1}}(l_{1})\delta
\eta _{\mu _{2}}(l_{2})\delta \eta _{\nu _{2}}(l_{2})}{\small A}^{\alpha
\beta }(\underline{q}_{1},\underline{q}_{2}).\   \notag \\
&&  \label{corrc}
\end{eqnarray}
\end{subequations}

When the path integral over $\underline{R}(l)$ in Eq. (\ref{corr}) is
carried out, $\underline{R}(l)$ must be replaced by

\begin{equation}
\underline{R}(l)=\underline{R}(0)+\ \overset{l}{\underset{0}{\int }}%
dl^{\prime }\underline{u}(l^{\prime }),
\end{equation}%
because of the delta function $\delta (\underline{u}(l)-\underline{\overset{.%
}{R}}(l))$. The path integral over $\underline{u}(l)$ remains together with
an integral over the initial position $\underline{R}(0)$. Then Eq. (\ref%
{corr}) becomes

\begin{gather}
{\small A}^{\alpha \beta }(\underline{q}_{1},\underline{q}_{2})=\underset{%
l_{1},l_{2}}{\sum }\sigma _{l_{1}}^{\alpha }\sigma _{l_{2}}^{\beta }\int d%
\underline{\widetilde{U}}\int d\underline{U}\int d\underline{R}(0)\int D%
\underline{u}\text{ }\delta (\underline{u}(0)-\underline{\widetilde{U}})%
\text{ }\delta (\underline{u}(L)-\underline{U})\exp \left( -H_{0}\right)
\times  \notag \\
\exp \left( -i\underline{q}_{1}\cdot \underline{R}(0)-i\underline{q}%
_{2}\cdot \underline{R}(0)\right) \times  \notag \\
\exp \left( -i\underline{q}_{1}\cdot \overset{l_{1}}{\underset{0}{\int }}dl%
\underline{u}(l)-i\underline{q}_{2}\cdot \overset{l_{2}}{\underset{0}{\int }}%
dl\underline{u}(l)+\overset{L}{\underset{0}{\int }}dl\underline{\eta }%
(l)\cdot \underline{u}(l)\right) =\   \notag \\
\int d\underline{R}(0)\exp \left( -i\underline{q}_{1}\cdot \underline{R}(0)-i%
\underline{q}_{2}\cdot \underline{R}(0)\right) \times  \notag \\
\underset{l_{1},l_{2}}{\sum }\sigma _{l_{1}}^{\alpha }\sigma _{l_{2}}^{\beta
}\int d\underline{\widetilde{U}}\int d\underline{U}\int D\underline{u}\text{ 
}\delta (\underline{u}(0)-\underline{\widetilde{U}})\text{ }\delta (%
\underline{u}(L)-\underline{U})\exp \left( -H_{0}\right) \times {\small \ }\ 
\notag \\
\exp \left( -i\underline{q}_{1}\cdot \overset{L}{\underset{0}{\int }}%
\underline{u}(l)\theta (l_{1}-l)-i\underline{q}_{2}\cdot \overset{L}{%
\underset{0}{\int }}\underline{u}(l)\theta (l_{2}-l)+\overset{L}{\underset{0}%
{\int }}dl\underline{\eta }(l)\cdot \underline{u}(l)\right) ={\normalsize \ }%
\   \notag \\
V\delta \left( \underline{q}_{1}+\underline{q}_{2}\right) \underset{%
l_{1},l_{2}}{\sum }\sigma _{l_{1}}^{\alpha }\sigma _{l_{2}}^{\beta }\int d%
\underline{\widetilde{U}}\int d\underline{U}\int D\underline{u}\text{ }%
\delta (\underline{u}(0)-\underline{\widetilde{U}})\text{ }\delta (%
\underline{u}(L)-\underline{U}) \times {\small \ }\ 
\notag \\
\exp \left( -H_{0}+ i\overset{L}{\underset{0}{\int }}dl\underline{Q}(l)\cdot \underline{u}(l)\right) .  \label{corr2}
\end{gather}%
In Eq. (\ref{corr2}) $\underline{Q}=\underline{Q}(l)$ is given by, 
\begin{equation}
{\tiny {\normalsize \underline{Q}(l)=-\underline{q}_{1}\theta (l_{1}-l)-%
\underline{q}_{2}\theta (l_{2}-l)-i\underline{\eta }(l).}}
\end{equation}%
In a certain interval of $l$, $\underline{Q}$ takes a constant value if $%
\underline{\eta }(l)\rightarrow \underline{0}$. $\underline{Q}=-\underline{q}%
_{1}-\underline{q}_{2}$ if $l$ is smaller than both $l_{1}$ and $l_{2}$. $%
\underline{Q}=-\underline{q}_{1}$ or $\underline{Q}=-\underline{q}_{2}$ if $%
l_{2}<l<l_{1}$ or $l_{1}<l<l_{2}$, respectively. The persistence length $%
\lambda $ takes also constant values in certain intervals of $l$, because
the persistence length in every block is constant. To simplify the
calculations the interval $\left( 0,L\right) $ of the parameter $l$ is
divided into smaller parts. In every subinterval $\left(
L_{p-1},L_{p}\right) $ both $\underline{Q}$ and $\lambda $\ must be
constant, which will be denoted as $\underline{Q}_{p}$ and $\lambda _{p}$. $%
\underline{Q}_{p}$ is constant in $\left( L_{p-1},L_{p}\right) $ if $%
\underline{\eta }(l)$ is constant. $\underline{\eta }(l)$ is taken equal to
zero everywhere except in $(l_{1}-\Delta L,l_{1})$ and $(l_{2}-\Delta
L,l_{2})$. In these intervals $\underline{\eta }(l)$ is constant and equal
to $\underline{\eta }(l_{1})$ and $\underline{\eta }(l_{2}),$ respectively. $%
\Delta L$ is very small and goes to zero. In this way it is possible to
calculate the functional derivatives in Eq. (\ref{corrb}) and
(\ref{corrc}). For a certain $p=p_{1}$ or $p_{2}$, $L_{p}=l_{1}$ or $l_{2}$,
respectively. Then in\ $\left( L_{p-1},L_{p}\right) $, $\underline{Q}_{p}$
becomes 
\begin{equation}
\underline{Q}_{p}=-\underline{q}_{1}\theta (l_{1}-L_{p})-\underline{q}%
_{2}\theta (l_{2}-L_{p})-i\delta _{pp_{1}}{\tiny {\normalsize \underline{%
\eta }(l_{1})-i}}\delta _{pp_{2}}{\tiny {\normalsize \underline{\eta }%
(l_{2}).}}  \label{qp}
\end{equation}%
According to Eq. (\ref{H0}) the free Hamiltonian of each interval is 
\begin{equation}
H_{0}^{(p)}=\frac{3}{4}u_{0}^{2}\delta _{p1}+\frac{3}{4}u_{p^{\prime
}}^{2}\delta _{pp^{\prime }}+\frac{3}{4}\overset{L_{p}}{\underset{L_{p-1}}{%
\int }}dl[\lambda _{p}\overset{.}{u}^{2}(l)+\frac{1}{\lambda _{p}}u^{2}(l)].
\end{equation}%
So Eq. (\ref{corr2}) can be written as 
\begin{eqnarray}
{\small A}^{\alpha \beta }(\underline{q}_{1},\underline{q}_{2}) &=&V\delta
\left( \underline{q}_{1}+\underline{q}_{2}\right) \underset{l_{1},l_{2}}{%
\sum }\sigma _{l_{1}}^{\alpha }\sigma _{l_{2}}^{\beta }\int d\underline{%
\widetilde{U}}\int d\underline{U}\times  \notag \\
&&\int D\underline{u}\text{ }\delta (\underline{u}(0)-\underline{\widetilde{U%
}})\text{ }\delta (\underline{u}(L)-\underline{U})\exp \left( -\frac{3}{4}%
u_{0}^{2}-\frac{3}{4}u_{p^{\prime }}^{2}\right) \times  \notag \\
&&\exp \left( -\underset{p=1}{\overset{p^{\prime }}{\sum }}\overset{L_{p}}{%
\underset{L_{p-1}}{\int }}dl[\dfrac{3\lambda _{p}}{4}\overset{.}{u}^{2}(l)+%
\frac{3}{4\lambda _{p}}u^{2}(l)-i\underline{Q}_{p}\cdot \underline{u}%
(l)]\right) .  \notag \\
&&  \label{corr3}
\end{eqnarray}%
In Eq. (\ref{corr3}) the last interval $p$ is denoted as $p^{\prime }$.
Instead of integrating over the whole contour, one can integrate over $%
\underline{u}(l)$ separately for each interval $(L_{p-1},L_{p})$. This is
only allowed when the begin and end points of two subsequent pieces have the
same tangent vector. Then along the whole contour $\underline{u}(l)$ is
continuous. This is done in Eq. (\ref{corr4}), 
\begin{gather}
{\small A}^{\alpha \beta }(\underline{q}_{1},\underline{q}_{2})=V\delta
\left( \underline{q}_{1}+\underline{q}_{2}\right) \underset{l_{1},l_{2}}{%
\sum }\sigma _{l_{1}}^{\alpha }\sigma _{l_{2}}^{\beta }\left\langle \exp
\left( -\underset{p=1}{\overset{p^{\prime }}{\sum }}\overset{L_{p}}{\underset%
{L_{p-1}}{\int }}dl[-i\underline{Q}_{p}\cdot \underline{u}(l)]\right)
\right\rangle _{0}=  \notag \\
V\delta \left( \underline{q}_{1}+\underline{q}_{2}\right) \underset{%
l_{1},l_{2}}{\sum }\sigma _{l_{1}}^{\alpha }\sigma _{l_{2}}^{\beta }\int 
\underset{n=0}{\overset{p^{\prime }}{\prod }}d\underline{u}_{n}\times  \notag
\\
\exp \left( -\frac{3}{4}u_{0}^{2}-\frac{3}{4}u_{p^{\prime }}^{2}\right) 
\underset{p=1}{\overset{p^{\prime }}{\prod }}\ \overset{\underline{u}_{p}}{%
\underset{\underline{u}_{p-1}}{\int }}D\underline{u}\exp \left( -\overset{%
L_{p}}{\underset{L_{p-1}}{\int }}dl[\dfrac{3\lambda _{p}}{4}\overset{.}{u}%
^{2}(l)+\frac{3}{4\lambda _{p}}u^{2}(l)-i\underline{Q}_{p}\cdot \underline{u}%
(l)]\right) .  \label{corr4}
\end{gather}%
In this equation $\underline{u}_{p-1}$and $\underline{u}_{p}$ are the
tangent vectors at the end points of $(L_{p-1},L_{p})$. In this notation $%
\underline{\widetilde{U}}=\underline{u}_{0}$ and $\underline{U}=\underline{u}%
_{p^{\prime }}$. In the functional integration over $\underline{u}(l)$\ in
the interval $(L_{p-1},L_{p})$ the tangent vectors $\underline{u}_{p-1}$and $%
\underline{u}_{p}$\ are fixed. After the functional integration over $%
\underline{u}(l)$ integrations over the tangent vectors $\underline{u}_{p}$
are carried out. In Eq. (\ref{corr4}) the intergral $I_{p}$, 
\begin{equation}
I_{p}=\overset{\underline{u}_{p}}{\underset{\underline{u}_{p-1}}{\int }}D%
\underline{u}\exp \left( -\overset{L_{p}}{\underset{L_{p-1}}{\int }}dl[%
\dfrac{3\lambda _{p}}{4}\overset{.}{u}^{2}(l)+\frac{3}{4\lambda _{p}}%
u^{2}(l)-i\underline{Q}_{p}\cdot \underline{u}(l)]\right) ,
\end{equation}%
has the same form as Eq. (5.21) in \cite{Freed}. If $l$ is replaced by $it$
and $\hbar =1$, then $I_{p}$ is the Feynman path integral representation of
a quantum mechanical harmonic oscillator with a constant external force $%
\underline{Q}_{p}$ on each segment. Because the force $\underline{Q}_{p}$ is
constant, $I_{p}$ can be written in terms of a representation of a quantum
mechanical harmonic oscillator without external force, which is given by Eq.
(5.31) in \cite{Freed}, 
\begin{eqnarray}
I_{p}(\underline{u}_{p},\underline{u}_{p-1}) &=&\exp \left( -\frac{%
Q_{p}^{2}\Delta L_{p}\lambda _{p}}{3}\right) \overset{\underline{u}_{p}-%
\dfrac{2}{3}i\lambda _{p}\underline{Q}_{p}}{\underset{\underline{u}_{p-1}-%
\dfrac{2}{3}i\lambda _{p}\underline{Q}_{p}}{\int }}D\underline{u}\exp \left(
-\overset{L_{p}}{\underset{L_{p-1}}{\int }}dl[\dfrac{3\lambda _{p}}{4}%
\overset{.}{u}^{2}(l)+\frac{3}{4\lambda _{p}}u^{2}(l)]\right) =  \notag \\
&&\exp \left( -\frac{Q_{p}^{2}\Delta L_{p}\lambda _{p}}{3}\right) G(%
\underline{u}_{p}-\dfrac{2}{3}i\lambda _{p}\underline{Q}_{p},\underline{u}%
_{p-1}-\dfrac{2}{3}i\lambda _{p}\underline{Q}_{p},\Delta L_{p},0).
\label{Ip}
\end{eqnarray}%
In Eq. (\ref{Ip}) the propagator $G$ is given by Eq. (\ref{propagator}).
Quantum mechanically the propagator $G(\underline{u}_{p},\underline{u}%
_{p-1},\Delta L_{p},0)$ is the probability that a particle at $t=0$ and $%
\underline{x}=\underline{u}_{p-1}$ travels to $\underline{x}=\underline{u}%
_{p}$ after a time $\Delta t=-i\Delta L_{p}$. In our case it is regarded as
the probability that a semi-flexible chain of length $\Delta L_{p}$ has a
tangent vector $\underline{u}_{p}$ along the end point, if $\underline{u}%
_{p-1}$ is the tangent vector along the begin point. The result is well
known in quantum mechanics and given by Eq. (5.32) in \cite{Freed}, 
\begin{equation}
G(\underline{u}_{p},\underline{u}_{p-1},\Delta L_{p},0)=\left( \frac{b^{(p)}%
}{\pi \sinh a^{(p)}}\right) ^{\frac{3}{2}}\exp \left( -\frac{b^{(p)}}{\sinh
a^{(p)}}\left[ (u_{p}^{2}+u_{p-1}^{2})\cosh a^{(p)}-2\underline{u}_{p}\cdot 
\underline{u}_{p-1}\right] \right) ,  \label{propagator}
\end{equation}%
in which $a^{(p)}$ and $b^{(p)}$ are given by 
\begin{subequations}
\begin{eqnarray}
a^{(p)} &=&\frac{\Delta L_{p}}{\lambda _{p}} \\
&&\text{and}  \notag \\
b^{(p)} &=&\frac{3}{4}.\text{ \ }
\end{eqnarray}
\end{subequations}
Applying Eq. (\ref{propagator}) in Eq. (\ref{Ip}) yields 
\begin{gather}
I_{p}(\underline{u}_{p},\underline{u}_{p-1})=\exp \left( -\frac{%
Q_{p}^{2}\Delta L_{p}\lambda _{p}}{3}\right) \left( \frac{b^{(p)}}{\pi \sinh
a^{(p)}}\right) ^{\frac{3}{2}}\times  \notag \\
\exp \left( -\frac{b^{(p)}}{\sinh a^{(p)}}\left[ \left(
u_{p}^{2}+u_{p-1}^{2}-\frac{4i\lambda _{p}}{3}\underline{Q}_{p}\cdot (%
\underline{u}_{p}+\underline{u}_{p-1})-\frac{8\lambda _{p}^{2}Q_{p}^{2}}{9}%
\right) \cosh a^{(p)}\right] \right) \times  \notag \\
\exp \left( -\frac{b^{(p)}}{\sinh a^{(p)}}\left[ -2\underline{u}_{p}\cdot 
\underline{u}_{p-1}+\frac{4i\lambda _{p}}{3}\underline{Q}_{p}\cdot (%
\underline{u}_{p}+\underline{u}_{p-1})+\frac{8\lambda _{p}^{2}Q_{p}^{2}}{9}%
\right] \right) .
\end{gather}%
$I_{p}(\underline{u}_{p},\underline{u}_{p-1})$ can be written in the
following form 
\begin{eqnarray}
I_{p}(\underline{u}_{p},\underline{u}_{p-1}) &=&E_{p}\exp \left( -\frac{1}{2}%
A_{p}(u_{p-1}^{2}+u_{p}^{2})+B_{p}\underline{u}_{p-1}\cdot \underline{u}%
_{p}\right) \times  \notag \\
&&\exp \left( C_{p}\underline{Q}_{p}\cdot (\underline{u}_{p-1}+\underline{u}%
_{p})-\frac{1}{2}D_{p}Q_{p}^{2}\right) ,
\end{eqnarray}%
in which $A_{p}$, $B_{p}$, $C_{p}$, $D_{p}$ and $E_{p}$ are given by 
\begin{subequations}
\begin{eqnarray}
A_{p} &=&2b^{(p)}\coth a^{(p)}=\frac{3}{2}\coth (\frac{\Delta L_{p}}{\lambda
_{p}}), \\
B_{p} &=&\frac{2b^{(p)}}{\sinh a^{(p)}}=\frac{3}{2\sinh (\frac{\Delta L_{p}}{%
\lambda _{p}})}, \\
C_{p} &=&\frac{-4i\lambda _{p}b^{(p)}(1-\cosh (a^{(p)}))}{3\sinh a^{(p)}}=%
\frac{-i\lambda _{p}\left( 1-\cosh (\frac{\Delta L_{p}}{\lambda _{p}}%
)\right) }{\sinh (\frac{\Delta L_{p}}{\lambda _{p}})}, \\
D_{p} &=&\frac{2\Delta L_{p}\lambda _{p}}{3}+\frac{16\lambda
_{p}^{2}b^{(p)}(1-\cosh a^{(p)})}{9\sinh a^{(p)}}=\frac{2\Delta L_{p}\lambda
_{p}}{3}+\frac{4\lambda _{p}^{2}\left( 1-\cosh (\frac{\Delta L_{p}}{\lambda
_{p}})\right) }{3\sinh (\frac{\Delta L_{p}}{\lambda _{p}})}  \notag \\
&& \\
&&\text{and}  \notag \\
E_{p} &=&\left( \frac{b^{(p)}}{\pi \sinh a^{(p)}}\right) ^{\frac{3}{2}%
}=\left( \frac{3}{4\pi \sinh (\frac{\Delta L_{p}}{\lambda _{p}})}\right) ^{%
\frac{3}{2}}.
\end{eqnarray}%
\end{subequations}
To calculate the second order correlation function we have to perform in Eq.
(\ref{corr4}) the integral 
\begin{eqnarray}
&&\int d\underline{u}_{p^{\prime }}\underset{p=1}{\overset{p^{\prime }}{%
\prod }}\int d\underline{u}_{p-1}I_{p}(\underline{u}_{p},\underline{u}%
_{p-1})\exp \left( -\frac{3}{4}u_{0}^{2}-\frac{3}{4}u_{p^{\prime
}}^{2}\right)  \notag \\
&=&\int d\underline{u}_{p^{\prime }}\underset{p=1}{\overset{p^{\prime }}{%
\prod }}\int d\underline{u}_{p-1}E_{p}\exp \left( -\frac{3}{4}u_{0}^{2}-%
\frac{3}{4}u_{p^{\prime }}^{2}\right) \times  \notag \\
&&\exp \left( -\frac{1}{2}A_{p}(u_{p-1}^{2}+u_{p}^{2})+B_{p}\underline{u}%
_{p-1}\cdot \underline{u}_{p}\right) \times  \notag \\
&&\exp \left( C_{p}\underline{Q}_{p}\cdot (\underline{u}_{p-1}+\underline{u}%
_{p})-\frac{1}{2}D_{p}Q_{p}^{2}\right) .  \label{intIp}
\end{eqnarray}%
The integral over $\underline{u}_{p-1}$ is also a Gaussian integral of the
form, 
\begin{equation}
\int d\underline{x}\exp (-\frac{1}{2}Cx^{2}+\underline{\eta }\cdot 
\underline{x})=\left( \frac{2\pi }{C}\right) ^{\frac{3}{2}}\exp \left( \frac{%
\eta ^{2}}{2C}\right) .
\end{equation}%
First we evaluate the integral over $\underline{u}_{0}$, 
\begin{eqnarray}
&&\int d\underline{u}_{0}I_{1}(\underline{u}_{0},\underline{u}_{1})\exp (-%
\frac{3}{4}u_{0}^{2})  \notag \\
&=&\int d\underline{u}_{0}E_{1}\exp \left( -\frac{1}{2}(A_{1}+\frac{3}{2}%
)u_{0}^{2}-\frac{1}{2}A_{1}u_{1}^{2}+B_{1}\underline{u}_{0}\cdot \underline{u%
}_{1}\right) \times  \notag \\
&&\exp \left( C_{1}\underline{Q}_{1}\cdot (\underline{u}_{0}+\underline{u}%
_{1})-\frac{1}{2}D_{1}Q_{1}^{2}\right)  \notag \\
&=&E_{1}\left( \frac{2\pi }{A_{1}+\frac{3}{2}}\right) ^{\frac{3}{2}}\exp
\left( \frac{B_{1}^{2}}{2(A_{1}+\frac{3}{2})}u_{1}^{2}+\frac{B_{1}C_{1}%
\underline{Q}_{1}}{A_{1}+\frac{3}{2}}\cdot \underline{u}_{1}+\frac{%
C_{1}^{2}Q_{1}^{2}}{2(A_{1}+\frac{3}{2})}\right) \times  \notag \\
&&\exp \left( -\frac{1}{2}A_{1}u_{1}^{2}+C_{1}\underline{Q}_{1}\cdot 
\underline{u}_{1}-\frac{1}{2}D_{1}Q_{1}^{2}\right) .  \label{intI1}
\end{eqnarray}%
Using Eq. (\ref{intIp}) and (\ref{intI1}) the integration over $\underline{u}%
_{1}$ can be written in the following form 
\begin{eqnarray}
&&\int d\underline{u}_{1}\widetilde{E}_{2}\exp \left( -\frac{1}{2}\widetilde{%
A}_{2}u_{1}^{2}-\frac{1}{2}A_{2}u_{2}^{2}+B_{2}\underline{u}_{1}\cdot 
\underline{u}_{2}\right) \times  \notag \\
&&\exp \left( \underline{\widetilde{C}}_{2}\cdot \underline{u}_{1}+C_{2}%
\underline{Q}_{2}\cdot \underline{u}_{2}-\frac{1}{2}\widetilde{D}_{2}\right)
,  \label{intI2}
\end{eqnarray}%
which is a similar form as Eq. (\ref{intIp}). 
$\widetilde{A}_{2}$, $%
\underline{\widetilde{C}}_{2}$, $\widetilde{D}_{2}$ and $\widetilde{E}_{2}$
are given by 
\begin{eqnarray}
\widetilde{A}_{2} &=&A_{2}+A_{1}-\frac{B_{1}^{2}}{A_{1}+\frac{3}{2}},
\label{A2thilde} \\
\widetilde{\underline{C}}_{2} &=&C_{2}\underline{Q}_{2}+\left( \frac{B_{1}}{%
A_{1}+\frac{3}{2}}+1\right) C_{1}\underline{Q}_{1},  \label{D2thilde} \\
\widetilde{D}_{2} &=&D_{2}Q_{2}^{2}+D_{1}Q_{1}^{2}-\frac{C_{1}^{2}Q_{1}^{2}}{%
A_{1}+\frac{3}{2}}  \label{E2thilde} \\
&&\text{and}  \notag \\
\widetilde{E}_{2} &=&E_{2}E_{1}\left( \frac{2\pi }{A_{1}+\frac{3}{2}}\right)
^{\frac{3}{2}}.  \label{F2thilde}
\end{eqnarray}%
In Eq. (\ref{intI2}) we can again carry out the integral over $\underline{u}%
_{1}$. By means of Eq. (\ref{intIp}) an integral over $\underline{u}_{2}$
and next tangent vectors can be derived in the same form as Eq. (\ref{intI2}%
). This can be done in the same way for $\underline{u}_{3}$ and next tangent
vectors. After the integration over $\underline{u}_{p^{\prime }-2}$ and
previous tangent vectors we have to carry out the integration over $%
\underline{u}_{p^{\prime }-1}$ and $\underline{u}_{p^{\prime }}$, 
\begin{eqnarray}
&&\int d\underline{u}_{p^{\prime }}\exp (-\frac{3}{4}u_{p^{\prime
}}^{2})\int d\underline{u}_{p^{\prime }-1}\widetilde{E}_{p^{\prime }}\times 
\notag \\
&&\exp \left( -\frac{1}{2}\widetilde{A}_{p^{\prime }}u_{p^{\prime }-1}^{2}-%
\frac{1}{2}A_{p^{\prime }}u_{p^{\prime }}^{2}+B_{p^{\prime }}\underline{u}%
_{p^{\prime }-1}\cdot \underline{u}_{p^{\prime }}\right) \times  \notag \\
&&\exp \left( \underline{\widetilde{C}}_{p^{\prime }}\cdot \underline{u}%
_{p^{\prime }-1}+C_{p^{\prime }}\underline{Q}_{p^{\prime }}\cdot \underline{u%
}_{p^{\prime }}-\frac{1}{2}\widetilde{D}_{p^{\prime }}\right)  \notag \\
&=&\int d\underline{u}_{p^{\prime }}\widetilde{E}_{p^{\prime }+1}\exp \left(
-\frac{1}{2}\widetilde{A}_{p^{\prime }+1}u_{p^{\prime }}^{2}+\underline{%
\widetilde{C}}_{p^{\prime }+1}\cdot \underline{u}_{p^{\prime }}-\frac{1}{2}%
\widetilde{D}_{p^{\prime }+1}\right)  \notag \\
&=&\widetilde{E}_{p^{\prime }+2}\exp \left( -\frac{1}{2}\widetilde{D}%
_{p^{\prime }+2}\right) .  \label{intIp'}
\end{eqnarray}%
For $p=1,2,...,p^{\prime }$, the coeffients $\widetilde{A}_{p}$, $\underline{%
\widetilde{C}}_{p}$, $\widetilde{D}_{p}$ and $\widetilde{E}_{p}$ are equal
to 
\begin{subequations}
\begin{eqnarray}
\widetilde{A}_{p} &=&A_{p}+A_{p-1}-\frac{B_{p-1}^{2}}{\widetilde{A}_{p-1}},
\label{Apthilde} \\
\widetilde{\underline{C}}_{p} &=&C_{p}\underline{Q}_{p}+C_{p-1}\underline{Q}%
_{p-1}+\frac{B_{p-1}\underline{\widetilde{C}}_{p-1}}{\widetilde{A}_{p-1}},
\label{Dpthilde} \\
\widetilde{D}_{p} &=&D_{p}Q_{p}^{2}+\widetilde{D}_{p-1}-\frac{\widetilde{C}%
_{p-1}^{2}}{\widetilde{A}_{p-1}},  \label{Epthilde} \\
&&\text{and}  \notag \\
\widetilde{E}_{p} &=&E_{p}\widetilde{E}_{p-1}\left( \frac{2\pi }{\widetilde{A%
}_{p-1}}\right) ^{\frac{3}{2}}.  \label{Fpthilde}
\end{eqnarray}%
However, $p=p^{\prime }+1$ and $p=p^{\prime }+2$ give different coeffients, 
\end{subequations}
\begin{subequations}
\begin{eqnarray}
\widetilde{A}_{p^{\prime }+1} &=&A_{p^{\prime }}-\frac{B_{p^{\prime }}^{2}}{%
\widetilde{A}_{p^{\prime }}}+\frac{3}{2},  \label{Ap'+1thilde} \\
\widetilde{\underline{C}}_{p^{\prime }+1} &=&C_{p^{\prime }}\underline{Q}%
_{p^{\prime }}+\frac{B_{p^{\prime }}\underline{\widetilde{C}}_{p^{\prime }}}{%
\widetilde{A}_{p^{\prime }}},  \label{Dp'+1thilde} \\
\widetilde{D}_{p^{\prime }+1} &=&\widetilde{D}_{p^{\prime }}-\frac{%
\widetilde{C}_{p^{\prime }}^{2}}{\widetilde{A}_{p^{\prime }}},
\label{Ep'+1thilde} \\
\widetilde{E}_{p^{\prime }+1} &=&\widetilde{E}_{p^{\prime }}\left( \frac{%
2\pi }{\widetilde{A}_{p^{\prime }}}\right) ^{\frac{3}{2}},
\label{Fp'+1thilde} \\
\widetilde{D}_{p^{\prime }+2} &=&\widetilde{D}_{p^{\prime }+1}-\frac{%
\widetilde{C}_{p^{\prime }+1}^{2}}{\widetilde{A}_{p^{\prime }+1}},
\label{Ep'+2thilde} \\
&&\text{and}  \notag \\
\widetilde{E}_{p^{\prime }+2} &=&\widetilde{E}_{p^{\prime }+1}\left( \frac{%
2\pi }{\widetilde{A}_{p^{\prime }+1}}\right) ^{\frac{3}{2}}.
\label{Fp'+2thilde}
\end{eqnarray}%
The factor $\widetilde{E}_{p^{\prime }+2}$ in Eq. (\ref{intIp'}) must be
omitted to normalize the average $\left\langle {}\right\rangle _{0}$ in Eq. (%
\ref{corr4}) over all possible chain configurations, because $\left\langle
1\right\rangle _{0}\equiv 1$. Then the second order correlation function
given by Eq. (\ref{corr4}) becomes, 
\end{subequations}
\begin{equation}
{\small A}^{\alpha \beta }(\underline{q}_{1},\underline{q}_{2})=V\delta
\left( \underline{q}_{1}+\underline{q}_{2}\right)\underset{l_{1},l_{2}}{\sum }\sigma _{l_{1}}^{\alpha }\sigma
_{l_{2}}^{\beta }\exp \left( -\frac{1}{2}\widetilde{D}_{p^{\prime
}+2}\right) .  \label{corr5}
\end{equation}%
By means of this result the generalised second order correlation function defined by Eq. (\ref{corr}) at the begining of this supplemental material is calculated. This generalised correlation function is applied in Eq. (\ref{corra}), (\ref{corrb}) and (\ref{corrc}). In the tensorial correlation functions (\ref{corrb}) and (\ref{corrc}) functional derivatives are carried out over components of path  
$\underline{\eta }(l)$ which result in,
\begin{subequations}
\begin{gather}
\frac{\delta^{2}{\small A}^{\alpha \beta }(\underline{q}_{1},\underline{q}_{2})
}{\delta \eta _{\mu _{1}}(l_{1})\delta \eta _{\nu _{1}}(l_{1})}=V\delta \left( \underline{q}_{1}+\underline{q}_{2}\right) \underset{%
l_{1},l_{2}}{\sum }\sigma _{l_{1}}^{\alpha }\sigma _{l_{2}}^{\beta
}G_{2}\exp \left( -\frac{1}{2}\widetilde{D}_{p^{\prime }+2}\right)
\label{corrb2} \\
\text{and}  \notag \\
\frac{\delta^{4}{\small A}^{\alpha \beta }(\underline{q}_{1},\underline{q}_{2})
}{\delta \eta _{\mu _{1}}(l_{1})\delta \eta _{\nu _{1}}(l_{1})\delta \eta _{\mu _{2}}(l_{2})\delta \eta _{\nu _{2}}(l_{2})}=V\delta \left( \underline{q}_{1}+\underline{q}_{2}\right) \underset{%
l_{1},l_{2}}{\sum }\sigma _{l_{1}}^{\alpha }\sigma _{l_{2}}^{\beta
}G_{4}\exp \left( -\frac{1}{2}\widetilde{D}_{p^{\prime }+2}\right),
\label{corrc2}
\end{gather} 
\end{subequations}
in which the multiplication factor $G_{2}$ is given by, 
\begin{equation}
G_{2} =-\frac{1}{2}\frac{\delta ^{2}\widetilde{D}_{p^{\prime }+2}}{\delta \eta
_{\mu _{1}}(l_{1})\delta \eta _{\nu _{1}}(l_{1})}+\frac{1}{4}\frac{\delta 
\widetilde{D}_{p^{\prime }+2}}{\delta \eta _{\mu _{1}}(l_{1})}\frac{\delta 
\widetilde{D}_{p^{\prime }+2}}{\delta \eta _{\nu _{1}}(l_{1})}. \label{G2.1}
\end{equation}
By again applying functional derivatives with respect to $\eta _{\mu _{2}}(l_{2})$ and $\eta _{\nu _{2}}(l_{2})$ in Eq. (\ref{corrb2}) we obtain Eq. (\ref{corrc2}). In this way the multiplication factor $G_{4}$ in Eq. (\ref{corrc2}) can be expressed as,
\begin{equation}
G_{4} =\left( \frac{\delta }{\delta \eta _{\nu _{2}}(l_{2})}-\frac{1}{2}%
\frac{\delta \widetilde{D}_{p^{\prime }+2}}{\delta \eta _{\nu _{2}}(l_{2})}%
\right) \left( \frac{\delta }{\delta \eta _{\mu _{2}}(l_{2})}-\frac{1}{2}%
\frac{\delta \widetilde{D}_{p^{\prime }+2}}{\delta \eta _{\mu _{2}}(l_{2})}%
\right) G_{2}.  \label{G4.1} \\
\end{equation}
In Eq. (\ref{G2.1}) we see that $G_{2}$ dependends on the first and second order functional derivatives of $\widetilde{D}_{p^{\prime }+2}$ which are calculated in detail in Supplemental material IV. According to Supplemental material IV third and higher order functional derivatives of $\widetilde{D}_{p^{\prime }+2}$ are zero so that $G_{4}$ in Eq. (\ref{G4.1}) can also be written in terms of first and second order functional derivatives of $\widetilde{D}_{p^{\prime }+2}$,  

\begin{gather}
{\normalsize G}_{4}={\normalsize \underset{perm(\mu _{1},\nu
_{1},\mu _{2},\nu _{2})}{\sum }(\frac{1}{32}\frac{\delta ^{2}\widetilde{D}_{p^{\prime }+2}%
}{\delta \eta _{\mu _{1}}\delta \eta _{\nu _{1}}}\frac{\delta ^{2}\widetilde{%
D}_{p^{\prime }+2}}{\delta \eta _{\mu _{2}}\delta \eta _{\nu _{2}}}+}  \notag
\\
{\normalsize -\frac{1}{32}\frac{\delta ^{2}\widetilde{D}_{p^{\prime }+2}}{\delta \eta _{\mu
_{1}}\delta \eta _{\nu _{1}}}\frac{\delta \widetilde{D}_{p^{\prime }+2}}{%
\delta \eta _{\mu _{2}}}\frac{\delta \widetilde{D}_{p^{\prime }+2}}{\delta
\eta _{\nu _{2}}}+\ }  \notag \\
{\normalsize \frac{1}{4!16}\frac{\delta \widetilde{D}_{p^{\prime }+2}}{\delta \eta _{\mu
_{1}}}\frac{\delta \widetilde{D}_{p^{\prime }+2}}{\delta \eta _{\nu _{1}}}%
\frac{\delta \widetilde{D}_{p^{\prime }+2}}{\delta \eta _{\mu _{2}}}\frac{%
\delta \widetilde{D}_{p^{\prime }+2}}{\delta \eta _{\nu _{2}}}),}
\label{G4.4}
\end{gather}%
in which a summation is carried out over all possible permutations of $\mu _{1}$, $\mu _{2}$, $\nu _{1}$ and $\nu _{2}$. The expressions of the multiplication factors $G_{2}$ and $G_{4}$ in Eq. (\ref{G2.1}) and (\ref{G4.1}) and the generalised second order correlation function according to Eq. (\ref{corr5}) are used to calculate $\left\langle \rho ^{\alpha }(\underline{q}_{1})\rho ^{\beta }(\underline{q}_{2})\right\rangle _{0}$, $\left\langle \widehat{S}_{\mu _{1}\nu _{1}}^{\alpha }(\underline{q}_{1})\rho
^{\beta }(\underline{q}_{2})\right\rangle _{0}$ and $\left\langle \widehat{S}_{\mu _{1}\nu _{1}}^{\alpha }(\underline{q}_{1})\widehat{S}_{\mu _{2}\nu _{2}}^{\beta }(\underline{q}_{2})\right\rangle _{0}$ in Eq. (\ref{corra}), (\ref{corrb}) and (\ref{corrc}). The tensorial correlation functions have to be rewritten to the form $\left\langle \widehat{Q}_{\mu _{1}\nu _{1}}^{\alpha }(\underline{q}_{1})\rho^{\beta }(\underline{q}_{2})\right\rangle _{0}$ and $\left\langle \widehat{Q}_{\mu _{1}\nu _{1}}^{\alpha }(\underline{q}_{1})\widehat{Q}_{\mu _{2}\nu _{2}}^{\beta }(\underline{q}_{2})\right\rangle _{0}$ in Eq. (\ref{corrbQ}) and (\ref{corrcQ}). The scalar correlation function $\left\langle \rho ^{\alpha }(\underline{q}_{1})\rho ^{\beta }(\underline{q}_{2})\right\rangle _{0}$ and tensorial correlation functions $\left\langle \widehat{Q}_{\mu _{1}\nu _{1}}^{\alpha }(\underline{q}_{1})\rho^{\beta }(\underline{q}_{2})\right\rangle _{0}$ and $\left\langle \widehat{Q}_{\mu _{1}\nu _{1}}^{\alpha }(\underline{q}_{1})\widehat{Q}_{\mu _{2}\nu _{2}}^{\beta }(\underline{q}_{2})\right\rangle _{0}$ are the $A$'s applied in Eq. (\ref{74A}) en (\ref{75A}) in Appendix A. These $A$'s are the final second order single-chain correlation functions which we want to derive in this supplemental material. Besides $A$'s Eq. (\ref{74A}) and (\ref{75A}) also contains $B$'s and $C$'s which are third and fourth order single-chain correlation functions, respectively. The $A$'s, $B$'s and $C$'s are necessary to calculate the vertices ($\Gamma$'s) in the final form of the Landau free energy (\ref{89A}) derived in Appendix A. This supplemental material is rescricted to the derivation of the $A$'s. The $B$'s and $C$'s can be derived in a similar way and are more complicated. In Eq. (\ref{corr5}), (\ref{corrb2}) and (\ref{corrc2}) the summations
over $l_{1}$ and $l_{2}$ can be replaced by an integration over the $\alpha $%
- and $\beta $-block, respectively, so,
\begin{subequations}
\begin{gather}
{\small A}^{\alpha \beta }(\underline{q}_{1},\underline{q}_{2})=V\delta
\left( \underline{q}_{1}+\underline{q}_{2}\right) \int_{\alpha}dl_{1}\int_{\beta }dl_{2}\exp \left( -\frac{1}{2}\widetilde{D}_{p^{\prime}+2}\right), \\
\frac{\delta^{2}{\small A}^{\alpha \beta }(\underline{q}_{1},\underline{q}_{2})
}{\delta \eta _{\mu _{1}}(l_{1})\delta \eta _{\nu _{1}}(l_{1})}=V\delta \left( \underline{q}_{1}+\underline{q}_{2}\right)  \int_{\alpha}dl_{1}\int_{\beta }dl_{2}G_{2}\exp \left( -\frac{1}{2}\widetilde{D}_{p^{\prime }+2}\right) \\
\text{and}  \notag \\
\frac{\delta^{4}{\small A}^{\alpha \beta }(\underline{q}_{1},\underline{q}_{2})
}{\delta \eta _{\mu _{1}}(l_{1})\delta \eta _{\nu _{1}}(l_{1})\delta \eta _{\mu _{2}}(l_{2})\delta \eta _{\nu _{2}}(l_{2})}=V\delta \left( \underline{q}_{1}+\underline{q}_{2}\right) \int_{\alpha}dl_{1}\int_{\beta }dl_{2}G_{4}\exp \left( -\frac{1}{2}\widetilde{D}_{p^{\prime }+2}\right).
\end{gather} 
\end{subequations}

\section*{Supplemental material II}
\markboth{II}{}\renewcommand{\theequation}{II.\arabic{equation}}%
\setcounter{equation}{0}

For large values of the
system's volume $V$, $\widetilde{Z}[\underline{\Psi },\overline{\Upsilon }]$
can be evaluated with the well-known \textit{saddle-point method}, i.e.
approximating $\widetilde{Z}[\underline{\Psi },\overline{\Upsilon }]$ by, 
\begin{equation}
\widetilde{Z}[\underline{\Psi },\overline{\Upsilon }]\simeq e^{V\,\Phi
\lbrack \underline{\Psi },\overline{\Upsilon }]},  \label{73}
\end{equation}%
where $\Phi \lbrack \underline{\Psi },\overline{\Upsilon }]$ is the
stationary value of $i\,[v_{a}\Psi ^{a}+w_{\overline{a}}\Upsilon ^{\overline{%
a}}]+\frac{\Lambda \lbrack \underline{v},\overline{w}]}{V}$ with respect to
the set of $v$'s and $w$'s for which its absolute value is the smallest.
Using Eq. (\ref{44}), (\ref{2eordeomega}), (\ref{3eordeomega}) and (\ref{4eordeomega}) in Supplemental material III this stationary point is a solution of the following set of equations, 
\begin{eqnarray}
i\,\Psi ^{a} &=&A^{ab}\,v_{b}+A^{a\overline{b}}\,w_{\overline{b}}-\frac{i}{2}%
\,B^{abc}\,v_{b}v_{c}-i\,B^{a\overline{b}c}\,w_{\overline{b}}v_{c}-\frac{i}{2%
}\,B^{a\overline{b}\overline{c}}\,w_{\overline{b}}w_{\overline{c}}+  \notag
\\
&&-\frac{1}{6}\,C^{abcd}\,v_{b}v_{c}v_{d}-\frac{1}{2}\,C^{a\overline{b}%
cd}\,w_{\overline{b}}v_{c}v_{d}-\frac{1}{2}\,C^{a\overline{b}\overline{c}%
d}\,w_{\overline{b}}w_{\overline{c}}v_{d}-\frac{1}{6}\,C^{a\overline{b}%
\overline{c}\overline{d}}\,w_{\overline{b}}w_{\overline{c}}w_{\overline{d}} 
\notag \\
&&,\forall a  \label{74}
\end{eqnarray}%
and 
\begin{eqnarray}
i\,\Upsilon ^{\overline{a}} &=&A^{\overline{a}\overline{b}}\,w_{\overline{b}%
}+A^{\overline{a}b}\,v_{b}-\frac{i}{2}\,B^{\overline{a}bc}\,v_{b}v_{c}-i\,B^{%
\overline{a}\overline{b}c}\,w_{\overline{b}}v_{c}-\frac{i}{2}\,B^{\overline{a%
}\overline{b}\overline{c}}\,w_{\overline{b}}w_{\overline{c}}+  \notag \\
&&-\frac{1}{6}\,C^{\overline{a}bcd}\,v_{b}v_{c}v_{d}-\frac{1}{2}\,C^{%
\overline{a}\overline{b}cd}\,w_{\overline{b}}v_{c}v_{d}-\frac{1}{2}\,C^{%
\overline{a}\overline{b}\overline{c}d}\,w_{\overline{b}}w_{\overline{c}%
}v_{d}-\frac{1}{6}\,C^{\overline{a}\overline{b}\overline{c}\overline{d}}\,w_{%
\overline{b}}w_{\overline{c}}w_{\overline{d}}  \notag \\
&&,\forall \overline{a}.  \label{75}
\end{eqnarray}%
As we ultimately want to arrive at a Landau free energy as an expansion
up-to fourth order in the $\Psi ^{a}$ - and the $\Upsilon ^{\overline{a}}$
fields, we only need to solve these last two vector-equations iteratively
for $v_{a}$ and $w_{\overline{a}}$ up-to third order in the $\Psi $'s and
the $\Upsilon $'s. Now, the set of equations (\ref{74}) and (\ref{75}) can
be written in the following matrix form,

\bigskip 
\begin{equation}
\left[ 
\begin{array}{cc}
I_{a}^{b} & A_{ac}^{-1}A^{c\overline{b}} \\ 
A_{\overline{a}\overline{c}}^{-1}A^{\overline{c}b} & I_{\overline{a}}^{%
\overline{b}}%
\end{array}%
\right] \left[ 
\begin{array}{c}
v_{b} \\ 
w_{\overline{b}}%
\end{array}%
\right] =\left[ 
\begin{array}{c}
iA_{ab}^{-1}\Psi ^{b}+... \\ 
iA_{\overline{a}\overline{b}}^{-1}\Upsilon ^{\overline{b}}+...%
\end{array}%
\right] .  \label{76}
\end{equation}%
where $I_{a}^{b}$ and $I_{\overline{a}}^{\overline{b}}$ are appropriate
''unit'' matrices. By defining the matrix $\mathbf{T}$ to be the inverse of
the matrix on the lhs of this equation, i.e.,

\begin{equation}
\mathbf{T}\equiv \left[ 
\begin{array}{cc}
I_{a}^{b} & A_{ac}^{-1}A^{c\overline{b}} \\ 
A_{\overline{a}\overline{c}}^{-1}A^{\overline{c}b} & I_{\overline{a}}^{%
\overline{b}}%
\end{array}%
\right] ^{-1}  \label{77}
\end{equation}%
it follows that the ''first order'' solution is given by,

\begin{equation}
\left[ 
\begin{array}{c}
v_{b} \\ 
w_{\overline{b}}%
\end{array}%
\right] =\mathbf{T}\left[ 
\begin{array}{c}
iA_{ab}^{-1}\Psi ^{b} \\ 
iA_{\overline{a}\overline{b}}^{-1}\Upsilon ^{\overline{b}}%
\end{array}%
\right] .  \label{78}
\end{equation}%
The contributions of terms second order in the $\Psi $'s and the $\Upsilon $%
's are then obtained by substitution of this first order result in the terms
which are quadratic in the $v$'s and the $w$'s on the rhs of (\ref{76}) and
so on. The final result for $v_{a}$ and $w_{\overline{a}}$ up-to third
order\ in the $\Psi $'s and the $\Upsilon $'s can then be written as, 
\begin{eqnarray}
v_{a}\equiv M_{ae}^{(1)}\Psi ^{e}+M_{a\overline{e}}^{(1)}\Upsilon ^{%
\overline{e}}+\frac{i}{2}M_{aef}^{(2)}\Psi ^{e}\Psi ^{f}+iM_{ae\overline{f}%
}^{(2)}\Psi ^{e}\Upsilon \overline{^{f}}+\frac{i}{2}M_{a\overline{e}%
\overline{f}}^{(2)}\Upsilon ^{\overline{e}}\Upsilon \overline{^{f}}+  \notag
\label{79} \\
\frac{1}{6}M_{aefg}^{(3)}\Psi ^{e}\Psi ^{f}\Psi ^{g}+\frac{1}{2}M_{aef%
\overline{g}}^{(3)}\Psi ^{e}\Psi ^{f}\Upsilon ^{\overline{g}}+\frac{1}{2}%
M_{ae\overline{f}\overline{g}}^{(3)}\Psi ^{e}\Upsilon \overline{^{f}}%
\Upsilon ^{\overline{g}}+\frac{1}{6}M_{a\overline{e}\overline{f}\overline{g}%
}^{(3)}\Upsilon ^{\overline{e}}\Upsilon \overline{^{f}}\Upsilon ^{\overline{g%
}}  \notag \\
\end{eqnarray}%
and 
\begin{eqnarray}
w_{\overline{a}} &\equiv &M_{\overline{a}e}^{(1)}\Psi ^{e}+M_{\overline{a}%
\overline{e}}^{(1)}\Upsilon ^{\overline{e}}+\frac{i}{2}M_{\overline{a}%
ef}^{(2)}\Psi ^{e}\Psi ^{f}+iM_{\overline{a}e\overline{f}}^{(2)}\Psi
^{e}\Upsilon \overline{^{f}}+\frac{i}{2}M_{\overline{a}\overline{e}\overline{%
f}}^{(2)}\Upsilon ^{\overline{e}}\Upsilon \overline{^{f}}+  \notag \\
&&\frac{1}{6}M_{\overline{a}efg}^{(3)}\Psi ^{e}\Psi ^{f}\Psi ^{g}+\frac{1}{2}%
M_{\overline{a}ef\overline{g}}^{(3)}\Psi ^{e}\Psi ^{f}\Upsilon ^{\overline{g}%
}+\frac{1}{2}M_{\overline{a}e\overline{f}\overline{g}}^{(3)}\Psi
^{e}\Upsilon \overline{^{f}}\Upsilon ^{\overline{g}}+\frac{1}{6}M_{\overline{%
a}\overline{e}\overline{f}\overline{g}}^{(3)}\Upsilon ^{\overline{e}%
}\Upsilon \overline{^{f}}\Upsilon ^{\overline{g}}  \notag \\
&&  \label{80}
\end{eqnarray}%
with the various coefficients given by,

\begin{subequations}
\begin{eqnarray}
M_{ae}^{(1)} &\equiv &i\,T_{a}^{b}A_{be}^{-1}  \label{81} \\
M_{\overline{a}e}^{(1)} &\equiv &i\,T_{\overline{a}}^{b}A_{be}^{-1}
\label{82} \\
M_{a\overline{e}}^{(1)} &\equiv &i\,T_{a}^{\overline{b}}A_{\overline{b}%
\overline{e}}^{-1}  \label{83} \\
M_{\overline{a}\overline{e}}^{(1)} &\equiv &i\,T_{\overline{a}}^{\overline{b}%
}A_{\overline{b}\overline{e}}^{-1}  \label{84s}
\end{eqnarray}%
and 
\end{subequations}
\begin{subequations}
\begin{equation}
M_{\widetilde{a}\widetilde{e}\widetilde{f}}^{(2)}\equiv -i\,M_{\widetilde{a}%
\widetilde{d}}^{(1)}B^{\widetilde{d}\widetilde{b}\widetilde{c}}M_{\widetilde{%
b}\widetilde{e}}^{(1)}M_{\widetilde{c}\widetilde{f}}^{(1)}  \label{85}
\end{equation}%
and 
\end{subequations}
\begin{equation}
M_{\widetilde{a}\widetilde{e}\widetilde{f}\widetilde{g}}^{(3)}\equiv -iM_{%
\widetilde{a}\widetilde{h}}^{(1)}\left[ C^{\widetilde{h}\widetilde{b}%
\widetilde{c}\widetilde{d}}+3i\,B^{\widetilde{h}\widetilde{b}\widetilde{k}%
}M_{\widetilde{k}\widetilde{l}}^{(1)}B^{\widetilde{l}\widetilde{c}\widetilde{%
d}}\right] M_{\widetilde{b}\widetilde{e}}^{(1)}M_{\widetilde{c}\widetilde{f}%
}^{(1)}M_{\widetilde{d}\widetilde{g}}^{(1)}.  \label{86}
\end{equation}%
In these last two expressions a label such as $\widetilde{a}$ denotes both $%
a $ and $\overline{a}$ and if $\widetilde{a}$ appears both as a subscript
and a superscript, such as in $x_{\widetilde{a}}\,y^{\widetilde{a}}$, then
the following summation convention is implied $x_{\widetilde{a}}\,y^{%
\widetilde{a}}\equiv x_{a}\,y^{a}+x_{\overline{a}}\,y^{\overline{a}}$.

If we now substitute the $\Psi -\Upsilon $ expansions for $v_{a}$ (\ref{79})
and $w_{a}$ (\ref{80}) into $i\,[v_{a}\Psi ^{a}+w_{\overline{a}}\Upsilon ^{%
\overline{a}}]+\frac{\Lambda \lbrack \underline{v},\overline{w}]}{V}$ we
obtain $\Phi \lbrack \underline{\Psi },\overline{\Upsilon }]$ and hence the
partition function according to Eq. (\ref{71}), (\ref{72}) and (\ref{73}), 
\begin{equation}
Z\simeq \prod_{c}\text{\/}^{\prime }\prod_{\overline{d}}\text{\/}\int D\Psi
^{c}\int D\Upsilon ^{\overline{d}}\,e^{V\,\{\tilde{\chi}_{ab}\Psi ^{a}\Psi
^{b}+\frac{1}{2}\omega _{\overline{a}\overline{b}}\Upsilon ^{\overline{a}%
}\Upsilon ^{\overline{b}}+\Phi \lbrack \underline{\Psi },\overline{\Upsilon }%
]\}}.\,  \label{87}
\end{equation}%
The Landau free energy, that is the free energy of the system within the 
\textit{mean field approximation}, can be obtained by again applying the
saddle-point method, but now to approximately evaluate this last set of
functional integrals. If we write the result as, 
\begin{equation}
Z\simeq e^{-F_{L}},  \label{88}
\end{equation}%
then this Landau free energy $F_{L}$ (in units of $k_{B}T$) is given by, 
\begin{gather}
\frac{F_{L}}{V}=\underset{\underline{\Psi },\overline{\Upsilon }}{\min }%
\{(\Gamma _{ab}^{(2)}-\widetilde{\chi }_{ab})\Psi ^{a}\Psi ^{b}+2\Gamma _{a%
\overline{b}}^{(2)}\Psi ^{a}\Upsilon ^{\overline{b}}+  \notag \\
(\Gamma _{\overline{a}\overline{b}}^{(2)}-\frac{1}{2}\omega _{\overline{a}%
\overline{b}})\Upsilon ^{\overline{a}}\Upsilon ^{\overline{b}}-\frac{1}{3}%
\omega _{ab}\Upsilon ^{a,ij}\delta _{ij}(\Psi ^{b}+f^{b})+  \notag \\
\Gamma _{abc}^{(3)}\Psi ^{a}\Psi ^{b}\Psi ^{c}+3\Gamma _{ab\overline{c}%
}^{(3)}\Psi ^{a}\Psi ^{b}\Upsilon ^{\overline{c}}+3\Gamma _{a\overline{b}%
\overline{c}}^{(3)}\Psi ^{a}\Upsilon ^{\overline{b}}\Upsilon ^{\overline{c}%
}+\Gamma _{\overline{a}\overline{b}\overline{c}}^{(3)}\Upsilon ^{\overline{a}%
}\Upsilon ^{\overline{b}}\Upsilon ^{\overline{c}}+  \notag \\
\Gamma _{abcd}^{(4)}\Psi ^{a}\Psi ^{b}\Psi ^{c}\Psi ^{d}+4\Gamma _{abc%
\overline{d}}^{(4)}\Psi ^{a}\Psi ^{b}\Psi ^{c}\Upsilon ^{\overline{d}%
}+6\Gamma _{ab\overline{c}\overline{d}}^{(4)}\Psi ^{a}\Psi ^{b}\Upsilon ^{%
\overline{c}}\Upsilon ^{\overline{d}}+  \notag \\
4\Gamma _{a\overline{b}\overline{c}\overline{d}}^{(4)}\Psi ^{a}\Upsilon ^{%
\overline{b}}\Upsilon ^{\overline{c}}\Upsilon ^{\overline{d}}+\Gamma _{%
\overline{a}\overline{b}\overline{c}\overline{d}}^{(4)}\Upsilon ^{\overline{a%
}}\Upsilon ^{\overline{b}}\Upsilon ^{\overline{c}}\Upsilon ^{\overline{d}}\}.
\label{89}
\end{gather}%

The coefficient functions ($\Gamma $'s) in this expression are called 
\textit{vertices} and are defined by, 
\begin{equation}
\Gamma _{\widetilde{a}\widetilde{b}}^{(2)}\equiv \frac{1}{2}\underset{perm(%
\widetilde{a},\widetilde{b})}{\sum }[-iM_{\widetilde{a}\widetilde{b}%
}^{(1)}+\ \frac{1}{2}\ A^{\widetilde{c}\widetilde{d}}M_{\widetilde{c}%
\widetilde{a}}^{(1)}M_{\widetilde{d}\widetilde{b}}^{(1)}]  \label{vertex2}
\end{equation}%
and 
\begin{equation}
\Gamma _{\widetilde{a}\widetilde{b}\widetilde{c}}^{(3)}\equiv \dfrac{1}{6}%
\underset{perm(\widetilde{a},\widetilde{b},\widetilde{c})}{\sum }[\frac{1}{2}%
M_{\widetilde{a}\widetilde{b}\widetilde{c}}^{(2)}+\frac{i}{2}\ A^{\widetilde{%
d}\widetilde{e}}M_{\widetilde{d}\widetilde{a}\widetilde{b}}^{(2)}M_{%
\widetilde{e}\widetilde{c}}^{(1)}-\frac{i}{6}\ B^{\widetilde{d}\widetilde{e}%
\widetilde{f}}M_{\widetilde{d}\widetilde{a}}^{(1)}M_{\widetilde{e}\widetilde{%
b}}^{(1)}M_{\widetilde{f}\widetilde{c}}^{(1)}]  \label{vertex3}
\end{equation}%
and finally 
\begin{gather}
\Gamma _{\widetilde{a}\widetilde{b}\widetilde{c}\widetilde{d}}^{(4)}\equiv 
\frac{1}{24}\underset{perm(\widetilde{a},\widetilde{b},\widetilde{c},%
\widetilde{d})}{\sum }[-\frac{i}{6}M_{\widetilde{a}\widetilde{b}\widetilde{c}%
\widetilde{d}}^{(3)}+\dfrac{1}{6}A^{\widetilde{e}\widetilde{f}}M_{\widetilde{%
e}\widetilde{a}\widetilde{b}\widetilde{c}}^{(3)}M_{\widetilde{f}\widetilde{d}%
}^{(1)}-\frac{1}{8}A^{\widetilde{e}\widetilde{f}}M_{\widetilde{e}\widetilde{a%
}\widetilde{b}}^{(2)}M_{\widetilde{f}\widetilde{c}\widetilde{d}}^{(2)}+ 
\notag \\
\dfrac{1}{4}B^{\widetilde{e}\widetilde{f}\widetilde{g}}M_{\widetilde{e}%
\widetilde{a}\widetilde{b}}^{(2)}M_{\widetilde{f}\widetilde{c}}^{(1)}M_{%
\widetilde{g}\widetilde{d}}^{(1)}\ \ \ -\frac{1}{24}C^{\widetilde{e}%
\widetilde{f}\widetilde{g}\widetilde{h}}M_{\widetilde{e}\widetilde{a}%
}^{(1)}M_{\widetilde{f}\widetilde{b}}^{(1)}M_{\widetilde{g}\widetilde{c}%
}^{(1)}M_{\widetilde{h}\widetilde{d}}^{(1)}].  \label{vertex4}
\end{gather}%
The fourth order vertex is symmetric with respect to a permutation of the
indices $\widetilde{a}$, $\widetilde{b}$, $\widetilde{c}$ and $\widetilde{d}$%
. The second and third order vertex contain the same kind of symmetry. The
terms in the vertices are not always permutational symmetric. For example if 
$\widetilde{a}=a$ and $\widetilde{b}=\overline{b}$ \ the vertex $\Gamma _{a%
\overline{b}}^{(2)}$ contains the terms $-iM_{a\overline{b}}^{(1)}$ and $%
-iM_{\overline{b}a}^{(1)}$. From Eq.(\ref{82}) and (\ref{83}) it is clear
that in general $M_{a\overline{b}}^{(1)}\neq M_{\overline{b}a}^{(1)}$.

\section*{Supplemental material III}
\markboth{III}{}\renewcommand{\theequation}{III.\arabic{equation}}%
\setcounter{equation}{0}

In order to be able to extract a Landau free energy from (\ref{35A}), we will
need to rework $\Lambda $ somewhat more. This, however, turns out to be the
most essential step in the whole derivation of this free energy. Using the
decompositions 
\begin{equation}
\hat{\rho}^{\alpha }(\underline{x})\equiv \sum\limits_{sm}\hat{\rho}%
_{sm}^{\alpha }(\underline{x})\quad \text{where}\quad \hat{\rho}%
_{sm}^{\alpha }(\underline{x})\equiv \int_{0}^{L_{s}}dl\ \sigma _{s}^{\alpha
}(l)\,\delta (\underline{x}-\underline{R}_{m}^{s}(l))  \label{37}
\end{equation}%
\begin{eqnarray}
\underline{\underline{\hat{S}}}^{\beta }(\underline{x}) &\equiv
&\sum\limits_{sm}\underline{\underline{\hat{S}}}_{sm}^{\beta }(\underline{x}%
)\quad \text{where}\quad  \notag \\
\underline{\underline{\hat{S}}}_{sm}^{\beta }(\underline{x}) &\equiv
&\int_{0}^{L_{s}}dl\,\sigma _{s}^{\alpha }(l)\,\underline{u}_{m}^{s}(l)\,%
\underline{u}_{m}^{s}(l)\,\delta (\underline{x}-\underline{R}_{m}^{s}(l))
\label{38}
\end{eqnarray}%
and 
\begin{eqnarray}
\hat{H}_{0} &\equiv &\sum\limits_{sm}\hat{H}_{0}^{sm}\quad \text{where} 
\notag \\
\hat{H}_{0}^{sm} &\equiv &\frac{3}{4}\left\{ \left[ \underline{u}%
_{m}^{s}(0)\,\right] ^{2}+\left[ \underline{u}_{m}^{s}(L_{s})\,\right]
^{2}\right\} +  \notag \\
&&+\frac{3}{4}\int_{0}^{L_{s}}dl\,\left\{ \frac{1}{\lambda _{s}(l)}\,\left[ 
\underline{u}_{m}^{s}(l)\,\right] ^{2}+\,\lambda _{s}(l)\,\left[ \underline{%
\dot{u}}_{m}^{s}(l)\,\right] ^{2}\right\}  \label{39}
\end{eqnarray}%
it follows that (\ref{36}) can be written as, 
\begin{eqnarray}
\Lambda &\equiv &\sum\limits_{sm}\ln \int d^{3}\widetilde{U}_{m}^{s}\int
d^{3}U_{m}^{s}\int D\underline{R}_{m}^{s}\,\int D\underline{u}_{m}^{s}\delta
(\,\underline{u}_{m}^{s}(0)-\widetilde{\underline{U}}_{m}^{s})\,\delta (\,%
\underline{u}_{m}^{s}(L_{s})-\underline{U}_{m}^{s})\,\times  \notag \\
&&\times \delta \left[ \underline{R}_{m}^{s}-\int dl\,\underline{u}%
_{m}^{s}(l)\right] \,e^{-\hat{H}_{0}^{sm}-\,\,i\int_{V}d^{3}x\,\{\sum%
\limits_{\alpha }\text{\/}\tilde{J}^{\alpha }(\underline{x})\,\hat{\rho}%
_{sm}^{\alpha }(\underline{x})+\sum\limits_{\beta }\underline{\underline{K}}%
^{\beta }(\underline{x})\,\,:\,\underline{\underline{\widehat{Q}}}%
_{sm}^{\beta }(\underline{x})\}}.  \notag \\
&&  \label{40}
\end{eqnarray}%
From a closer inspection of this last expression, it becomes clear that each
term in (\ref{40}) in the sum over $m$ for a given chain type $s$, i.e. each
term in the sum over all chains of a given type in the system, gives the
same contribution to $\Lambda $. Therefore $\Lambda $ can be simplified to, 
\begin{eqnarray}
\Lambda &=&\sum\limits_{s}n_{s}\ln \int d^{3}\widetilde{U}^{s}\int
d^{3}U^{s}\int D\underline{R}^{s}\,\int D\underline{u}^{s}\delta (\,%
\underline{u}^{s}(0)-\widetilde{\underline{U}}^{s})\,\delta (\,\underline{u}%
^{s}(L_{s})-\underline{U}^{s})\,\times  \notag \\
&&\times \delta \left[ \underline{R}^{s}-\int dl\,\underline{u}^{s}(l)\right]
\,e^{-\hat{H}_{0}^{s}-\,\,i\int_{V}d^{3}x\,\{\sum\limits_{\alpha }\text{\/}%
\tilde{J}^{\alpha }(\underline{x})\,\hat{\rho}_{s}^{\alpha }(\underline{x}%
)+\sum\limits_{\beta }\underline{\underline{K}}^{\beta }(\underline{x}%
)\,\,:\,\underline{\underline{\widehat{Q}}}_{s}^{\beta }(\underline{x})\}}.
\label{41}
\end{eqnarray}%
Note that the dummy index $m$ has been dropped in all quantities. By
introducing the number density of chains of type $s$, i.e. $\rho _{s}\equiv 
\frac{n_{s}}{V}$, and defining 
\begin{equation}
\hat{\Omega}\equiv -\,\,i\sum\limits_{\alpha }\int_{V}d^{3}x\,\{\text{\/}%
\tilde{J}^{\alpha }(\underline{x})\,\hat{\rho}_{s}^{\alpha }(\underline{x})+%
\underline{\underline{K}}^{\alpha }(\underline{x})\,\,:\,\underline{%
\underline{\widehat{Q}}}_{s}^{\alpha }(\underline{x})\}  \label{42}
\end{equation}%
(\ref{41}) can finally be written as, 
\begin{equation}
\Lambda \equiv V\sum\limits_{s}\rho _{s}\,\ln \,\langle \,e^{^{\hat{\Omega}%
_{s}}}\rangle _{0}\equiv V\,\langle \,\ln \,\langle \,e^{\hat{\Omega}%
}\rangle _{0}\rangle _{d}.  \label{43}
\end{equation}%
In this last expression the second average with subscript $d$ is a \textit{%
disorder average, i.e.} an average over the quenched disorder in the
copolymer chains. More important this quenched average involves the
logarithm of a quantity proportional to the partition function, as can be
seen from (\ref{35A}), and therefore it is the free energy that is being
averaged over the disorder. To calculate the average of the logarithm of the
partition function one can resort to the replica method \cite{Parisi}, but
this is not necessary for the kind of quenched disorder one encounters in
statistical copolymer systems, as will be shown now.

As the Landau free energy for this system involves an expansion up-to fourth
order in the order-parameter fields $\{\psi ^{\alpha }(\underline{x})\}$ and 
$\{\underline{\underline{Q}}^{\beta }(\underline{x})\}$ , the thing to do is
to expand $\Lambda $ up-to fourth order in $\hat{\Omega}$. The reason for
this step will become clear in the process. The result is, 
\begin{equation}
\frac{\Lambda }{V}\simeq \frac{1}{2}\,\langle \langle \,\hat{\Omega}%
^{2}\,\rangle _{0}\rangle _{d}+\frac{1}{6}\,\langle \langle \,\hat{\Omega}%
^{3}\,\rangle _{0}\rangle _{d}+\frac{1}{24}\,\langle \langle \,\hat{\Omega}%
^{4}\,\rangle _{0}\rangle _{d}-\frac{1}{8}\,\langle \langle \,\hat{\Omega}%
^{2}\,\rangle _{0}^{2}\rangle _{d}.  \label{44}
\end{equation}%
where we have used the fact that $\langle \,\hat{\Omega}\,\rangle _{0}=0$, a
result that is easily obtained. Let's first consider the second-order term.
By using (\ref{42}) it follows that, 
\begin{eqnarray}
\frac{1}{2}\,\langle \langle \,\hat{\Omega}^{2}\,\rangle _{0}\rangle _{d}
&\equiv &\frac{(-i)^{2}}{2}\sum_{\alpha \beta
}\int_{V}d^{3}x\int_{V}d^{3}y\,\{\tilde{J}^{\alpha }(\underline{x})\,\,%
\tilde{J}^{\beta }(\underline{y})\sum_{s}\rho _{s}\langle \,\hat{\rho}%
_{s}^{\alpha }(\underline{x})\,\hat{\rho}_{s}^{\beta }(\underline{y}%
)\,\rangle _{0}+  \notag \\
&&+2\sum_{ij}\,K_{ij}^{\alpha }(\underline{x})\,\,\tilde{J}^{\beta }(%
\underline{y})\sum_{s}\rho _{s}\langle \,\widehat{Q}_{s,ij}^{\alpha }(%
\underline{x})\,\hat{\rho}_{s}^{\beta }(\underline{y})\,\rangle _{0}+  \notag
\\
&&+\,\sum_{ij}\sum_{i^{\prime }j^{\prime }}K_{ij}^{\alpha }(\underline{x}%
)\,K_{i^{\prime }j^{\prime }}^{\beta }(\underline{y})\sum_{s}\rho
_{s}\langle \widehat{Q}_{s,ij}^{\alpha }(\underline{x})\,\,\widehat{Q}%
_{s,i^{\prime }j^{\prime }}^{\beta }(\underline{y})\,\rangle _{0}\}.
\label{45}
\end{eqnarray}%
Now by invoking the Fourier-representations of both $\hat{\rho}_{s}^{\alpha
}(\underline{x})$ and $\widehat{Q}_{s,ij}^{\alpha }(\underline{x})$, i.e., 
\begin{equation}
\hat{\rho}_{s}^{\alpha
}(\underline{x})\equiv \frac{1}{V}\sum_{\underline{q}%
}\int_{0}^{L_{s}}dl\,\sigma _{s}^{\alpha }(l)\,e^{i\underline{q}\cdot (%
\underline{x}-\underline{R}^{s}(l))}  \label{46}
\end{equation}%
and 
\begin{equation}
\widehat{Q}_{s,ij}^{\alpha }(\underline{x})\equiv \frac{1}{V}\sum_{%
\underline{q}}\int_{0}^{L_{s}}dl\,\sigma _{s}^{\alpha }(l)\,(\underline{u}%
^{s}(l)\,\underline{u}^{s}(l)\,\,-\frac{1}{3}\underline{\underline{I}})e^{i%
\underline{q}\cdot (\underline{x}-\underline{R}^{s}(l))}  \label{47}
\end{equation}%
the three correlation functions appearing in (\ref{45}) can be written as, 
\begin{eqnarray}
\langle \langle \,\hat{\rho}^{\alpha }(\underline{x})\,\hat{\rho}^{\beta }(%
\underline{y})\,\rangle _{0}\rangle _{d} &=&\frac{1}{V^{2}}\sum_{\underline{q%
}\underline{q}^{^{\prime }}}e^{i(\underline{q}\cdot \underline{x}+\underline{%
q^{\prime }}\cdot \underline{y})}\sum_{s}\rho
_{s}\int_{0}^{L_{s}}dl\,\int_{0}^{L_{s}}dl^{\prime }\,\sigma _{s}^{\alpha
}(l)\,\sigma _{s}^{\beta }(l^{\prime })\,\times  \notag \\
&&\times \langle \,e^{-\,i(\underline{q}\cdot \underline{R}^{s}(l)+%
\underline{q}^{^{\prime }}\cdot \underline{R}^{s}(l^{\prime }))}\,\rangle
_{0}  \notag \\
&\equiv &\frac{1}{V^{2}}\sum_{\underline{q}\underline{q}^{^{\prime }}}e^{i(%
\underline{q}\cdot \underline{x}+\underline{q^{\prime }}\cdot \underline{y}%
)}\,\sum_{s}\rho _{s}A_{s}^{JJ,\alpha \beta }(\underline{q},\underline{q}%
^{^{\prime }})  \notag \\
&\equiv &\frac{1}{V^{2}}\sum_{\underline{q}\underline{q}^{^{\prime }}}e^{i(%
\underline{q}\cdot \underline{x}+\underline{q^{\prime }}\cdot \underline{y}%
)}\,A^{JJ,\alpha \beta }(\underline{q},\underline{q}^{^{\prime }}),
\label{48}
\end{eqnarray}%
\begin{eqnarray}
\langle \langle \,\widehat{Q}_{s,ij}^{\alpha }(\underline{x})\,\hat{\rho}%
_{s}^{\beta }(\underline{y})\,\rangle _{0}\rangle _{d} &=&\frac{1}{V^{2}}%
\sum_{\underline{q}\underline{q}^{^{\prime }}}e^{i(\underline{q}\cdot 
\underline{x}+\underline{q^{\prime }}\cdot \underline{y})}\sum_{s}\rho
_{s}\int_{0}^{L_{s}}dl\,\int_{0}^{L_{s}}dl^{\prime }\,\sigma _{s}^{\alpha
}(l)\,\sigma _{s}^{\beta }(l^{\prime })\,\times  \notag \\
&&\times \langle \,(u_{i}^{s}(l)\,u_{j}^{s}(l)-\frac{1}{3}\delta
_{ij})\,e^{-\,i(\underline{q}\cdot \underline{R}^{s}(l)+\underline{q}%
^{^{\prime }}\cdot \underline{R}^{s}(l^{\prime }))}\,\rangle _{0}  \notag \\
&\equiv &\frac{1}{V^{2}}\sum_{\underline{q}\underline{q}^{^{\prime }}}e^{i(%
\underline{q}\cdot \underline{x}+\underline{q^{\prime }}\cdot \underline{y}%
)}\,\sum_{s}\rho _{s}A_{s,ij}^{KJ,\alpha \beta }(\underline{q},\underline{q}%
^{^{\prime }})  \notag \\
&\equiv &\frac{1}{V^{2}}\sum_{\underline{q}\underline{q}^{^{\prime }}}e^{i(%
\underline{q}\cdot \underline{x}+\underline{q^{\prime }}\cdot \underline{y}%
)}\,A_{ij}^{KJ,\alpha \beta }(\underline{q},\underline{q}^{^{\prime }})
\label{49}
\end{eqnarray}%
and 
\begin{eqnarray}
\langle \langle \,\widehat{Q}_{s,ij}^{\alpha }(\underline{x})\,\,\widehat{Q}%
_{s,i^{\prime }j^{\prime }}^{\beta }(\underline{y})\,\rangle _{0}\rangle
_{d} &=&\frac{1}{V^{2}}\sum_{\underline{q}\underline{q}^{^{\prime }}}e^{i(%
\underline{q}\cdot \underline{x}+\underline{q^{\prime }}\cdot \underline{y}%
)}\sum_{s}\rho _{s}\int_{0}^{L_{s}}dl\,\int_{0}^{L_{s}}dl^{\prime }\,\sigma
_{s}^{\alpha }(l)\,\sigma _{s}^{\beta }(l^{\prime })\,\times  \notag \\
&&\times \langle \,(u_{i}^{s}(l)\,u_{j}^{s}(l)-\frac{1}{3}\delta
_{ij})\,(u_{i^{\prime }}^{s}(l^{\prime })\,u_{j^{\prime }}^{s}(l^{\prime })-%
\frac{1}{3}\delta _{i^{\prime }j^{\prime }}\,)e^{-\,i(\underline{q}\cdot 
\underline{R}^{s}(l)+\underline{q}^{^{\prime }}\cdot \underline{R}%
^{s}(l^{\prime }))}\,\rangle _{0}  \notag \\
&\equiv &\frac{1}{V^{2}}\sum_{\underline{q}\underline{q}^{^{\prime }}}e^{i(%
\underline{q}\cdot \underline{x}+\underline{q^{\prime }}\cdot \underline{y}%
)}\,\sum_{s}\rho _{s}A_{s,iji^{\prime }j^{\prime }}^{KK,\alpha \beta }(%
\underline{q},\underline{q}^{^{\prime }})  \notag \\
&\equiv &\frac{1}{V^{2}}\sum_{\underline{q}\underline{q}^{^{\prime }}}e^{i(%
\underline{q}\cdot \underline{x}+\underline{q^{\prime }}\cdot \underline{y}%
)}\,A_{iji^{\prime }j^{\prime }}^{KK,\alpha \beta }(\underline{q},\underline{%
q}^{^{\prime }}).  \label{50}
\end{eqnarray}%
The functions $A^{JJ,\alpha \beta }(\underline{q},\underline{q}^{^{\prime
}}) $, $A_{ij}^{KJ,\alpha \beta }(\underline{q},\underline{q}^{^{\prime }})$
and $A_{iji^{\prime }j^{\prime }}^{KK,\alpha \beta }(\underline{q},%
\underline{q}^{^{\prime }})$ that appear in respectively (\ref{48}), (\ref%
{49}) and (\ref{50}) are  second order single-chain correlation functions and will be calculated in Supplemental material I. Now, with the help of
these last three expressions this second-order contribution to $\frac{%
\Lambda }{V}$ becomes, 
\begin{eqnarray}
\frac{1}{2}\,\langle \langle \,\hat{\Omega}^{2}\,\rangle _{0}\rangle _{d}
&\equiv &\frac{(-i)^{2}}{2V^{2}}\sum_{\alpha \beta }\sum_{\underline{q}_{1},%
\underline{q}_{2}}\{A^{JJ,\alpha \beta }(\underline{q}_{1},\underline{q}%
_{2})\,\,\tilde{J}^{\alpha }(\underline{q}_{1})\,\tilde{J}^{\beta }(%
\underline{q}_{2})+  \notag \\
&&+2\sum_{ij}A_{ij}^{KJ,\alpha \beta }(\underline{q}_{1},\underline{q}%
_{2})\,\,K_{ij}^{\alpha }(\underline{q}_{1})\,\tilde{J}^{\beta }(\underline{q%
}_{2})+  \notag \\
&&+\sum_{i_{1}j_{1}}\sum_{i_{2}j_{2}}A_{i_{1}j_{1}i_{2}j_{2}}^{KK,\alpha
\beta }(\underline{q}_{1},\underline{q}_{2})\,K_{i_{1}j_{1}}^{\alpha }(%
\underline{q}_{1})\,K_{i_{2}j_{2}}^{\beta }(\underline{q}_{2})\},  \label{51}
\end{eqnarray}%
where $f(\underline{q})$ denotes the Fourier-transform of $f(\underline{x})$%
, i.e., 
\begin{equation}
f(\underline{q})\equiv \int_{V}d^{3}x\,e^{i\,\underline{q}\cdot \underline{x}%
}\,f(\underline{x}).  \label{52}
\end{equation}

As the partition function $Z$ (\ref{32}) involves a functional
integration over the $J$ fields, while $\frac{1}{2}\,\langle \langle \,\hat{%
\Omega}^{2}\,\rangle _{0}\rangle _{d}$ involves the $\tilde{J}$ fields, we
need to transform $\frac{1}{2}\,\langle \langle \,\hat{\Omega}^{2}\,\rangle
_{0}\rangle _{d}$ to the former kind of fields. This is most easily done by
recalling that from the definition of $\tilde{J}^{\alpha }(\underline{x})$ (%
\ref{JKthilde}), it follows that, 
\begin{equation}
\tilde{J}^{\alpha }(\underline{q})\equiv J^{\alpha }(\underline{q}%
)-J^{\alpha }(\underline{0})\,\delta (\underline{q}).  \label{JKthildeq}
\end{equation}%
In other words $\tilde{J}^{\alpha }(\underline{q})=J^{\alpha }(\underline{q}%
) $ for $\underline{q}\neq \underline{0}$ and if $\underline{q}=0$ then $%
\tilde{J}^{\alpha }(\underline{0})=0$. Therefore as the sums over the $%
\underline{q}$'s belonging to the fields $\tilde{J}^{\alpha }(\underline{q})$
in Eq. (\ref{51}) are anyhow restricted to non-zero $\underline{q}$'s, we
can simply change the $\tilde{J}$'s herein to $J$'s. Symbolically we now can
write the expression for $\frac{1}{2}\,\langle \langle \,\hat{\Omega}%
^{2}\,\rangle _{0}\rangle _{d}$ as, 
\begin{equation}
\frac{1}{2}\,\langle \langle \,\hat{\Omega}^{2}\,\rangle _{0}\rangle _{d}=-%
\frac{1}{2}\,[A^{ab}\,v_{a}v_{b}+2\,A^{\overline{a}b}\,w_{\overline{a}%
}v_{b}+A^{\overline{a}\overline{b}}\,w_{\overline{a}}w_{\overline{b}}],
\label{2eordeomega}
\end{equation}%
where we have introduced two sets of \textit{composite labels} $a\equiv (%
\underline{q}\neq \underline{0},\alpha )$, $b\equiv (\underline{q}\neq 
\underline{0},\beta )$ etc. and $\overline{a}\equiv (\underline{q},ij,\alpha
)$, $\overline{b}\equiv (\underline{q^{\prime }},i^{\prime }j^{\prime
},\beta )$ etc. and where $v_{a}\equiv \frac{J^{\alpha }(\underline{q})}{V}$
and $w_{\overline{a}}\equiv \frac{K_{ij}^{\alpha }(\underline{q})}{V}$.
Furthermore the Einstein summation convention has been used. The third and
fourth order contributions of $\frac{\Lambda }{V}$ are given by, 
\begin{subequations}
\begin{gather}
\frac{1}{6}\,\langle \langle \,\hat{\Omega}^{3}\,\rangle _{0}\rangle _{d}=+%
\frac{i}{6}\,[B^{abc}\,v_{a}v_{b}v_{c}+3\,B^{\overline{a}bc}\,w_{\overline{a}%
}v_{b}v_{c}+  \notag \\
+3\,B^{\overline{a}\overline{b}c}\,w_{\overline{a}}w_{\overline{b}}v_{c}+B^{%
\overline{a}\overline{b}\overline{c}}\,w_{\overline{a}}w_{\overline{b}}w_{%
\overline{c}}]  \label{3eordeomega} \\
\text{and}  \notag \\
\frac{1}{24}\,\langle \langle \,\hat{\Omega}^{4}\,\rangle _{0}\rangle _{d}-%
\frac{1}{8}\,\langle \langle \,\hat{\Omega}^{2}\,\rangle _{0}^{2}\rangle
_{d}=+\frac{1}{24}\,[C^{abcd}\,v_{a}v_{b}v_{c}v_{d}+  \notag \\
+4\,C^{\overline{a}bcd}\,w_{\overline{a}}v_{b}v_{c}v_{d}+6\,C^{\overline{a}%
\overline{b}cd}\,w_{\overline{a}}w_{\overline{b}}v_{c}v_{d}+  \notag \\
+4\,C^{\overline{a}\overline{b}\overline{c}d}\,w_{\overline{a}}w_{\overline{b%
}}w_{\overline{c}}v_{d}+C^{\overline{a}\overline{b}\overline{c}\overline{d}%
}\,w_{\overline{a}}w_{\overline{b}}w_{\overline{c}}w_{\overline{d}}].
\label{4eordeomega}
\end{gather}%
In the second order contribution (\ref{2eordeomega}) the $A$'s are second
order single-chain correlation functions. These are defined in Eq. (\ref{48}%
), (\ref{49}) and (\ref{50}). In the third order term the $B$'s are third
order single-chain correlation functions defined in a similar way. The $C$'s
in Eq. (\ref{4eordeomega}) contain a regular and a non-local contribution, $%
C^{abcd}=C_{reg}^{abcd}-C_{nl}^{abcd}$, $C^{\overline{a}bcd}=C_{reg}^{%
\overline{a}bcd}-C_{nl}^{\overline{a}bcd}$, etc. The regular contributions
are fourth order single-chain correlation functions which are similar to the 
$A$'s and $B$'s. The non-local correlation functions of a chain of kind $s$
follow from the term $-\frac{1}{8}\,\langle \langle \,\hat{\Omega}%
^{2}\,\rangle _{0}^{2}\rangle _{d}$ and are given by, 
\end{subequations}
\begin{gather}
C_{nl,s}^{abcd}=A_{s}^{ab}A_{s}^{cd}+A_{s}^{ac}A_{s}^{bd}+A_{s}^{ad}A_{s}^{bc},
\notag \\
C_{nl,s}^{\overline{a}bcd}=4A_{s}^{\overline{a}b}A_{s}^{cd}+4A_{s}^{%
\overline{a}c}A_{s}^{bd}+4A_{s}^{\overline{a}d}A_{s}^{bc},  \notag \\
C_{nl,s}^{\overline{a}\overline{b}cd}=6A_{s}^{\overline{a}\overline{b}%
}A_{s}^{cd}+6A_{s}^{\overline{a}c}A_{s}^{\overline{b}d}+6A_{s}^{\overline{a}%
d}A_{s}^{\overline{b}c},  \notag \\
C_{nl,s}^{\overline{a}\overline{b}\overline{c}d}=4A_{s}^{\overline{a}%
\overline{b}}A_{s}^{\overline{c}d}+4A_{s}^{\overline{a}\overline{c}}A_{s}^{%
\overline{b}d}+4A_{s}^{\overline{a}d}A_{s}^{\overline{b}\overline{c}}  \notag
\\
\text{and}  \notag \\
C_{nl,s}^{\overline{a}\overline{b}\overline{c}\overline{d}}=A_{s}^{\overline{%
a}\overline{b}}A_{s}^{\overline{c}\overline{d}}+A_{s}^{\overline{a}\overline{%
c}}A_{s}^{\overline{b}\overline{d}}+A_{s}^{\overline{a}\overline{d}}A_{s}^{%
\overline{b}\overline{c}}.
\end{gather}%
The quenched average is not taken over the $A$'s separately, but over the
terms in the non-local correlation functions, i.e. $\underset{s}{\sum }\rho
_{s}A_{s}^{ab}A_{s}^{cd}$ etc. These contributions to the fourth-order
coefficients are the so-called \textit{non-local terms}, which are typical
for polydisperse copolymer melts and which vanish once the number of segment
types $M$ exceeds the number of chain types in the system \cite{PanyKu1}.

By Fourier-transforming all the integrals involving the $\psi $ fields and
the $\underline{\underline{Q}}$ fields and making use of the fact that
according to the definition of $\tilde{\chi}_{\alpha \beta }$ (see (\ref{16}%
)) $\tilde{\chi}_{MM}\equiv 0$, the partition function $Z$ (\ref{35A}) can be
written as,
\begin{equation}
Z\equiv \prod_{c}\text{\/}^{\prime }\prod_{\overline{d}}\text{\/}\int D\Psi
^{c}\int D\Upsilon ^{\overline{d}}\,e^{V\,\{\tilde{\chi}_{ab}\Psi ^{a}\Psi
^{b}+\frac{1}{2}\omega _{\overline{a}\overline{b}}\Upsilon ^{\overline{a}%
}\Upsilon ^{\overline{b}}\}}\,\widetilde{Z}[\underline{\Psi },\overline{%
\Upsilon }]  \label{71}
\end{equation}%
with $\tilde{\chi}_{ab}\equiv \tilde{\chi}_{\alpha \beta }\,\delta (%
\underline{q}_{1}+\underline{q}_{2})$, $\omega _{\overline{a}\overline{b}%
}\equiv \omega _{\alpha \beta }\,|2\delta _{ii^{\prime }}\delta _{jj^{\prime
}}-\delta _{ij}\delta _{i^{\prime }j^{\prime }}|\,\delta (\underline{q}_{1}+%
\underline{q}_{2})$, $\Psi ^{a}$ $\equiv \frac{\psi ^{\alpha }(-\underline{q}%
)}{V}$, $\Upsilon ^{\overline{a}}\equiv \frac{Q_{ij}^{\alpha }(-\underline{q}%
)}{V}$ and 
\begin{equation}
\widetilde{Z}[\underline{\Psi },\overline{\Upsilon }]\equiv \prod_{g}\prod_{%
\overline{h}}\int Dv_{g}\int Dw_{\overline{h}}\,\,e^{V\,\{i\,[v_{a}\Psi
^{a}+w_{\overline{a}}\Upsilon ^{\overline{a}}]+\frac{\Lambda \lbrack 
\underline{v},\overline{w}]}{V}\,\}},  \label{72}
\end{equation}%
where $\underline{\Psi }\equiv \{\Psi ^{a}\}_{a}$ and $\overline{\Upsilon }%
\equiv \{\Upsilon ^{\overline{a}}\}_{\overline{a}}$.

\section*{Supplemental material IV}
\markboth{Supplemental material IV}{}\renewcommand{\theequation}{IV.\arabic{equation}}%
\setcounter{equation}{0}

In this supplemental material the first and second order
derivatives of $\widetilde{D}_{p^{\prime }+2}$ are calculated. These can be
used to calculate the factors $G_{2}$ and $G_{4}$ that are
given by Eq. (\ref{G2.1}) and (\ref{G4.4}). First
the limit $\Delta L\rightarrow 0$ in the interval $(l_{1}-\Delta L,l_{1})$
is considered. The index $p=p_{1}$ corresponds to this interval. It can be
derived that in this limit $A_{p_{1}}$, $B_{p_{1}}$\ ,$C_{p_{1}}$\ ,$%
D_{p_{1}}$\ and $E_{p_{1}}$\ become 
\begin{eqnarray}
A_{{\normalsize _{p_{1}}}} &\rightarrow &B_{{\normalsize _{p_{1}}}%
}\rightarrow \frac{3\lambda _{{\normalsize _{p_{1}}}}}{2\Delta L}, 
\label{Ap1} \\
C_{{\normalsize _{p_{1}}}} &=&\frac{i}{2}\Delta L,
\label{Cp1} \\
D_{{\normalsize _{p_{1}}}} &\rightarrow &\frac{-(\Delta L)^{3}}{18\lambda _{%
{\normalsize _{p_{1}}}}}\label{Dp1} \\
&&\text{and}  \notag \\
E_{{\normalsize _{p_{1}}}} &\rightarrow &\left( \frac{3\lambda _{%
{\normalsize _{p_{1}}}}}{4\pi \Delta L}\right) ^{\frac{3}{2}}.
\label{Ep1}
\end{eqnarray}%
To derive $\widetilde{A}_{p_{1}+1}$, $\widetilde{\underline{C}}_{p_{1}+1}$, $%
\widetilde{D}_{p_{1}+1}$ and $\widetilde{E}_{p_{1}+1}$, Eq.(\ref{intIp'}) is
applied in which $p^{\prime }$ is replaced by $p_{1}$ and the limit $\Delta
L\rightarrow 0$ is taken, 
\begin{gather}
\underset{{\normalsize \Delta L\rightarrow 0}}{\lim }{\normalsize \int d%
\underline{u}_{p_{1}}\underset{p=1}{\overset{p_{1}}{\prod }}\int d\underline{%
u}_{p-1}I_{p}\exp \left( -\frac{3}{4}u_{p-1}^{2}-\frac{3}{4}u_{p}^{2}\right)
=}  \notag \\
\underset{{\normalsize \Delta L\rightarrow 0}}{\lim }{\normalsize \int d%
\underline{u}_{p_{1}}\int d\underline{u}_{p_{1}-1}\widetilde{E}_{p_{1}}\exp
(-\frac{1}{2}\widetilde{A}_{p_{1}}u_{p_{1}-1}^{2}-\frac{1}{2}%
A_{p_{1}}u_{p_{1}}^{2}+B_{p_{1}}\underline{u}_{p_{1}-1}\cdot \underline{u}%
_{p_{1}})\times }  \notag \\
\exp ({\normalsize \underline{\widetilde{C}}_{p_{1}}\cdot \underline{u}%
_{p_{1}-1}+C_{p_{1}}\underline{Q}_{p_{1}}\cdot \underline{u}_{p_{1}}-\frac{1%
}{2}\widetilde{D}_{p_{1}}).}  \label{intIp1}
\end{gather}%
For every tangent vector $\underline{u}_{p}$ the integrand $I_{p}\exp \left(
-\frac{3}{4}u_{p-1}^{2}-\frac{3}{4}u_{p}^{2}\right) $ in\ Eq. (\ref{intIp1})
is continuous with respect to $\Delta L\neq 0$. Therefore it is allowed to
interchange the limit $\Delta L\rightarrow 0$ and the integrals over the
tangent vectors $\underline{u}_{p}$.\ Then Eq. (\ref{Apthilde}),(\ref%
{Dpthilde}),(\ref{Epthilde}) and (\ref{Fpthilde}) can be inserted in Eq. (%
\ref{intIp1}), which gives, 
\begin{gather}
{\normalsize \int d\underline{u}_{p_{1}}\underset{p=1}{\overset{p_{1}}{\prod 
}}\int d\underline{u}_{p-1}}\underset{{\normalsize \Delta L\rightarrow 0}}{%
{\normalsize \lim }}I_{p}\exp \left( -\frac{3}{4}u_{p-1}^{2}-\frac{3}{4}%
u_{p}^{2}\right) {\normalsize =}  \notag \\
{\normalsize \int d\underline{u}_{p_{1}}\int d\underline{u}_{p_{1}-1}%
\underset{\Delta L\rightarrow 0}{\lim }\widetilde{E}_{p_{1}-1}\left( \dfrac{%
2\pi }{\widetilde{A}_{p_{1}-1}}\right) ^{\tfrac{3}{2}}\times }  \notag \\
{\normalsize E_{p_{1}}\exp \left( -\frac{1}{2}A_{p_{1}}u_{p_{1}-1}^{2}-\frac{%
1}{2}A_{p_{1}}u_{p_{1}}^{2}+B_{p_{1}}\underline{u}_{p_{1}-1}\cdot \underline{%
u}_{p_{1}}\right) \times \ }  \notag \\
{\normalsize \exp (-\frac{1}{2}A_{p_{1}-1}u_{p_{1}-1}^{2}+\frac{%
B_{p_{1}-1}^{2}}{2\widetilde{A}_{p_{1}-1}}u_{p_{1}-1}^{2})\times }  \notag \\
{\normalsize \exp (\underline{\widetilde{C}}_{p_{1}}\cdot \underline{u}%
_{p_{1}-1}+C_{p_{1}}\underline{Q}_{p_{1}}\cdot \underline{u}_{p_{1}}-\frac{1%
}{2}\widetilde{D}_{p_{1}}).}  \label{intIp1.2}
\end{gather}%
From Eq.(\ref{Ap1}) and (\ref{Ep1}) it follows that in the limit $\Delta
L\rightarrow 0$, 
\begin{equation}
{\normalsize E_{p_{1}}\exp \left( -\frac{1}{2}A_{p_{1}}u_{p_{1}-1}^{2}-\frac{%
1}{2}A_{p_{1}}u_{p_{1}}^{2}+B_{p_{1}}\underline{u}_{p_{1}-1}\cdot \underline{%
u}_{p_{1}}\right) \rightarrow \delta (\underline{u}_{p_{1}}-\underline{u}%
_{p_{1}-1}).}  
\end{equation}%
Then integrating over $\underline{u}_{p_{1}-1}$ in Eq. (\ref{intIp1.2})
yields, 
\begin{gather}
{\normalsize \int d\underline{u}_{p_{1}}\underset{p=1}{\overset{p_{1}}{\prod 
}}\int d\underline{u}_{p-1}\underset{\Delta L\rightarrow 0}{\lim }I_{p}\exp
\left( -\frac{3}{4}u_{p-1}^{2}-\frac{3}{4}u_{p}^{2}\right) =}  \notag \\
{\normalsize \int d\underline{u}_{p_{1}}\widetilde{E}_{p_{1}-1}\left( \dfrac{%
2\pi }{\widetilde{A}_{p_{1}-1}}\right) ^{\tfrac{3}{2}}\times }  \notag \\
{\normalsize \exp (-\frac{1}{2}A_{p_{1}-1}u_{p_{1}}^{2}+\frac{B_{p_{1}-1}^{2}%
}{2\widetilde{A}_{p_{1}-1}}u_{p_{1}}^{2})\times \ }  \notag \\
\underset{{\normalsize \Delta L\rightarrow 0}}{{\normalsize \lim }}%
{\normalsize \exp (\underline{\widetilde{C}}_{p_{1}}\cdot \underline{u}%
_{p_{1}}+C_{p_{1}}\underline{Q}_{p_{1}}\cdot \underline{u}_{p_{1}}-\frac{1}{2%
}\widetilde{D}_{p_{1}}).}  \label{intIp1.3}
\end{gather}%
To derive $\widetilde{A}_{p_{1}+1}$, $\widetilde{\underline{C}}_{p_{1}+1}$, $%
\widetilde{D}_{p_{1}+1}$ and $\widetilde{E}_{p_{1}+1}$, Eq. (\ref{intIp1.3})
can be used in combination with $I_{p_{1}+1}\exp \left( -\frac{3}{4}%
u_{p_{1}}^{2}-\frac{3}{4}u_{p_{1}+1}^{2}\right) $, given by 
\begin{gather}
{\normalsize I_{p_{1}+1}\exp \left( -\frac{3}{4}u_{p_{1}}^{2}-\frac{3}{4}%
u_{p_{1}+1}^{2}\right) =E_{p_{1}+1}\exp (-\frac{1}{2}%
A_{p_{1}+1}(u_{p_{1}}^{2}+u_{p_{1}+1}^{2})+B_{p_{1}+1}\underline{u}%
_{p_{1}}\cdot \underline{u}_{p_{1}+1})}  \notag \\
\exp ({\normalsize C_{p_{1}+1}\underline{Q}_{p_{1}+1}\cdot (\underline{u}%
_{p_{1}}+\underline{u}_{p_{1}+1})-\frac{1}{2}D_{p_{1}+1}Q_{p_{1}+1}^{2}).} 
\label{Ip1+1}
\end{gather}%
It follows that $\widetilde{A}_{p_{1}+1}$, $\widetilde{\underline{C}}%
_{p_{1}+1}$, $\widetilde{D}_{p_{1}+1}$ and $\widetilde{E}_{p_{1}+1}$ are
equal to 
\begin{eqnarray}
{\normalsize \widetilde{A}_{p_{1}+1}} &=&{\normalsize %
A_{p_{1}+1}+A_{p_{1}-1}-\frac{B_{p_{1}-1}^{2}}{\widetilde{A}_{p_{1}-1}},} 
\label{Athildep1+1} \\
{\normalsize \widetilde{\underline{C}}_{p_{1}+1}} &=&{\normalsize \widetilde{%
\underline{C}}_{p_{1}}+C_{p_{1}}\underline{Q}_{p_{1}}+C_{p_{1}+1}\underline{Q%
}_{p_{1}+1},}\label{Dthildep1+1} \\
{\normalsize \widetilde{D}_{p_{1}+1}} &=&{\normalsize \widetilde{D}%
_{p_{1}}+D_{p_{1}+1}Q_{p_{1}+1}^{2},}\label{Ethildep1+1} \\
&&\text{and}  \notag \\
{\normalsize \widetilde{E}_{p_{1}+1}} &=&{\normalsize E_{p_{1}+1}\widetilde{E%
}_{p_{1}-1}\left( \frac{2\pi }{\widetilde{A}_{p_{1}-1}}\right) ^{\frac{3}{2}%
}.}\label{Fthildep1+1}
\end{eqnarray}%
For $p=p_{2}+1$ the same relations can be applied in which $p_{2}$ corresponds to the interval $(l_{2}-\Delta L,l_{2})$.
For other $p$'s we use Eq. (\ref{Apthilde}), (\ref{Dpthilde}), (\ref%
{Epthilde}) and (\ref{Fpthilde}). Eq. (\ref{Athildep1+1}) and (\ref%
{Fthildep1+1}) have the same form as Eq. (\ref{Apthilde}) and (\ref{Fpthilde}%
). In the limit $\Delta L\rightarrow 0$ the contribution of the $p_{1}$-th
interval has disappeared in $\widetilde{A}_{p_{1}+1}$ and $\widetilde{E}%
_{p_{1}+1}$. In Eq. (\ref{Dthildep1+1}) and (\ref{Ethildep1+1}) the limit $%
\Delta L\rightarrow 0$ has not been taken yet. Taking the limit and applying
Eq. (\ref{Dpthilde}), (\ref{Epthilde}), (\ref{Cp1}) and (\ref{Dp1}), it
follows that,
\begin{eqnarray}
\widetilde{\underline{C}}_{p_{1}+1} &=&C_{p_{1}+1}\underline{Q}%
_{p_{1}+1}+C_{p_{1}-1}\underline{Q}_{p_{1}-1}+\frac{B_{p_{1}-1}\underline{%
\widetilde{C}}_{p_{1}-1}}{\widetilde{A}_{p_{1}-1}},
\label{Dthildep1+1.2} \\
&&\text{and}  \notag \\
\widetilde{D}_{p_{1}+1} &=&D_{p_{1}+1}Q_{p_{1}+1}^{2}+\widetilde{D}%
_{p_{1}-1}-\frac{\widetilde{C}_{p_{1}-1}^{2}}{\widetilde{A}_{p_{1}-1}}. 
\label{Ethildep1+1.2}
\end{eqnarray}%
These equations have also the same form as Eq. (\ref{Dpthilde}) and (\ref%
{Epthilde}). In this form $\widetilde{\underline{C}}_{p_{1}+1}$ and $%
\widetilde{\underline{D}}_{p_{1}+1}$ do not depend on $\underline{Q}_{p_{1}}$
any more. So the $\underline{\eta }(l)$-dependence of $\widetilde{\underline{%
C}}_{p_{1}+1}$ and $\widetilde{\underline{D}}_{p_{1}+1}$ is lost.\ To derive
the first and second order functional derivatives of $\widetilde{\underline{C%
}}_{p_{1}+1}$ and $\widetilde{D}_{p_{1}+1}$ with respect to $\underline{\eta 
}(l_{1})$ and/or $%
\underline{\eta }(l_{2})$, Eq. (\ref{Dthildep1+1}) and (\ref{Ethildep1+1})
are used instead of Eq. (\ref{Dthildep1+1.2}) and (\ref{Ethildep1+1.2}). In
these expressions the limit $\Delta L\rightarrow 0$ has not been taken yet
so that the $\underline{\eta }(l)$-dependence is conserved. The
functional derivative is defined by Eq. (\ref{afgl-Qp}). In this definition
the limit $\Delta L\rightarrow 0$ is included.

The derivatives of $\widetilde{\underline{C}}_{p_{1}+1}$ and $\widetilde{D}%
_{p_{1}+1}$ must be known to calculate the first and second order functional
derivatives. According to Eq. (\ref{Dthildep1+1}) the first order functional
derivative of $\widetilde{\underline{C}}_{p_{1}+1}$ is, 
\begin{equation}
{\normalsize \dfrac{\delta \widetilde{\underline{C}}_{p_{1}+1}}{\delta \eta
_{\mu _{1}}(l_{1})}=\dfrac{\delta \widetilde{\underline{C}}_{p_{1}}}{\delta
\eta _{\mu _{1}}(l_{1})}+C_{p_{1}}\dfrac{\delta \underline{Q}_{p_{1}}}{%
\delta \eta _{\mu _{1}}(l_{1})}+C_{p_{1}+1}\dfrac{\delta \underline{Q}%
_{p_{1}+1}}{\delta \eta _{\mu _{1}}(l_{1})}.}
\label{1e-afgl-Dthildep1+1}
\end{equation}%
In Eq. (\ref{1e-afgl-Dthildep1+1}), $\widetilde{\underline{C}}_{p_{1}}$ is
equal to, 
\begin{equation}
{\normalsize \widetilde{\underline{C}}_{p_{1}}=C_{p_{1}}\underline{Q}%
_{p_{1}}+C_{p_{1}-1}\underline{Q}_{p_{1}-1}+\frac{B_{p_{1}-1}\underline{%
\widetilde{C}}_{p_{1}-1}}{\widetilde{A}_{p_{1}-1}},}
\label{Dthildep1}
\end{equation}%
according to Eq. (\ref{Dpthilde}). Now Eq. (\ref{1e-afgl-Dthildep1+1}) can
be written as, 
\begin{gather}
{\normalsize \dfrac{\delta \widetilde{\underline{C}}_{p_{1}+1}}{\delta \eta
_{\mu _{1}}(l_{1})}=\frac{B_{p_{1}-1}}{\widetilde{A}_{p_{1}-1}}\dfrac{\delta 
\widetilde{\underline{C}}_{p_{1}-1}}{\delta \eta _{\mu _{1}}(l_{1})}%
+C_{p_{1}-1}\dfrac{\delta \underline{Q}_{p_{1}-1}}{\delta \eta _{\mu
_{1}}(l_{1})}+}  \notag \\
{\normalsize 2C_{p_{1}}\dfrac{\delta \underline{Q}_{p_{1}}}{\delta \eta
_{\mu _{1}}(l_{1})}+C_{p_{1}+1}\dfrac{\delta \underline{Q}_{p_{1}+1}}{\delta
\eta _{\mu _{1}}(l_{1})}.}  \label{1e-afgl-Dthildep1+1.2}
\end{gather}%
In Eq.(\ref{1e-afgl-Dthildep1+1.2}) the functional derivative of $\underline{%
Q}_{p}$ is defined as 
\begin{eqnarray}
{\normalsize \dfrac{\delta \underline{Q}_{p}}{\delta \eta _{\mu _{1}}(l_{1})}%
} &\equiv &{\normalsize \underset{\Delta L\rightarrow 0}{\lim }\dfrac{1}{%
\Delta L}\dfrac{\partial \underline{Q}_{p}}{\partial \eta _{\mu _{1}}(l_{1})}%
=}  \notag \\
{\normalsize -i\underline{e}_{\mu _{1}}\underset{\Delta L\rightarrow 0}{\lim 
}\dfrac{1}{\Delta L}\delta _{pp_{1}}} &=&{\normalsize -i\underline{e}_{\mu
_{1}}\delta (l_{1}-L_{p}),}\label{afgl-Qp}
\end{eqnarray}%
using Eq. (\ref{qp}). Therefore in Eq. (\ref{1e-afgl-Dthildep1+1.2}) the
first, second and fourth terms are zero, because $\widetilde{\underline{C}}%
_{p_{1}-1}$, $\underline{Q}_{p_{1}-1}$ and $\underline{Q}_{p_{1}+1}$ do not
depend on $\eta _{\mu _{1}}(l_{1})$. In the third term Eq. (\ref{Cp1}) and (%
\ref{afgl-Qp}) can be inserted which yields, 
\begin{equation}
{\normalsize \dfrac{\delta \widetilde{\underline{C}}_{p_{1}+1}}{\delta \eta
_{\mu _{1}}(l_{1})}=-2\dfrac{i^{2}}{2}\underline{e}_{\mu _{1}}\underset{%
\Delta L\rightarrow 0}{\lim }\Delta L\dfrac{1}{\Delta L}=\underline{e}_{\mu
_{1}}.}  \label{1e-afgl-Dthildep1+1.3}
\end{equation}%
In the the same way it follows that 
\begin{eqnarray}
{\normalsize \dfrac{\delta \widetilde{\underline{C}}_{p_{1}}}{\delta \eta
_{\mu _{1}}(l_{1})}} &=&{\normalsize \dfrac{1}{2}\underline{e}_{\mu _{1}}} 
\label{1e-afgl-Dthildep1} \\
&&\text{and}  \notag \\
{\normalsize \dfrac{\delta \widetilde{\underline{C}}_{p}}{\delta \eta _{\mu
_{1}}(l_{1})}} &=&{\normalsize 0\ \ \ }\text{for}{\normalsize \ p<p_{1}.} 
\label{1e-afgl-Dthildep}
\end{eqnarray}%
If $p>p_{1}+1$, then the first order derivative is not zero. For $p\leq
p^{\prime }$ Eq. (\ref{Dpthilde}) can be applied to calculate this first
order derivative, 
\begin{equation}
{\normalsize \frac{\delta \underline{\widetilde{C}}_{p}}{\delta \eta _{\mu
_{1}}(l_{1})}=C_{p}\frac{\delta \underline{Q}_{p}}{\delta \eta _{\mu
_{1}}(l_{1})}+C_{p-1}\frac{\delta \underline{Q}_{p-1}}{\delta \eta _{\mu
_{1}}(l_{1})}+\frac{B_{p-1}}{\widetilde{A}_{p-1}}\frac{\delta \underline{%
\widetilde{C}}_{p-1}}{\delta \eta _{\mu _{1}}(l_{1})}.}  
\label{1e-afgl-Dthildep.2}
\end{equation}%
The first and second term cancel out, because $\underline{Q}_{p}$ and $%
\underline{Q}_{p-1}$ are independent of $\eta _{\mu _{1}}(l_{1})$. The same
terms cancel out in the first order derivative of $\underline{\widetilde{C}}%
_{p-1}$, $\underline{\widetilde{C}}_{p-2}$, $\underline{\widetilde{C}}_{p-3}$
until $\underline{\widetilde{C}}_{p_{1}+2}$. Then it follows that, 
\begin{gather}
\frac{\delta \underline{\widetilde{C}}_{p}}{\delta \eta _{\mu _{1}}(l_{1})}=%
\frac{B_{p-1}}{\widetilde{A}_{p-1}}\frac{B_{p-2}}{\widetilde{A}_{p-2}}......%
\frac{B_{p_{1}+1}}{\widetilde{A}_{p_{1}+1}}\frac{\delta \underline{%
\widetilde{C}}_{p_{1}+1}}{\delta \eta _{\mu _{1}}(l_{1})}=  \notag \\
\frac{\delta \underline{\widetilde{C}}_{p_{1}+1}}{\delta \eta _{\mu
_{1}}(l_{1})}\overset{p-p_{1}-1}{\underset{n=1}{\prod }}\frac{B_{p-n}}{%
\widetilde{A}_{p-n}}=\underline{e}_{\mu _{1}}\overset{p-p_{1}-1}{\underset{%
n=1}{\prod }}\frac{B_{p-n}}{\widetilde{A}_{p-n}},  
\label{1e-afgl-Dthildep.3}
\end{gather}%
if $p_{1}+1<p<p_{2}+1$. If $p>p_{2}+1$, then 
\begin{equation}
\frac{\delta \underline{\widetilde{C}}_{p}}{\delta \eta _{\mu _{1}}(l_{1})}=%
\frac{B_{p-1}}{\widetilde{A}_{p-1}}\frac{B_{p-2}}{\widetilde{A}_{p-2}}......%
\frac{B_{p_{2}+1}}{\widetilde{A}_{p_{2}+1}}\frac{\delta \underline{%
\widetilde{C}}_{p_{2}+1}}{\delta \eta _{\mu _{1}}(l_{1})},  
\label{1e-afgl-Dthildep.4}
\end{equation}%
in which the derivative of $\underline{\widetilde{C}}_{p_{2}+1}$ is the same
as Eq. (\ref{1e-afgl-Dthildep1+1.2}), 
\begin{gather}
{\normalsize \dfrac{\delta \widetilde{\underline{C}}_{p_{2}+1}}{\delta \eta
_{\mu _{1}}(l_{1})}=\frac{B_{p_{2}-1}}{\widetilde{A}_{p_{2}-1}}\dfrac{\delta 
\widetilde{\underline{C}}_{p_{2}-1}}{\delta \eta _{\mu _{1}}(l_{1})}%
+C_{p_{2}-1}\dfrac{\delta \underline{Q}_{p_{2}-1}}{\delta \eta _{\mu
_{1}}(l_{1})}+}  \notag \\
{\normalsize 2C_{p_{2}}\dfrac{\delta \underline{Q}_{p_{2}}}{\delta \eta
_{\mu _{1}}(l_{1})}+C_{p_{2}+1}\dfrac{\delta \underline{Q}_{p_{2}+1}}{\delta
\eta _{\mu _{1}}(l_{1})}.}  \label{1e-afgl-Dthildep2+1}
\end{gather}%
\ If $l_{1}\neq l_{2}$, then $p_{1}$ is not equal to $p_{2}-1$, $p_{2}$ and $%
p_{2}+1.$ Therefore the second, third and fourth term are zero and only the
first term remains in Eq. (\ref{1e-afgl-Dthildep2+1}). Combining Eq.(\ref%
{1e-afgl-Dthildep.3}), Eq. (\ref{1e-afgl-Dthildep.4}) and Eq. (\ref%
{1e-afgl-Dthildep2+1}) yields, 
\begin{equation}
\frac{\delta \underline{\widetilde{C}}_{p}}{\delta \eta _{\mu _{1}}(l_{1})}=%
\underline{e}_{\mu _{1}}\overset{p-p_{2}-1}{\underset{n=1}{\prod }}\frac{%
B_{p-n}}{\widetilde{A}_{p-n}}\overset{p_{2}-p_{1}-1}{\underset{n=1}{\prod }}%
\frac{B_{p_{2}-n}}{\widetilde{A}_{p_{2}-n}}=\underline{e}_{\mu _{1}}\overset{%
p-p_{1}-1}{\underset{n\neq p-p_{2}}{\underset{n=1,}{\prod }}}\frac{B_{p-n}}{%
\widetilde{A}_{p-n}},  \label{1e-afgl-Dthildep.5}
\end{equation}%
In this form Eq. (\ref{1e-afgl-Dthildep.5}) has the same form as Eq. (\ref%
{1e-afgl-Dthildep.3}). The factor in which $n=p-p_{2}$ has been omitted. If $p=p^{\prime
}+1 $, then Eq. (\ref{1e-afgl-Dthildep.5}) can also be applied. This follows
from Eq.(\ref{Dp'+1thilde}). For every $p$ the second order derivative is
always zero, 
\begin{equation}
{\normalsize \frac{\delta ^{2}\underline{\widetilde{C}}_{p}}{\delta \eta
_{\mu _{1}}(l_{1})\delta \eta _{\nu _{1}}(l_{1})}=0,\ \ \ \forall p.} 
\label{2e-afgl-Dthildep}
\end{equation}%
The first and second order derivatives of $\underline{\widetilde{C}}$ are
used to calculate and simplify the derivatives of $\widetilde{D}$ for every $%
p$. If $p=p_{1}+1$, the first order functional derivative of $\widetilde{D}%
_{p_{1}+1}$ is, 
\begin{equation}
{\normalsize \dfrac{\delta \widetilde{D}_{p_{1}+1}}{\delta \eta _{\mu
_{1}}(l_{1})}=\dfrac{\delta \widetilde{D}_{p_{1}}}{\delta \eta _{\mu
_{1}}(l_{1})}+2D_{p_{1}+1}\underline{Q}_{p_{1}+1}\cdot \ \dfrac{\delta 
\underline{Q}_{p_{1}+1}}{\delta \eta _{\mu _{1}}(l_{1})},}  
\label{1e-afgl-Ethildep1+1}
\end{equation}%
using Eq. (\ref{Ethildep1+1}). Because $\underline{Q}_{p_{1}+1}$ does not
depend on $\eta _{\mu _{1}}(l_{1})$, Eq. (\ref{1e-afgl-Ethildep1+1})
becomes, 
\begin{equation}
{\normalsize \dfrac{\delta \widetilde{D}_{p_{1}+1}}{\delta \eta _{\mu
_{1}}(l_{1})}=\dfrac{\delta \widetilde{D}_{p_{1}}}{\delta \eta _{\mu
_{1}}(l_{1})}.}  \label{1e-afgl-Ethildep1+1.2}
\end{equation}%
Using Eq. (\ref{Epthilde}), Eq. (\ref{1e-afgl-Ethildep1+1.2}) becomes 
\begin{equation}
{\normalsize \dfrac{\delta \widetilde{D}_{p_{1}+1}}{\delta \eta _{\mu
_{1}}(l_{1})}=2D_{p_{1}}\underline{Q}_{p_{1}}\cdot \ \dfrac{\delta 
\underline{Q}_{p_{1}}}{\delta \eta _{\mu _{1}}(l_{1})}+\dfrac{\delta 
\widetilde{D}_{p_{1}-1}}{\delta \eta _{\mu _{1}}(l_{1})}-\frac{2\widetilde{%
\underline{C}}_{p_{1}-1}}{\widetilde{A}_{p_{1}-1}}\cdot \dfrac{\delta 
\widetilde{\underline{C}}_{p_{1}-1}}{\delta \eta _{\mu _{1}}(l_{1})}.} 
\label{1e-afgl-Ethildep1+1.3}
\end{equation}%
The derivatives of $\widetilde{D}_{p_{1}-1}$ and $\widetilde{\underline{C}}%
_{p_{1}-1}$ are zero, so only the first term of Eq. (\ref%
{1e-afgl-Ethildep1+1.3}) remains. In the first term Eq. (\ref{Dp1}) and (\ref%
{afgl-Qp}) can be inserted so that 
\begin{equation}
{\normalsize \dfrac{\delta \widetilde{D}_{p_{1}+1}}{\delta \eta _{\mu
_{1}}(l_{1})}=\underset{\Delta L\rightarrow 0}{\lim }2}\left( {\normalsize -}%
\dfrac{1}{18}\dfrac{(\Delta L)^{3}}{\lambda _{p_{1}}}\right) {\normalsize 
\underline{Q}_{p_{1}}\cdot }\left( -i{\normalsize \underline{e}_{\mu _{1}}%
\dfrac{1}{\Delta L}}\right) {\normalsize =\underset{\Delta L\rightarrow 0}{%
\lim }\dfrac{i(\Delta L)^{2}}{9\lambda _{p_{1}}}\underline{e}_{\mu
_{1}}\cdot \underline{Q}_{p_{1}}.\ }  
\label{1e-afgl-Ethildep1+1.4}
\end{equation}%
If we take the limit $\Delta L\rightarrow 0$, then the first order
derivative of $\widetilde{D}_{p_{1}+1}$ goes to zero. The second order
derivative of $\widetilde{D}_{p_{1}+1}$ becomes, 
\begin{gather}
\dfrac{\delta ^{2}\widetilde{D}_{p_{1}+1}}{\delta \eta _{\mu
_{1}}(l_{1})\delta \eta _{\nu _{1}}(l_{1})}=\underset{\Delta L\rightarrow 0}{%
\lim }\dfrac{\delta }{\delta \eta _{\nu _{1}}(l_{1})}\left( {\normalsize 
\dfrac{i(\Delta L)^{2}}{9\lambda _{p_{1}}}\underline{e}_{\mu _{1}}\cdot 
\underline{Q}_{p_{1}}}\right) =  \notag \\
\underset{\Delta L\rightarrow 0}{\lim }\left( {\normalsize \dfrac{i(\Delta
L)^{2}}{9\lambda _{p_{1}}}\underline{e}_{\mu _{1}}}\right) \cdot \left( 
{\normalsize -i\underline{e}_{\nu _{1}}\dfrac{1}{\Delta L}}\right) 
{\normalsize =\underset{\Delta L\rightarrow 0}{\lim }}\dfrac{{\normalsize %
\Delta L}}{9\lambda _{p_{1}}}{\normalsize \delta }_{\mu _{1}\nu _{1}}%
{\normalsize ,}  \label{2e-afgl-Ethildep1+1}
\end{gather}%
which is also going to zero. In general if $p\leq p_{1}+1$, then the first
and second order derivatives of $\widetilde{D}_{p}$ are zero, 
\begin{eqnarray}  \label{Dthilde_1st_order_derivative}
{\normalsize \frac{\delta \widetilde{D}_{p}}{\delta \eta _{\mu _{1}}(l_{1})}}
&=&{\normalsize 0,\ \ \ }\\   
&&\text{and}  \notag \\
{\normalsize \dfrac{\delta ^{2}\widetilde{D}_{p}}{\delta \eta _{\mu
_{1}}(l_{1})\delta \eta _{\nu _{1}}(l_{1})}} &=&{\normalsize \ 0,\ \ \ }%
\text{for}{\normalsize \ p\leq p_{1}+1.}  \label{Dthilde_2th_order_derivative}
\end{eqnarray}%
However, these derivatives are in general nonzero for $p>p_{1}+1$. From Eq. (%
\ref{Epthilde}) it follows that, 
\begin{equation}
{\normalsize \frac{\delta \widetilde{D}_{p}}{\delta \eta _{\mu _{1}}(l_{1})}%
=2D_{p}\underline{Q}_{p}\cdot \frac{\delta \underline{Q}_{p}}{\delta \eta
_{\mu _{1}}(l_{1})}-\frac{2\underline{\widetilde{C}}_{p-1}}{\widetilde{A}%
_{p-1}}\cdot \frac{\delta \underline{\widetilde{C}}_{p-1}}{\delta \eta _{\mu
_{1}}(l_{1})}+\frac{\delta \widetilde{D}_{p-1}}{\delta \eta _{\mu
_{1}}(l_{1})}.}  \label{1e-afgl-Ethildep}
\end{equation}%
If $p>p_{1}+1$, then the first term cancels out. In the second term Eq. (\ref%
{1e-afgl-Dthildep.3}) can be applied if $p>p_{1}+2$. 
\begin{equation}
{\normalsize \frac{\delta \widetilde{D}_{p}}{\delta \eta _{\mu _{1}}(l_{1})}=%
\frac{\delta \widetilde{D}_{p-1}}{\delta \eta _{\mu _{1}}(l_{1})}-\frac{2%
\underline{\widetilde{C}}_{p-1}}{\widetilde{A}_{p-1}}\cdot \underline{e}%
_{\mu _{1}}\underset{n=2}{\overset{p-p_{1}-1}{\prod }}\frac{B_{p-n}}{%
\widetilde{A}_{p-n}}.\ }  \label{1e-afgl-Ethildep.2}
\end{equation}%
If $p=p_{1}+2$, then the first order derivative of $\widetilde{D}_{p}$ is, 
\begin{eqnarray}
{\normalsize \frac{\delta \widetilde{D}_{p_{1}+2}}{\delta \eta _{\mu
_{1}}(l_{1})}} &=&{\normalsize \frac{\delta \widetilde{D}_{p_{1}+1}}{\delta
\eta _{\mu _{1}}(l_{1})}-\frac{2\underline{\widetilde{C}}_{p_{1}+1}}{%
\widetilde{A}_{p_{1}+1}}\cdot \underline{e}_{\mu _{1}}}  \notag \\
&=&{\normalsize -\frac{2\underline{\widetilde{C}}_{p_{1}+1}}{\widetilde{A}%
_{p_{1}+1}}\cdot \underline{e}_{\mu _{1}},}
\label{1e-afgl-Dthildep.6}
\end{eqnarray}%
using Eq. (\ref{1e-afgl-Dthildep1+1.3}) and (\ref{Dthilde_1st_order_derivative}). If $p=p^{\prime }+1$ or $p^{\prime
}+2$ it becomes more difficult to calculate the first order derivative of $%
\widetilde{D}_{p}$. The following equations can be derived, 
\begin{gather}
{\normalsize \frac{\delta \widetilde{D}_{p^{\prime }+1}}{\delta \eta _{\mu
_{1}}(l_{1})}}={\normalsize \frac{\delta \widetilde{D}_{p^{\prime }-1}}{%
\delta \eta _{\mu _{1}}(l_{1})}-\frac{2\underline{\widetilde{C}}_{p^{\prime
}-1}}{\widetilde{A}_{p^{\prime }-1}}\cdot \underline{e}_{\mu _{1}}\underset{%
n=2}{\overset{p^{\prime }-p_{1}-1}{\prod }}\frac{B_{p^{\prime }-n}}{%
\widetilde{A}_{p^{\prime }-n}},}  \notag \\
\text{{\normalsize if }}p_{2}=p^{\prime }>p_{1}+2\text{,}  
\label{1e-afgl-Ethildep'+1.1} \\
{\normalsize \frac{\delta \widetilde{D}_{p^{\prime }+1}}{\delta \eta _{\mu
_{1}}(l_{1})}}={\normalsize -\frac{2\underline{\widetilde{C}}_{p^{\prime }-1}%
}{\widetilde{A}_{p^{\prime }-1}}\cdot \underline{e}_{\mu _{1}},}\text{ if }%
p_{2}=p^{\prime }=p_{1}+2  \label{1e-afgl-Ethildep'+1.2} \\
{\normalsize \frac{\delta \widetilde{D}_{p^{\prime }+1}}{\delta \eta _{\mu
_{1}}(l_{1})}}={\normalsize 0,}\text{ if }p_{2}=p^{\prime }=p_{1}\text{,} 
\label{1e-afgl-Ethildep'+1.3} \\
\frac{\delta \widetilde{D}_{p^{\prime }+1}}{\delta \eta _{\mu _{1}}(l_{1})}=%
{\normalsize \frac{2\underline{\widetilde{C}}_{p^{\prime }}}{\widetilde{A}%
_{p^{\prime }}}\cdot \underline{e}_{\mu _{1}},}\text{ if }p_{2}=p^{\prime
}-1=p_{1}\text{,}  \label{1e-afgl-Ethildep'+1.5} \\
\frac{\delta \widetilde{D}_{p^{\prime }+1}}{\delta \eta _{\mu _{1}}(l_{1})}=%
\frac{\delta \widetilde{D}_{p^{\prime }}}{\delta \eta _{\mu _{1}}(l_{1})}%
{\normalsize -\frac{2\underline{\widetilde{C}}_{p^{\prime }}}{\widetilde{A}%
_{p^{\prime }}}\cdot \underline{e}_{\mu _{1}}\underset{n=2}{\overset{%
p^{\prime }-p_{1}-1}{\prod }}\frac{B_{p^{\prime }-n}}{\widetilde{A}%
_{p^{\prime }-n}},}  \notag \\
\text{ if }p_{2}=p^{\prime }-1,p_{2}>p_{1}\text{,}  
\label{1e-afgl-Ethildep'+1.6} \\
\frac{\delta \widetilde{D}_{p^{\prime }+1}}{\delta \eta _{\mu _{1}}(l_{1})}=%
\frac{\delta \widetilde{D}_{p^{\prime }}}{\delta \eta _{\mu _{1}}(l_{1})}%
{\normalsize -\frac{2\underline{\widetilde{C}}_{p^{\prime }}}{\widetilde{A}%
_{p^{\prime }}}\cdot \underline{e}_{\mu _{1}}\underset{n=1}{\overset{%
p^{\prime }-p_{1}-1}{\prod }}\frac{B_{p^{\prime }-n}}{\widetilde{A}%
_{p^{\prime }-n}},}  \notag \\
\text{ if }p^{\prime }>p_{2}+1\text{,}  
\label{1e-afgl-Ethildep'+1.7} \\
{\normalsize \frac{\delta \widetilde{D}_{p^{\prime }+2}}{\delta \eta _{\mu
_{1}}(l_{1})}}={\normalsize -\frac{2\underline{\widetilde{C}}_{p^{\prime }}}{%
\widetilde{A}_{p^{\prime }+1}}\cdot \underline{e}_{\mu _{1}},}\text{ if }%
p_{2}=p^{\prime }=p_{1}\text{,\ }  \label{1e-afgl-Ethildep'+2.1}
\\
{\normalsize \frac{\delta \widetilde{D}_{p^{\prime }+2}}{\delta \eta _{\mu
_{1}}(l_{1})}}={\normalsize -\frac{2\underline{\widetilde{C}}_{p^{\prime }}}{%
\widetilde{A}_{p^{\prime }+1}}\cdot \underline{e}_{\mu _{1}}\underset{n=1}{%
\overset{p^{\prime }-p_{1}-1}{\prod }}\frac{B_{p^{\prime }-n}}{\widetilde{A}%
_{p^{\prime }-n}},}  \notag \\
\text{ if }p_{2}=p^{\prime }>p_{1}\text{ and}  
\label{1e-afgl-Ethildep'+2.2} \\
\frac{\delta \widetilde{D}_{p^{\prime }+2}}{\delta \eta _{\mu _{1}}(l_{1})}=%
\frac{\delta \widetilde{D}_{p^{\prime }+1}}{\delta \eta _{\mu _{1}}(l_{1})}%
{\normalsize -\frac{2\underline{\widetilde{C}}_{p^{\prime }+1}}{\widetilde{A}%
_{p^{\prime }+1}}\cdot \underline{e}_{\mu _{1}}\underset{n=1}{\overset{%
p^{\prime }-p_{1}}{\prod }}\frac{B_{p^{\prime }+1-n}}{\widetilde{A}%
_{p^{\prime }+1-n}},}  \notag \\
\text{ if }p^{\prime }>p_{2}\text{.}  
\label{1e-afgl-Ethildep'+2.3}
\end{gather}%
In Eq. (\ref{1e-afgl-Ethildep'+2.1}) and (\ref{1e-afgl-Ethildep'+2.2}) $%
\underline{\widetilde{C}}_{p^{\prime }}\rightarrow C_{p^{\prime }-1}%
\underline{Q}_{p^{\prime }-1}+\dfrac{B_{p^{\prime }-1}\underline{\widetilde{C%
}}_{p^{\prime }-1}}{\widetilde{A}_{p^{\prime }-1}}$ in the limit $\Delta
L\rightarrow 0$. In Eq. (\ref{1e-afgl-Ethildep.2}) a differentiation can be
carried out a second time with respect to $\eta _{\nu _{1}}(l_{1})$, which
leads to 
\begin{gather}
{\normalsize \dfrac{\delta ^{2}\widetilde{D}_{p}}{\delta \eta _{\mu
_{1}}(l_{1})\delta \eta _{\nu _{1}}(l_{1})}=}  \notag \\
{\normalsize \dfrac{\delta ^{2}\widetilde{D}_{p-1}}{\delta \eta _{\mu
_{1}}(l_{1})\delta \eta _{\nu _{1}}(l_{1})}-\frac{2}{\widetilde{A}_{p-1}}%
\delta _{\mu _{1}\nu _{1}}\underset{n=2}{\overset{p-p_{1}-1}{\prod }}\left( 
\frac{B_{p-n}}{\widetilde{A}_{p-n}}\right) ^{2}}\text{, if }p>p_{1}+2\text{%
{\normalsize ,}}  \label{2e-afgl-Ethildep.1.1} \\
\text{and}  \notag \\
{\normalsize \dfrac{\delta ^{2}\widetilde{D}_{p}}{\delta \eta _{\mu
_{1}}(l_{1})\delta \eta _{\nu _{1}}(l_{1})}=-\frac{2%
}{\widetilde{A}_{p-1}}\delta _{\mu _{1}\nu _{1}}}\text{, if }p=p_{1}+2\text{.%
}  \label{2e-afgl-Ethildep.1.2}
\end{gather}%
If Eq. (\ref{1e-afgl-Ethildep.2}) is differentiated with respect to $\eta
_{\nu _{2}}(l_{2})$, then there are several possibilities, 
\begin{gather}
{\normalsize \dfrac{\delta ^{2}\widetilde{D}_{p}}{\delta \eta _{\mu
_{1}}(l_{1})\delta \eta _{\nu _{2}}(l_{2})}=0,\ \ \ }\text{if}{\normalsize \
p\leq p_{2}+1}  \label{2e-afgl-Ethildep.3} \\
{\normalsize \dfrac{\delta ^{2}\widetilde{D}_{p_{2}+2}}{\delta \eta _{\mu
_{1}}(l_{1})\delta \eta _{\nu _{2}}(l_{2})}=}-{\normalsize \frac{2}{%
\widetilde{A}_{p_{2}+1}}\delta _{\mu _{1}\nu _{2}}\underset{n=2}{\overset{%
p_{2}-p_{1}}{\prod }}\frac{B_{p_{2}+1-n}}{\widetilde{A}_{p_{2}+1-n}},\ \ \ }
\notag \\
\text{if}{\normalsize \ p=p_{2}+2}\text{ and }p_{1}<p_{2}-1  
\label{2e-afgl-Ethildep.4.1} \\
{\normalsize \dfrac{\delta ^{2}\widetilde{D}_{p_{2}+2}}{\delta \eta _{\mu
_{1}}(l_{1})\delta \eta _{\nu _{2}}(l_{2})}=}-{\normalsize \frac{2}{%
\widetilde{A}_{p_{2}+1}}\delta _{\mu _{1}\nu _{2}},\ \ \ }\text{if}%
{\normalsize \ p=p_{2}+2}\text{ and }p_{1}=p_{2}  
\label{2e-afgl-Ethildep.4.2} \\
\dfrac{\delta ^{2}\widetilde{D}_{p}}{\delta \eta _{\mu _{1}}(l_{1})\delta
\eta _{\nu _{2}}(l_{2})}={\normalsize \dfrac{\delta ^{2}\widetilde{D}_{p-1}}{%
\delta \eta _{\mu _{1}}(l_{1})\delta \eta _{\nu _{2}}(l_{2})}-\frac{2}{\widetilde{A}_{p-1}}\delta _{\mu _{1}\nu _{2}}{\times}} \notag \\  
{\normalsize \underset{n=2}{\overset{p-p_{1}-1}{\prod }}\frac{B_{p-n}}{\widetilde{A}_{p-n}} \underset{n=2}{\overset{p-p_{2}-1}{\prod }}\frac{B_{p-n}}{%
\widetilde{A}_{p-n}},\ \ \ }\text{if}{\normalsize \ p>p_{2}+2.}  
\label{2e-afgl-Ethildep.5}
\end{gather}%
It can be verified that Eq. (\ref{2e-afgl-Ethildep.1.1}), (\ref%
{2e-afgl-Ethildep.1.2}), (\ref{2e-afgl-Ethildep.3}), (\ref%
{2e-afgl-Ethildep.4.1}), (\ref{2e-afgl-Ethildep.4.2}) and (\ref%
{2e-afgl-Ethildep.5}) are the same if $p=p^{\prime }+1$ or $p^{\prime }+2$.
So now we also have found the second order derivatives of $\widetilde{D}%
_{p^{\prime }+2}$. The first order derivatives are given by Eq. (\ref%
{1e-afgl-Ethildep'+2.1}), (\ref{1e-afgl-Ethildep'+2.2}) and (\ref%
{1e-afgl-Ethildep'+2.3}). These can be used to calculate the factors $G_{2}$ and $G_{4}$ given by Eq. (\ref{G2.1}) and (\ref{G4.4}). The second order derivatives do not depend on $%
\underline{Q}_{p}$ any more. So third and higher order derivatives are
zero.



\end{document}